\documentclass[12pt]{article}
\usepackage{amsmath}
\usepackage{amsthm}
\usepackage{graphicx}
\usepackage{amssymb}
\usepackage{cite}
\usepackage{xcolor}

\numberwithin{equation}{section} \allowdisplaybreaks

\newtheorem{theorem}{Theorem}[section]
\newtheorem{lemma}{Lemma}[section]
\newtheorem{definition}{Definition}[section]

\renewcommand{\eqref}[1]{(\ref{#1})}

\begin{document}
	\title{On the fault diameter and wide diameter of the exchanged 3-ary $n$-cube}
	\author{Rongshuan Geng, Wantao Ning \thanks{Corresponding author}\\
		{\scriptsize  School of Mathematics and Statistics, Xidian University, Xi'an, Shaanxi, 710071,  P.R.  China}\\
		\mbox{}\hspace{0.15cm}{\scriptsize E-mails: rsgeng@stu.xidian.edu.cn, wtning@xidian.edu.cn}}
	\date{}
	
	\maketitle
	
	\begin{abstract}
	Fault diameter and wide diameter are two critical parameters for evaluating communication performance in interconnection networks. They measure the fault tolerance and transmission efficiency of networks. The exchanged 3-ary $n$-cube is a recently proposed variant of the hypercube, denoted by $E3C(r,s,t)$. In this work, we obtain that the ($2r+1$)-fault diameter and ($2r+2$)-wide diameter of $E3C(r,s,t)$ are bounded between $n+3$ and $n+5$ for $1\leq r\leq s\leq t$.\\
		

	\end{abstract}
	
	\section{Introduction}
	
	Interconnection networks are fundamental to the architecture of modern computing systems, playing a vital role in multiprocessor systems. The probability of system failure increases dramatically as the number of processors increases. Processor failures directly impact the system's fault tolerance. Therefore, fault tolerance should be a key consideration when designing a network topology for multiprocessor systems. For instance, in network design, vertices can represent computers, routers, or other devices, while edges represent the connections between them. This aids in optimizing network topologies, efficiently routing data, and detecting faults. Let $G = (V(G), E(G))$ be a graph, where $V(G)$ and $E(G)$ denote the vertex set and edge set respectively. For $v\in V (G)$, the neighborhood of $v$ in $G$ is denoted by $N(v)$. $d(v)=|N(v)|$ is the degree of $v$. $\delta(G)$ = min $\{d(v)\,|v \in V(G)\}$. For graphs $G_{1}$ and $G_{2}$, $G_{1}\cong G_{2}$ indicates that they are isomorphic. A path $P$ of length $k$ from $v_{0}$ to $v_{k}$ in $G$ can be represented as a sequence of distinct vertices, denoted by $v_{0}\rightarrow v_{1}\rightarrow v_{2}\rightarrow \cdots \rightarrow v_{k}$, where $v_{i-1}v_{i}\in E(G)$ for $1\leq i\leq k$. For simplicity, $P$ can also be represented as $v_{0}\rightarrow v_{1}\rightarrow v_{2}\rightarrow \cdots \rightarrow v_{i}\rightarrow H\rightarrow v_{j}\rightarrow \cdots \rightarrow v_{k}$, where $H$ is a subpath from $v_{i}$ to $v_{j}$ in $P$. The length of $P$ is denoted by $|P|$. The length of the shortest path between any two distinct vertices $u$ and $v$ in $G$ is denoted by $d(u,v)$. Then the diameter of $G$ can be defined as $D(G)$=max $\{d(u,v)|u,v\in V(G)\}$. Two paths from $u$ to $v$ in $G$ are internally disjoint if they have no other common vertices. For more graph theory notations, please refer to \cite{Bondy+Murty}.

Internally disjoint paths between distinct vertices enable multipath communication, ensuring secure and efficient data transmission while providing redundant routing alternatives in cases of vertex or link failures. These paths serve as the theoretical foundation for defining critical network robustness
metrics, such as connectivity, wide diameter, and fault diameter, which quantify network resilience based on the properties of vertex disjoint paths. A vertex cut of graph $G$ is defined as a set $S\subseteq V(G)$ such that $G-S$ is disconnected or has only one vertex. The connectivity of $G$ , denoted by $\kappa(G)$, is defined as the minimum size over all vertex cuts.  According to Menger's theorem \cite{Faudree+Ordman+Shelp+Jacobson}, there  exist at least $\kappa(G)$ internally disjoint paths between any two vertices in $G$, forming the theoretical foundation for multipath communication and fault tolerant routing.
	
	Hypercube $Q_{n}$ is one of the most popular interconnection networks in parallel computing and communication systems owing to its many excellent properties. Many variants have been proposed and studied, such as the crossed cube\cite{Efe}, exchanged crossed cube\cite{Li+Mu+Li+Min}, exchanged hypercube\cite{Lou+Hsu+Pan}, locally twisted cube\cite{Han+You+Lin+Fan}, twisted cube\cite{Hibers+Koopman+Snepscheut} and so on. The fault diameter is defined in \cite{Krishnamoorthy+Krishnamurthy} as the maximum diameter after removing fewer than $\kappa(G)$ vertices from $G$. It measures the worst-case impact on diameter under vertex faults.  Wide diameter was introduced by Flandrin and Li \cite{Flandrin+Li}, and Hsu and Lyuu \cite{Hsu+Lyuu} independently. In \cite{Hsu}, the author proposed the theoretical upper bound $D_{k}(G)\leq n-k+1$ for the wide diameter of $k$-connected graphs with $n$ vertices. Fault diameter measures the data transmission delay in a fault tolerant network, and wide diameter measures the fault tolerance and transmission efficiency of the network. Therefore,  fault diameter and wide diameter are important parameters for designing and evaluating the performance of networks, and their study is highly significant. Scholars have studied fault diameter and wide diameter for many hypercube variants \cite{Chang+Sung+Hsu,Day+Al-Ayyoub,Fu+Chen+Duh,Latifi,Niu+Zhou+Tian+Zhang,Qi+Zhu,Rouskov+Srimani,Tsai+Chen+ Tan,Wu+Chen+Kuo+Chang,Yin+Li+Chen+Zhong}.
	$E3C(r,s,t)$ is a recently proposed  hypercube variant. It is obtained by deleting different types of edges from the $3$-ary $n$-cube ($Q_{n}^{3}$). Therefore, $E3C(r,s,t)$ retains most topological properties of $Q_{n}^{3}$. Compared to $Q_{n}^{3}$ of the same dimension, $E3C(r,s,t)$ has fewer edges and a smaller degree for each vertex, and its diameter exceeds that of $Q_{n}^{3}$ by only 2. In this work, we focus on the fault diameter and wide diameter of $E3C(r,s,t)$ and obtain that the ($2r+1$)-fault diameter and ($2r+2$)-wide diameter of $E3C(r,s,t)$ are bounded between $n+3$ and $n+5$ for $1\leq r\leq s\leq t$.

	\section{Preliminaries}
	
	\subsection{Fault diameter and wide diameter}
	
	First, let's introduce the $\alpha$-fault diameter of graph $G$.

\begin{definition}(\cite{Krishnamoorthy+Krishnamurthy})\label{def1}
Let $\alpha$ be a positive integer satisfying $\alpha \leq \kappa(G)-1$. The $\alpha$-fault distance between distinct vertices $u$ and $v$, denoted by $d_{\alpha}^{f}(u,v)$, is defined as $d_{\alpha}^{f}(u,v)= max \{d_{G-F}(u,v)\mid F\subseteq V(G)\setminus \{u,v\}, |F|=\alpha\}$. The $\alpha$-fault diameter of $G$, denoted by $D_{\alpha}^{f}(G)$, is defined as $D_{\alpha}^{f}(G)= max \{d_{\alpha}^{f}(u,v)\, | \,u,v \in V(G)\}$.
	
\end{definition}

The $\beta$-wide diameter of graph $G$ is formally defined as follows.

\begin{definition}(\cite{Flandrin+Li,Hsu+Lyuu})\label{def2}
Let $\beta$ be a positive integer with $\beta \leq \kappa(G)$. For distinct vertices $u,v\in V(G)$, denote by $D_{\beta}(u,v)$ the set of all $\beta$ internally disjoint paths between $u$ and $v$. Let $l_{i}(u,v)$ be the longest length among the $\beta$ paths of the $i$-th element of $D_{\beta}(u,v)$, and $l_{D_{\beta}}(u,v)= min\{l_{i}(u,v) \, | \,i\leq |D_{\beta}(u,v)| \}$. The $\beta$-wide diameter of $G$, denoted by $D_{\beta}(G)$, is defined as $D_{\beta}(G)=max \{ l_{D_{\beta}}(u,v)| u,v \in V(G)\}$.

\end{definition}

From these two definitions, it is evident that $D_{1}^{f}(G)=D_{1}(G)=D(G)$. $D_{\kappa(G)-1}^{f}(G)$ and $D_{\kappa(G)}(G)$ represent the fault diameter and wide diameter of $G$, respectively. Moreover, we have $D(G)\leq D_{\kappa(G)-1}^{f}(G) \leq D_{\kappa(G)}(G)$.
	
	\subsection{The $k$-ary $n$-cube and its properties}

This section introduces the definition of the $k$-ary $n$-cube and presents its useful properties.

For any positive integers $k$ and $n$ with $k\geq 2$ and $n\geq 1$, the $k$-ary $n$-cube, denoted by $Q_{n}^{k}$, has $k^{n}$ vertices. Each vertex is represented by an $n$-bit string $u_{n-1}u_{n-2}\cdots u_{0}$, where $u_{i} \in \{ 0,1,2,\ldots ,k-1 \}$ for $0\leq i \leq n-1$. Two vertices $u=u_{n-1}u_{n-2}\cdots u_{0}$ and $v=v_{n-1}v_{n-2}\cdots v_{0}$ in $Q_{n}^{k}$ are adjacent if and only if there exists an integer $i$ such that $u_{i}=v_{i}\pm 1 \, (mod \, k)$, and $u_{j}=v_{j}$ for every $j\neq i$.

\begin{definition}(\cite{Bose+Broeg+Kwon+Ashir})\label{def3LeeDis}
Let $B=b_{n-1}b_{n-2}\cdots b_{0}$ be an $n$-bit $k$-ary string. The Lee weight of $B$ is defined as $W_{L}(B)=\sum _{i=0}^{n-1}|b_{i}|$, where $|b_{i}|=min\{b_{i},k-b_{i}\}$.
The Lee distance between two strings $B$ and $C$, denoted by $D_{L}(B,C)$, is defined as $W_{L}(B-C)$.

\end{definition}

Let $D_{H}(B,C)$ be the Hamming distance between two strings $B$ and $C$, i.e., the number of positions in which they differ.  Then $D_{L}(B,C)=D_{H}(B,C)$ when $k=2$ or 3.

For example, when $k=5$, $W_{L}(4321)=1+2+2+1=6$, and $D_{L}(1234,4321)=W_{L}(3113)=6$.

\begin{lemma}(\cite{Bose+Broeg+Kwon+Ashir})\label{lemQnkjiben} $Q_{n}^{k}$ has the following properties.

(1)	$D(Q_{n}^{k})=n\lfloor \frac{k}{2} \rfloor$.

(2)	$\kappa(Q_{n}^{k})=2n$.

(3) The length of the shortest path between any two vertices $u$ and $v$ in $Q_{n}^{k}$ is $D_{L}(u,v)$.	
	\end{lemma}

Suppose $u$ and $v$ are any two distinct vertices in $Q_{n}^{3}$. From Lemma \ref{lemQnkjiben} (3) and $k=3$, we have $D_{L}(u,v)\leq D(Q_{n}^{3})=n\lfloor \frac{3}{2} \rfloor=n$.

\begin{lemma}(\cite{Bose+Broeg+Kwon+Ashir})\label{lemQnklujuli}
Let $u=u_{n-1}u_{n-2}\cdots u_{0}$ and $v=v_{n-1}v_{n-2}\cdots v_{0}$ be two distinct vertices of $Q_{n}^{k}$. Define $l=D_{L}(u,v)$, $h=D_{H}(u,v)$, and $w_{i}=D_{L}(u_{i},v_{i})$ for $0\leq i\leq n-1$. Then there exist $2n$ internally disjoint paths between $u$ and $v$ of which

(1) $h$ paths have length $l$,

(2) $2(n-h)$ paths have length $l+2$, and

(3) For each $i$ satisfying $w_{i}>0$, there exists a path of length $l+k-2w_{i}$ (there are $h$ such paths).

\end{lemma}

Suppose $u$ and $v$ are any two distinct vertices in $Q_{n}^{3}$. By Lemma \ref{lemQnklujuli}, we can deduce that the length of any one of the $2n$ internally disjoint paths from $u$ to $v$ satisfies $|P|\leq D(Q_{n}^{3})+2=n+2$.

\subsection{The exchanged 3-ary $n$-cube}

Based on $Q_{n}^{k}$, the authors proposed the definition of exchanged 3-ary $n$-cube in \cite{Lv+Lin+Wang}.

\begin{definition}(\cite{Lv+Lin+Wang})\label{defE3C}
Let $h(a,b)=0$ if $a=b$, and $h(a,b)=1$ if $a\neq b$. The exchanged 3-ary $n$-cube, denoted by $E3C(r,s,t) = (V, E)$, is a connected graph with parameters $r, s, t \geq 1$ satisfying $n = r + s + t + 1$. $V= \{ a_{r-1} \cdots a_{0}b_{s-1}\cdots b_{0}c_{t-1}\cdots c_{0}d \,| \, a_{i},b_{j},c_{k},d\in \{0,1,2\}, 0\leq i\leq r-1, 0\leq j \leq s-1, 0\leq k \leq t-1\}$. The edge set $E$ partitions into four types $E_{0}$, $E_{1}$, $E_{2}$ and $E_{3}$, which are defined as follows.

(1) $E_{0}= \{(x,y) \, | \, H_{1}^{r+s+t}(x,y)=0 \,and\, x[0]\neq y[0] \}$,

(2) $E_{1}= \{(x,y) \, | \, H_{t+1}^{r+s+t}(x,y)=0,\, H_{1}^{t}=1 \,and\, x[0]= y[0]=0 \}$,

(3) $E_{2}= \{(x,y) \, | \, H_{s+t+1}^{r+s+t}(x,y)=0,\, H_{t+1}^{s+t}=1,\, H_{1}^{t}=0 \,and\, x[0]= y[0]=1 \}$,

(4) $E_{3}= \{(x,y) \, | \, H_{s+t+1}^{r+s+t}(x,y)=1,\, H_{1}^{s+t}=0 \,and\, x[0]= y[0]=2 \}$,

where $H(x,y)=\sum_{i=0}^{r+s+t}h(x[i],y[i])$ denotes the Hamming distance between vertices $x$ and $y$, and
$H_{p}^{q}(x,y)=\sum_{i=p}^{q}h(x[i],y[i])$. Here $x[i]$ denotes the value at dimension $i$ of vertex $x$.

\end{definition}

\begin{figure}[h]
		\centering
		\includegraphics[width=1.0\textwidth]{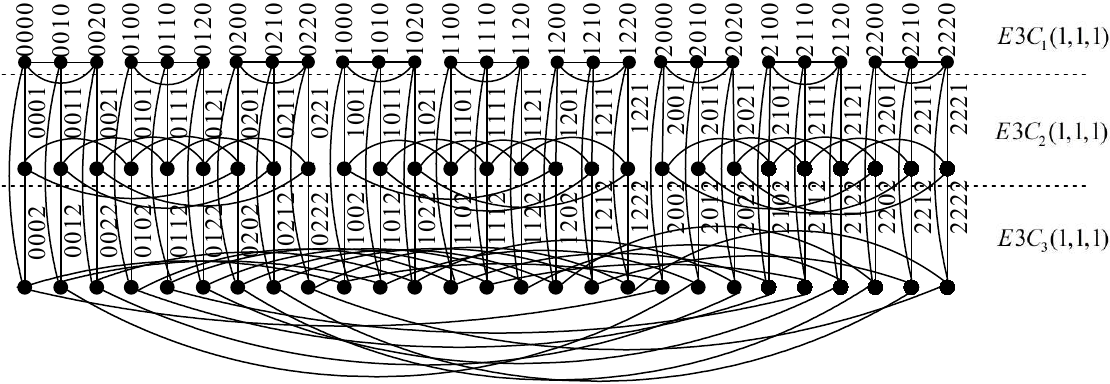}
		\renewcommand{\figurename}{Fig.}
		\caption{$E3C(1,1,1)$}
		\label{E3C}
	\end{figure}

To understand this definition, the authors provided $E3C(1,1,1)$ in \cite{Lv+Lin+Wang} (see Fig. \ref{E3C}). For vertex $a_{0}b_{0}c_{0}d=0000$, following Definition \ref{defE3C} (1), 0001, 0002 are two neighbors of 0000. By Definition \ref{defE3C} (2), 0010, 0020 are the remaining two neighbors of 0000 since $d=0$.

	From the definition of $E3C(r,s,t)$, there has the following lemma.

	\begin{lemma}(\cite{Lv+Lin+Wang})\label{lemjichu} $E3C(r,s,t)$ has the following properties.
		
		(1) $|V(E3C(r,s,t))|=3^{r+s+t+1}=3^{n}$ where $n=r+s+t+1$.
		
		(2) $|E(E3C(r,s,t))|=(r+s+t+3)\cdot 3^{r+s+t}=(n+2)\cdot 3^{n-1}$.

        (3) $D(E3C(r,s,t))=r+s+t+3=n+2$.

        (4) For $u\in V(E3C(r,s,t))$, then \[
\ d(u) =
\begin{cases}
    2t+2, & \text{if } u[0]=0\\
    2s+2, & \text{if } u[0]=1 \\
    2r+2, & \text{if } u[0]=2
\end{cases}
\]

Moreover, $\delta(E3C(r,s,t))=min\{ 2r+2,2s+2,2t+2\}$.
		
	\end{lemma}
	
	The following lemma represents the isomorphism of $E3C(r,s,t)$.
	
	\begin{lemma}(\cite{Lv+Lin+Wang,Ning+Guo})\label{lemisomorphism}
		Let $r,s,t$ be any three positive integers and $\{ r',s',t'\}=\{ r,s,t\}$. Then $E3C(r,s,t)\cong E3C(r',s',t')$.
		
	\end{lemma}

Based on this lemma, we can get the following conclusion. For $r\geq 2$, $E3C(r,s,t)$ can be decomposed into three vertex disjoint subgraphs $E3C_{i}$ for $i=0,1,2$, where each vertex $u \in V(E3C_{i})$ satisfies $u[r+s+t]=i$. Moreover, each subgraph $E3C_{i}$ is isomorphic to $E3C(r-1,s,t)$. Furthermore, $E3C(r,s,t)$ can be divided into $3^{s+t}$ disjoint subgraphs isomorphic to $Q_{r}^{3}$, $3^{r+t}$ disjoint subgraphs isomorphic to $Q_{s}^{3}$, and $3^{r+s}$ disjoint subgraphs isomorphic to $Q_{t}^{3}$.

For the sake of discussion, we divide $E3C(r,s,t)$ into three disjoint subgraphs $L$, $M$, $R$ and edges between them, where every vertex $u\in V(L)$ satisfies $u[0] = 0$, every vertex $v\in V(M)$ satisfies $v[0]=1$, and every vertex $w\in V(R)$ satisfies $w[0] = 2$. $L$ can be decomposed into $3^{r+s}$ subgraphs, denoted as $L_{i}$ for $i=1,2,\ldots,3^{r+s}$. Similarly, $M$ can be decomposed into $3^{r+t}$ subgraphs, denoted as $M_{j}$ for $j=1,2,\ldots,3^{r+t}$. We can also decompose $R$ into $3^{s+t}$ subgraphs, denoted as $R_{k}$ for $k=1,2,\ldots,3^{s+t}$.\\

In \cite{Ning+Guo}, the authors obtained the connectivity of $E3C(r,s,t)$.

\begin{lemma}(\cite{Ning+Guo})\label{lemE3Cconnectivity}
The connectivity of $E3C(r,s,t)$ is $\kappa(E3C(r,s,t))=2r+2$ for $r=min\{ r,s,t\}$.

\end{lemma}

Combining the definition of $E3C(r,s,t)$ and the above lemmas, the following conclusion can be obtained.

\begin{lemma}\label{lemE3Clianbian} For $Q_{n}^{3}$ and $E3C(r,s,t)$, the following results hold.

(1) Each $L_{i}\cong Q_{t}^{3}$, $M_{k}\cong Q_{s}^{3}$ and $R_{p}\cong Q_{r}^{3}$ for $i=1,2,\ldots,3^{r+s}$, $k=1,2,\ldots,3^{r+t}$, $p=1,2,\ldots,3^{s+t}$.

(2) For any two distinct $L_{i}$ and $L_{j}$ there are no edges between them for $i,j \in \{1,2,\ldots,3^{r+s}\}$. Similarly, there are no edges between any two distinct $M_{k}$ and $M_{h}$
for $k,h \in \{1,2,\ldots,3^{r+t}\}$. And there are no edges between any two distinct $R_{p}$ and $R_{q}$
for $p,q \in \{1,2,\ldots,3^{s+t}\}$.

(3) For any vertex $u\in V(L_{i})$ with $i\in \{1,2,\ldots,3^{r+s}\}$, there exists exactly one external neighbor $u'$ in some specific $M_{k}$ and exactly one external neighbor $u''$ in some specific $R_{p}$,  where $k\in \{1,2,\ldots,3^{r+t}\}$ and $p\in \{1,2,\ldots,3^{s+t}\}$, and $u'$ is adjacent to $u''$. For any vertex $v\in V(M_{h})$ with $h\in \{1,2,\ldots,3^{r+t}\}$, there exists exactly one external neighbor $v'$ in some specific $L_{j}$ and exactly one external neighbor $v''$ in some specific $R_{q}$, where $j\in \{1,2,\ldots,3^{r+s}\}$ and $q\in \{1,2,\ldots,3^{s+t}$\}, and $v'$ is adjacent to $v''$. A similar result holds for any vertex $w\in V(R_{i})$ with $i\in \{1,2,\ldots,3^{s+t}\}$.

(4) For two distinct vertices $u, v\in V(L_{i})$ where $i\in \{ 1,2,\ldots,3^{r+s}\}$, $u'$ and $v'$ belong to distinct $V(M_{k})$ and $V(M_{h})$ respectively, for some $k,h \in \{1,2,\ldots,3^{r+t}\}$, $u''$ and $v''$ belong to distinct $V(R_{p})$ and $V(R_{q})$ respectively, for some $p,q \in \{1,2,\ldots,3^{s+t}\}$. Similar results hold for distinct vertices $w$, $z$ in any $V(M_{h})$ $(h\in \{1,2,\ldots,3^{r+t}\})$ and for distinct vertices $x$, $y$ in any $V(R_{p})$ $(p\in \{1,2,\ldots,3^{s+t}\})$.

\end{lemma}

	\section{Fault diameter and wide diameter of $E3C$}
	
By Lemma \ref{lemisomorphism}, without loss of generality, we assume that $r\leq s\leq t$ in this paper.

First, we give the lower bound on the ($2r+1$)-fault diameter of $E3C(r,s,t)$.
	
	\begin{theorem}\label{th1}
		$D_{2r+1}^{f}(E3C(r,s,t))\geq r+s+t+4=n+3$ for $1\leq r\leq s\leq t$.
	\end{theorem}	
	\noindent \textbf{Proof.} Let $u=0^{r}0^{s}0^{t}2$ and $v=1^{r}1^{s}1^{t}0$ be two vertices of $E3C(r,s,t)$. Without loss of generality, assume $u\in V(R_{1})$ and $v\in V(L_{1})$. Then $u'=0^{r}0^{s}0^{t}0$ and $u''=0^{r}0^{s}0^{t}1$ are the two outside neighbors of $u$. $v'=1^{r}1^{s}1^{t}2$ is an outside neighbor of $v$. Assume $u'\in V(L_{2})$, $u''\in V(M_{1})$ and $v'\in V(R_{2})$. Suppose $F=(N(u)\bigcap V(R_{1}))\bigcup \{u''\}$. Then $|F|=2r+1$. The shortest path $P$ from $u$ to $v$ in $E3C-F$ must pass $uu'$. Thus $P$ can be represented as $u\rightarrow u' \rightarrow P'\rightarrow v$, where $P'$ is a shortest path from $u'$ to $v$ in $E3C-F$. Choose $x=0^{r}0^{s}1^{t}0$ in $L_{2}$. Let $\bar{x}=0^{r}0^{s}1^{t}1$, which is an outside neighbor of $x$ in $M_{2}$. Select $y=0^{r}1^{s}1^{t}1$ in $M_{2}$. $y$ has an outside neighbor $\bar{y}=0^{r}1^{s}1^{t}2$ in $R_{2}$. Then $P'$ can be further represented as $u'\rightarrow K\rightarrow x\rightarrow \bar{x}\rightarrow H\rightarrow y\rightarrow  \bar{y}\rightarrow N\rightarrow  v'\rightarrow v$  , where $K$, $H$ and $N$ are the shortest paths from $u'$ to $x$ in $L_{2}$, $\bar{x}$ to $y$ in $M_{2}$ and $\bar{y}$ to $v'$ in $R_{2}$, respectively. According to Lemma \ref{lemQnkjiben} and the analysis under Lemma \ref{lemisomorphism}, it is obtained that $|K|=D_{L}(0^{r}0^{s}0^{t}0,0^{r}0^{s}1^{t}0)=W_{L}(0^{r}0^{s}1^{t}0)=t$, $|H|=D_{L}(0^{r}0^{s}1^{t}1,0^{r}1^{s}1^{t}1)=W_{L}(0^{r}1^{s}0^{t}0)=s$ and $|N|=D_{L}(0^{r}1^{s}1^{t}2,1^{r}1^{s}1^{t}2)=r$. Hence $|P|=4+|K|+|H|+|N|=4+r+s+t=n+3$.

Since $P$ is a shortest path from $u$ to $v$ in $E3C-F$ and $|F|=2r+1$, then $D_{2r+1}^{f}(E3C(r,s,t))\geq r+s+t+4=n+3$ for $1\leq r\leq s\leq t$.   \qed
	\\
	
Based on Lemma \ref{lemE3Cconnectivity} and Menger's theorem, there exist $2r+2$ internally disjoint paths between any two vertices in $E3C(r,s,t)$.
In the following, we construct them in $E3C(r,s,t)$ for $1\leq r\leq s\leq t$. For simplicity, for any vertex $u$ in $E3C$ can be denoted as $u=a_{r-1} \cdots a_{0}b_{s-1}\cdots b_{0}c_{t-1}\cdots c_{0}d=ABCd$, where $A=a_{r-1} \cdots a_{0}$, $B=b_{s-1}\cdots b_{0}$ and $C=c_{t-1}\cdots c_{0}$. Similarly, $v=a'_{r-1} \cdots a'_{0}b'_{s-1}\cdots b'_{0}c'_{t-1}\cdots c'_{0}d'$ is denoted as $v=A'B'C'd'$.
	
	\begin{lemma} \label{fenlei1}
		Let $u=ABCd$ and $v=A'B'C'd'$ be two distinct vertices in $E3C(r,s,t)$ such that $A=A'$ and $B=B'$. We discuss three cases.

\item 1. If $d=d'=0$, there exist $2r+2$ internally disjoint paths between $u$ and $v$ with length at most $t+6$.

\item 2. If $d=d'=1$, there exist $2r+2$ internally disjoint paths between $u$ and $v$ with length at most $t+6$.	

\item 3. If $d=d'=2$, there exist $2r+2$ internally disjoint paths between $u$ and $v$ with length at most $t+6$.
	\end{lemma}
\noindent \textbf{Proof.}  We consider three cases based on the value of $d$.

Case 1. $d=d'=0$.

\begin{figure}[h]
		\centering
		\includegraphics[width=0.5\textwidth]{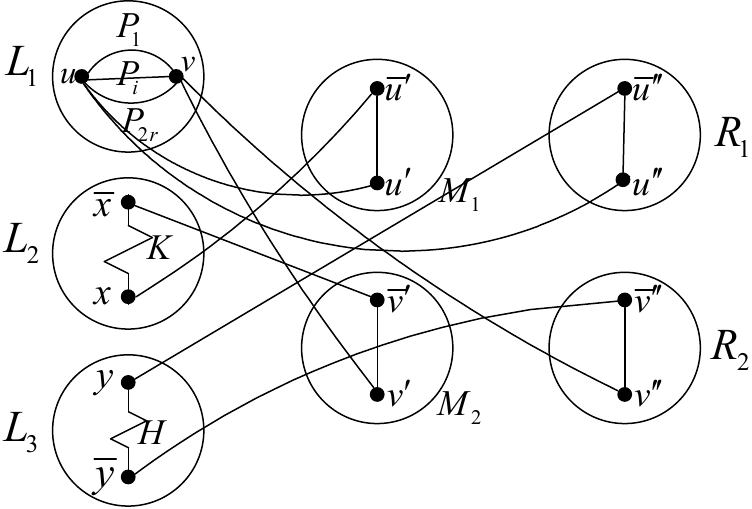}
		\renewcommand{\figurename}{Fig.}
		\caption{The illustration of Case 1 of Lemma \ref{fenlei1}}
		\label{1case1}
	\end{figure}

Let $u=ABC0$ and $v=ABC'0$ be in $L_{1}$ (see Fig. \ref{1case1}). According to Lemma \ref{lemQnklujuli} and its following analysis, there
exist $2r$ internally disjoint paths $P_{1},P_{2}, \ldots ,P_{2r}$ between $u$ and $v$ in $L_{1}$, satisfying
$|P_{i}| \leq D(L_{1})+2=D(Q_{t}^{3})+2=t+2$ for $1\leq i\leq 2r$. Assume that $u'=ABC1$ and $u''=ABC2$ are in $M_{1}$ and $R_{1}$, respectively. Let $v'=ABC'1$ and $v''=ABC'2$ be in $M_{2}$ and $R_{2}$, respectively.
$u'$ has a neighbor $\bar{u}'=AB_{1}C1$ in $M_{1}$, which has an outside neighbor $x=AB_{1}C0$ in $L_{2}$. $v'$ has a neighbor $\bar{v}'=AB_{1}C'1$ in $M_{2}$, which has an outside neighbor $\bar{x}=AB_{1}C'0$ also in $L_{2}$. Let $K$ be a shortest path between $x$ and $\bar{x}$ in $L_{2}$. $u''$ has a neighbor $\bar{u}''=A_{1}BC2$ in $R_{1}$, which has an outside neighbor $y=A_{1}BC0$ in $L_{3}$. $v''$ has a neighbor $\bar{v}''=A_{1}BC'2$ in $R_{2}$, which has an outside neighbor $\bar{y}=A_{1}BC'0$ also in $L_{3}$. Let $H$ be a shortest path between $y$ and $\bar{y}$ in $L_{3}$. Thus, we construct the path $P_{2r+1}$ as $u\rightarrow u'\rightarrow \bar{u}'\rightarrow x\rightarrow K\rightarrow \bar{x}\rightarrow \bar{v}'\rightarrow v'\rightarrow v$. Since $|K|\leq D(L_{2})= D(Q_{t}^{3})=t$, then $|P_{2r+1}|=|K|+6\leq t+6$. Define the path $P_{2r+2}$ as $u\rightarrow u''\rightarrow \bar{u}''\rightarrow y\rightarrow H\rightarrow \bar{y}\rightarrow \bar{v}''\rightarrow v''\rightarrow v$. Since $|H|\leq D(L_{3})= D(Q_{t}^{3})=t$, then $|P_{2r+2}|=|H|+6\leq t+6$.

Therefore, we obtain $2r+2$ internally disjoint paths $P_{1},P_{2}, \ldots ,P_{2r+2}$ between
$u$ and $v$ in $E3C(r,s,t)$, with lengths at most $t+6$.  \\

Case 2. $d=d'=1$.

\begin{figure}[h]
		\centering
		\includegraphics[width=0.5\textwidth]{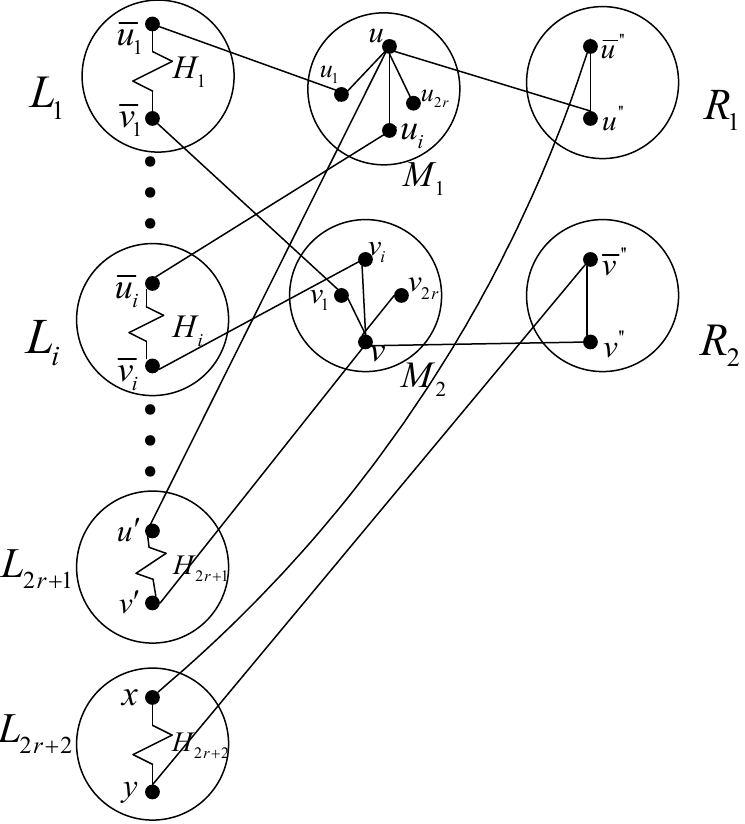}
		\renewcommand{\figurename}{Fig.}
		\caption{The illustration of Case 2 of Lemma \ref{fenlei1}}
		\label{1case2}
	\end{figure}

Let $u=ABC1\in V(M_{1})$ and $v=ABC'1\in V(M_{2})$ (see Fig. \ref{1case2}). Define $u_{i}$ as one of the
$2r$ neighbors of $u$ in $M_{1}$ , where $u_{i}=AB_{i}C1$ for $1\leq i\leq 2r$. Define $v_{i}$ as one of the
$2r$ neighbors of $v$ in $M_{2}$ , where $v_{i}=AB_{i}C'1$ for $1\leq i\leq 2r$. Let $\bar{u}_{i}=AB_{i}C0$ be an outside neighbor of $u_{i}$ in $L_{i}$ for $1\leq i\leq 2r$. $v_{i}$ has an outside neighbor $\bar{v}_{i}=AB_{i}C'0$ also in $L_{i}$ for $1\leq i\leq 2r$. Let $H_{i}$ be a shortest path between $\bar{u}_{i}$ and $\bar{v}_{i}$ in $L_{i}$ for $1\leq i\leq 2r$. Define the path $P_{i}$ as $u\rightarrow u_{i}\rightarrow \bar{u}_{i}\rightarrow H_{i}\rightarrow \bar{v}_{i}\rightarrow v_{i}\rightarrow v$ for $1\leq i\leq 2r$. Since $|H_{i}|\leq D(L_{i})= D(Q_{t}^{3})=t$, then $|P_{i}|=|H_{i}|+4\leq t+4$ for $1\leq i\leq 2r$. Let $u'=ABC0$ be in $L_{2r+1}$. Then $v'=ABC'0$ is also in $L_{2r+1}$. Let $H_{2r+1}$ be a shortest path between $u'$ and $v'$ in $L_{2r+1}$. Define the path $P_{2r+1}$ as $u\rightarrow u'\rightarrow H_{2r+1}\rightarrow v'\rightarrow v$. Since $|H_{2r+1}|\leq D(L_{2r+1})= D(Q_{t}^{3})=t$, then $|P_{2r+1}|=|H_{2r+1}|+2\leq t+2$. Assume that $u''= ABC2$ and $v''= ABC'2$ are in $R_{1}$ and $R_{2}$, respectively. $u''$ has a neighbor $\bar{u}''=A_{1}BC2$ in $R_{1}$, which has an outside neighbor  $x=A_{1}BC0$ in $L_{2r+2}$. $v''$ has a neighbor $\bar{v}''=A_{1}BC'2$ in $R_{2}$, which has an outside neighbor $y=A_{1}BC'0$ also in $L_{2r+2}$. Let $H_{2r+2}$ be a shortest path between $x$ and $y$ in $L_{2r+2}$. Define the path $P_{2r+2}$ as $u\rightarrow u''\rightarrow \bar{u}''\rightarrow x\rightarrow H_{2r+2}\rightarrow y\rightarrow \bar{v}''\rightarrow v''\rightarrow v$. Since $|H_{2r+2}|\leq D(L_{2r+2})= D(Q_{t}^{3})=t$, then $|P_{2r+2}|=|H_{2r+2}|+6\leq t+6$.

Therefore, we obtain $2r+2$ internally disjoint paths $P_{1},P_{2}, \ldots ,P_{2r+2}$ between
$u$ and $v$ in $E3C(r,s,t)$, with lengths at most $t+6$.  \\

Case 3. $d=d'=2$.

The proof for this case is analogous to Case 2.  \qed
\\

\begin{lemma} \label{fenlei2}
		Let $u=ABCd$ and $v=A'B'C'd'$ be two distinct vertices in $E3C(r,s,t)$ such that $A=A'$ and $C=C'$. We discuss three cases.

\item 1. If $d=d'=0$, there exist $2r+2$ internally disjoint paths between $u$ and $v$ with length at most $s+6$.

\item 2. If $d=d'=1$, there exist $2r+2$ internally disjoint paths between $u$ and $v$ with length at most $s+6$.	

\item 3. If $d=d'=2$, there exist $2r+2$ internally disjoint paths between $u$ and $v$ with length at most $s+6$.
	\end{lemma}
\noindent \textbf{Proof.} We consider three cases based on the value of $d$.

Case 1. $d=d'=0$.

\begin{figure}[h]
		\centering
		\includegraphics[width=0.5\textwidth]{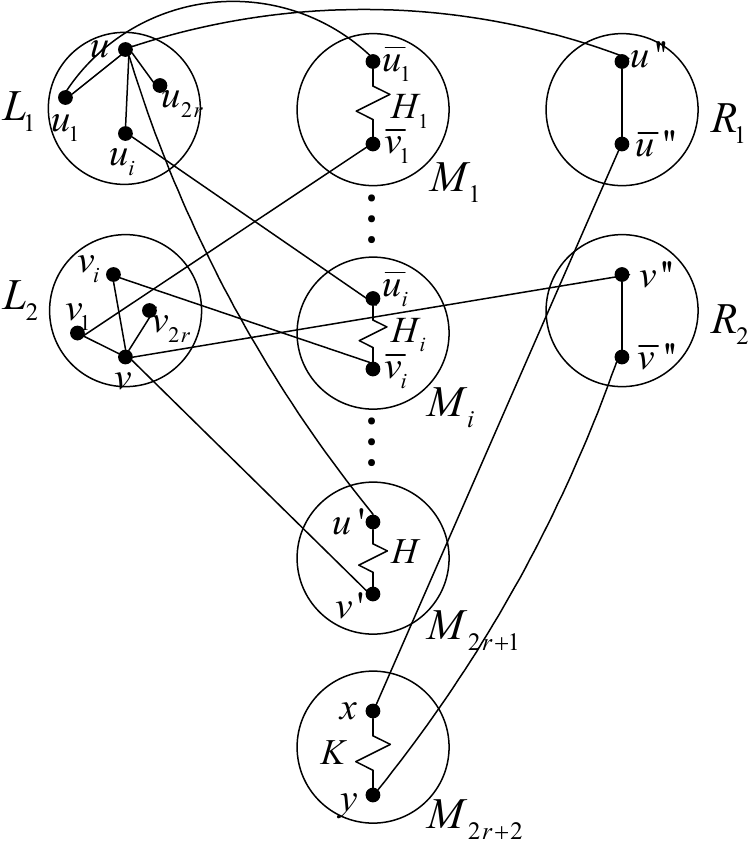}
		\renewcommand{\figurename}{Fig.}
		\caption{The illustration of Case 1 of Lemma \ref{fenlei2}}
		\label{2case1}
	\end{figure}

Let $u=ABC0\in V(L_{1})$ and $v=AB'C0\in V(L_{2})$ (see Fig. \ref{2case1}). Define $u_{i}$ as one of the
$2r$ neighbors of $u$ in $L_{1}$ , where $u_{i}=ABC_{i}0$ for $1\leq i\leq 2r$. Define $v_{i}$ as one of the
$2r$ neighbors of $v$ in $L_{2}$ , where $v_{i}=AB'C_{i}0$ for $1\leq i\leq 2r$. Let $\bar{u}_{i}=ABC_{i}1$ be an outside neighbor of $u_{i}$ in $M_{i}$ for $1\leq i\leq 2r$. $v_{i}$ has an outside neighbor $\bar{v}_{i}=AB'C_{i}1$ also in $M_{i}$ for $1\leq i\leq 2r$. Let $H_{i}$ be a shortest path between $\bar{u}_{i}$ and $\bar{v}_{i}$ in $M_{i}$ for $1\leq i\leq 2r$. Define the path $P_{i}$ as $u\rightarrow u_{i}\rightarrow \bar{u}_{i}\rightarrow H_{i}\rightarrow \bar{v}_{i}\rightarrow v_{i}\rightarrow v$ for $1\leq i\leq 2r$. Since $|H_{i}|\leq D(M_{i})= D(Q_{s}^{3})=s$, then $|P_{i}|=|H_{i}|+4\leq s+4$ for $1\leq i\leq 2r$. Let $u'=ABC1$ be in $M_{2r+1}$. Then $v'=AB'C1$ is also in $M_{2r+1}$. Let $H$ be a shortest path between $u'$ and $v'$ in $M_{2r+1}$. Define the path $P_{2r+1}$ as $u\rightarrow u'\rightarrow H\rightarrow v'\rightarrow v$. Since $|H|\leq D(M_{2r+1})= D(Q_{s}^{3})=s$, then $|P_{2r+1}|=|H|+2\leq s+2$. Assume that $u''= ABC2$ and $v''= AB'C2$ are in $R_{1}$ and $R_{2}$, respectively. $u''$ has a neighbor $\bar{u}''=A_{1}BC2$ in $R_{1}$, which has an outside neighbor $x=A_{1}BC1$ in $M_{2r+2}$. $v''$ has a neighbor $\bar{v}''=A_{1}B'C2$ in $R_{2}$, which has an outside neighbor $y=A_{1}B'C1$ also in $M_{2r+2}$. Let $K$ be a shortest path between $x$ and $y$ in $M_{2r+2}$. Define the path $P_{2r+2}$ as $u\rightarrow u''\rightarrow \bar{u}''\rightarrow x\rightarrow K\rightarrow y\rightarrow \bar{v}''\rightarrow v''\rightarrow v$. Since $|K|\leq D(M_{2r+2})= D(Q_{s}^{3})=s$, then $|P_{2r+2}|=|K|+6\leq s+6$.

Therefore, we obtain $2r+2$ internally disjoint paths $P_{1},P_{2}, \ldots ,P_{2r+2}$ between
$u$ and $v$ in $E3C(r,s,t)$, with lengths at most $s+6$.  \\

Case 2. $d=d'=1$.

\begin{figure}[h]
		\centering
		\includegraphics[width=0.5\textwidth]{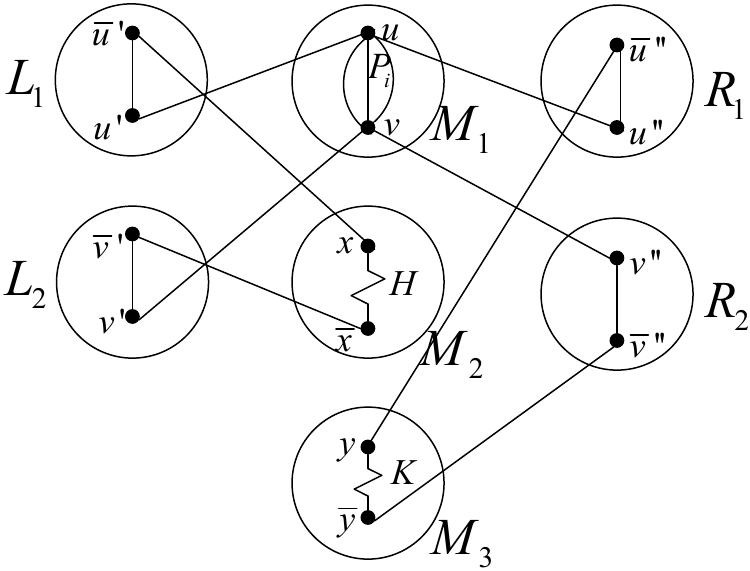}
		\renewcommand{\figurename}{Fig.}
		\caption{The illustration of Case 2 of Lemma \ref{fenlei2}}
		\label{2case2}
	\end{figure}

Let $u=ABC1$ and $v=AB'C1$ be in $M_{1}$ (see Fig. \ref{2case2}). According to Lemma \ref{lemQnklujuli} and its following analysis, there exist $2r$ internally disjoint paths $P_{1},P_{2}, \ldots ,P_{2r}$ between $u$ and $v$ in $M_{1}$, such that
$|P_{i}| \leq D(M_{1})+2=D(Q_{s}^{3})+2=s+2$ for $1\leq i\leq 2r$. Assume that $u'=ABC0$ and $u''=ABC2$ are in $L_{1}$ and $R_{1}$, respectively. Let $v'=AB'C0$ and $v''=AB'C2$ be in $L_{2}$ and $R_{2}$, respectively.
$u'$ has a neighbor $\bar{u}'=ABC_{1}0$ in $L_{1}$, which has an outside neighbor $x=ABC_{1}1$ in $M_{2}$. $v'$ has a neighbor $\bar{v}'=AB'C_{1}0$ in $L_{2}$, which has an outside neighbor $\bar{x}=AB'C_{1}1$ also in $M_{2}$. Let $H$ be a shortest path between $x$ and $\bar{x}$ in $M_{2}$. $u''$ has a neighbor $\bar{u}''=A_{1}BC2$ in $R_{1}$, which has an outside neighbor $y=A_{1}BC1$ in $M_{3}$. $v''$ has a neighbor $\bar{v}''=A_{1}B'C2$ in $R_{2}$, which has an outside neighbor $\bar{y}=A_{1}B'C1$ also in $M_{3}$. Let $K$ be a shortest path between $y$ and $\bar{y}$ in $M_{3}$. Thus, we construct the path $P_{2r+1}$ as $u\rightarrow u'\rightarrow \bar{u}'\rightarrow x\rightarrow H\rightarrow \bar{x}\rightarrow \bar{v}'\rightarrow v'\rightarrow v$. Since $|H|\leq D(M_{2})= D(Q_{s}^{3})=s$, then $|P_{2r+1}|=|H|+6\leq s+6$. Define the path $P_{2r+2}$ as $u\rightarrow u''\rightarrow \bar{u}''\rightarrow y\rightarrow K\rightarrow \bar{y}\rightarrow \bar{v}''\rightarrow v''\rightarrow v$. Since $|K|\leq D(M_{3})= D(Q_{s}^{3})=s$, then $|P_{2r+2}|=|K|+6\leq s+6$.

Therefore, we obtain $2r+2$ internally disjoint paths $P_{1},P_{2}, \ldots ,P_{2r+2}$ between
$u$ and $v$ in $E3C(r,s,t)$, with lengths at most $s+6$. \\

Case 3. $d=d'=2$.

The proof for this case is analogous to Case 1.  \qed
\\

\begin{lemma} \label{fenlei3}
		Let $u=ABCd$ and $v=A'B'C'd'$ be two distinct vertices in $E3C(r,s,t)$ such that $A=A'$, $B\neq B'$, $C\neq C'$ and $d=d'$. We discuss three cases.

\item 1. If $d=d'=0$, there exist $2r+2$ internally disjoint paths between $u$ and $v$ with length at most $s+t+7$.

\item 2. If $d=d'=1$, there exist $2r+2$ internally disjoint paths between $u$ and $v$ with length at most $s+t+7$.	

\item 3. If $d=d'=2$, there exist $2r+2$ internally disjoint paths between $u$ and $v$ with length at most $s+t+5$.
	\end{lemma}
\noindent \textbf{Proof.} We consider three cases based on the value of $d$.

Case 1. $d=d'=0$.

\begin{figure}[h]
		\centering
		\includegraphics[width=0.5\textwidth]{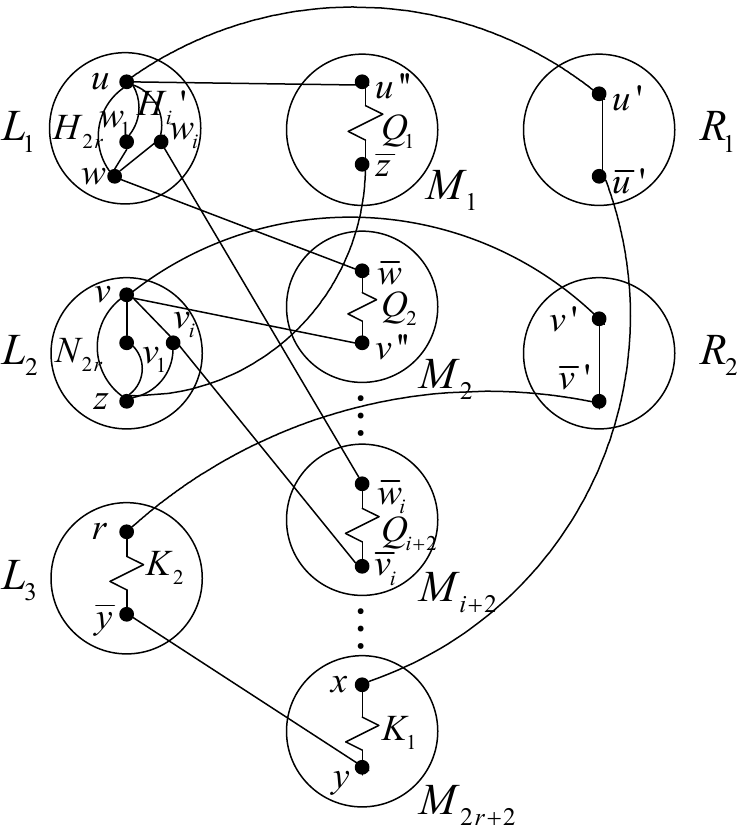}
		\renewcommand{\figurename}{Fig.}
		\caption{The illustration of Case 1 of Lemma \ref{fenlei3}}
		\label{3case1}
	\end{figure}

Let $u=ABC0\in V(L_{1})$ and $v=AB'C'0\in V(L_{2})$ (see Fig. \ref{3case1}). Choose $w=ABC'0$ in $L_{1}$. According to Lemma \ref{lemQnklujuli} and its following analysis, there
exist $2r$ internally disjoint paths $H_{1},H_{2}, \ldots ,H_{2r}$ between $u$ and $w$ in $L_{1}$, such that
$|H_{i}| \leq D(L_{1})+2=D(Q_{t}^{3})+2=t+2$ for $1\leq i\leq 2r$. Without loss of generality, we assume that $|H_{i}|\geq 2$ for $1\leq i\leq 2r-1$. Let $w_{i}=ABC'_{i}0$ be a neighbor of $w$ in $L_{1}$ with $ww_{i}\in E(H_{i})$ for $1\leq i\leq 2r-1$. Define $H'_{i}$ as the subpath from $u$ to $w_{i}$ in $H_{i}$
for $1\leq i\leq 2r-1$. Similarly, choose $z=AB'C0$ in $L_{2}$. There
exist $2r$ internally disjoint paths $N_{1},N_{2}, \ldots ,N_{2r}$ between $v$ and $z$ in $L_{2}$, such that
$|N_{i}| \leq D(L_{2})+2=D(Q_{t}^{3})+2=t+2$ for $1\leq i\leq 2r$. Assume $|N_{i}|\geq 2$ for
$1\leq i\leq 2r-1$. Let $v_{i}=AB'C'_{i}0$ be a neighbor of $v$ in $L_{2}$ with $vv_{i}\in E(N_{i})$ for $1\leq i\leq 2r-1$. Let $\bar{w}_{i}=ABC'_{i}1$, which is an outside neighbor of $w_{i}$ in $M_{i+2}$ for $1\leq i\leq 2r-1$. $v_{i}$ has an outside neighbor $\bar{v}_{i}=AB'C'_{i}1$ also in $M_{i+2}$ for $1\leq i\leq 2r-1$. Let $Q_{i+2}$ be a shortest path between $\bar{w}_{i}$ and $\bar{v}_{i}$ in $M_{i+2}$ for $1\leq i\leq 2r-1$. Let $\bar{z}=AB'C1$ be an outside neighbor of $z$ in $M_{1}$. Then $u''=ABC1$ is also in $M_{1}$. Let $Q_{1}$ be a shortest path between $\bar{z}$ and $u''$ in $M_{1}$. Let
$\bar{w}=ABC'1$ be an outside neighbor of $w$ in $M_{2}$. Then $v''=AB'C'1$ is also in $M_{2}$. Let $Q_{2}$ be a shortest path between $\bar{w}$ and $v''$ in $M_{2}$. Let $u'=ABC2 \in V(R_{1})$ be another outside neighbor of $u$. Then $u'$ has a neighbor $\bar{u}'=A_{1}BC2\in V(R_{1})$, whose outside neighbor is  $x=A_{1}BC1\in V(M_{2r+2})$. Select $\bar{x}=A_{1}B'C1$ in $M_{2r+2}$. Let $K_{1}$ be a shortest path between $x$ and $\bar{x}$ in $M_{2r+2}$. Note that $\bar{x}$ has an outside neighbor $y=A_{1}B'C0$ in $L_{3}$. Let $v'=AB'C'2 \in V(R_{2})$ be another outside neighbor of $v$. $v'$ has a neighbor $\bar{v}'=A_{1}B'C'2$ in $R_{2}$, whose outside neighbor is $\bar{y}=A_{1}B'C'0\in V(L_{3}$). Let $K_{2}$ be a shortest path between $y$ and $\bar{y}$ in $L_{3}$. Define the path $P_{i}$ as $u\rightarrow H'_{i}\rightarrow w_{i}\rightarrow \bar{w}_{i}\rightarrow Q_{i+2}\rightarrow \bar{v}_{i}\rightarrow v_{i}\rightarrow v$ for $1\leq i\leq 2r-1$. Since $|H'_{i}|=|H_{i}|-1\leq D(L_{1})-1=t-1$ and $|Q_{i+2}|\leq D(M_{i+2})=D(Q_{s}^{3})=s$, then $|P_{i}|=|H'_{i}|+|Q_{i+2}|+3=s+t+2$ for $1\leq i\leq 2r-1$. Construct the path $P_{2r}$ as $u\rightarrow u''\rightarrow Q_{1}\rightarrow \bar{z}\rightarrow z\rightarrow N_{2r}\rightarrow v$. Then $|P_{2r}|=2+|Q_{1}|+|N_{2r}|\leq 2+D(M_{1})+D(L_{2})=s+t+2$. Define the path $P_{2r+1}$ as $u\rightarrow H_{2r}\rightarrow w\rightarrow \bar{w}\rightarrow Q_{2}\rightarrow v''\rightarrow v$. Then $|P_{2r+1}|=|H_{2r}|+|Q_{2}|+2\leq D(L_{1})+D(M_{2})+2=s+t+2$. Construct the path $P_{2r+2}$ as $u\rightarrow u'\rightarrow \bar{u}'\rightarrow x\rightarrow K_{1}\rightarrow \bar{x}\rightarrow y \rightarrow K_{2}\rightarrow \bar{y}\rightarrow \bar{v}'\rightarrow v'\rightarrow v$. Then $|P_{2r+2}|=|K_{1}|+|K_{2}|+7\leq D(M_{2r+2})+D(L_{3})+7=s+t+7$.

Therefore, we obtain $2r+2$ internally disjoint paths $P_{1},P_{2}, \ldots ,P_{2r+2}$ between
$u$ and $v$ in $E3C(r,s,t)$, with lengths at most $s+t+7$. \\

Case 2. $d=d'=1$.

The proof for this case is analogous to Case 1. \\

Case 3. $d=d'=2$.

\begin{figure}[h]
		\centering
		\includegraphics[width=0.5\textwidth]{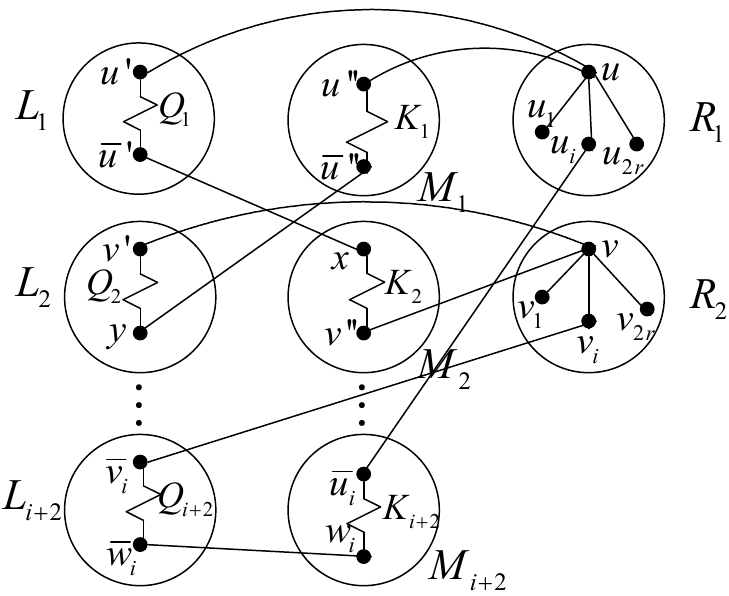}
		\renewcommand{\figurename}{Fig.}
		\caption{The illustration of Case 3 of Lemma \ref{fenlei3}}
		\label{3case3}
	\end{figure}

Let $u=ABC2\in V(R_{1})$ and $v=AB'C'2\in V(R_{2})$ (see Fig. \ref{3case3}). Define $u_{i}$ as one of the
$2r$ neighbors of $u$ in $R_{1}$ , where $u_{i}=A_{i}BC2$ for $1\leq i\leq 2r$. Define $v_{i}$ as one of the
$2r$ neighbors of $v$ in $R_{2}$ , where $v_{i}=A_{i}B'C'2$ for $1\leq i\leq 2r$. Let $\bar{u}_{i}=A_{i}BC1$ be an outside neighbor of $u_{i}$ in $M_{i+2}$ for $1\leq i\leq 2r$. Choose $w_{i}=A_{i}B'C1$ in $M_{i+2}$ for $1\leq i\leq 2r$. Let $K_{i+2}$ be a shortest path between $\bar{u}_{i}$ and $w_{i}$ in $M_{i+2}$ for $1\leq i\leq 2r$. Note that $w_{i}$ has an outside neighbor $\bar{w}_{i}=A_{i}B'C0$ in $L_{i+2}$ for $1\leq i\leq 2r$. Since $\bar{v}_{i}=A_{i}B'C'0$ is also in $L_{i+2}$, there exists a shortest path $Q_{i+2}$ between $\bar{w}_{i}$ and $\bar{v}_{i}$ in $L_{i+2}$ for $1\leq i\leq 2r$. Define the path $P_{i}$ as $u\rightarrow u_{i}\rightarrow \bar{u}_{i}\rightarrow K_{i+2}\rightarrow w_{i}\rightarrow \bar{w}_{i}\rightarrow Q_{i+2}\rightarrow \bar{v}_{i}\rightarrow v_{i}\rightarrow v$ for $1\leq i\leq 2r$. Then $|P_{i}|=5+|K_{i+2}|+|Q_{i+2}|\leq 5+D(M_{i+2})+D(L_{i+2})=s+t+5$ for $1\leq i\leq 2r$.
Assume that $u'= ABC0$ and $v''= AB'C'1$ are in $L_{1}$ and $M_{2}$, respectively. Choose $\bar{u}'=ABC'0$ in $L_{1}$. Let $Q_{1}$ be a shortest path between $u'$ and $\bar{u}'$ in $L_{1}$. $\bar{u}'$ has an outside neighbor $x=ABC'1$ also in $M_{2}$. Then there exists a shortest path $K_{2}$ between $x$ and $v''$ in $M_{2}$. Assume that $u''= ABC1$ and $v'= AB'C'0$ are in $M_{1}$ and $L_{2}$, respectively. Choose $\bar{u}''=AB'C1$ in $M_{1}$. Let $K_{1}$ be a shortest path between $u''$ and $\bar{u}''$ in $M_{1}$. $\bar{u}''$ has an outside neighbor $y=AB'C0$ also in $L_{2}$. Then there exists a shortest path $Q_{2}$ between $y$ and $v'$ in $L_{2}$. Construct the path $P_{2r+1}$ as $u\rightarrow u'\rightarrow Q_{1}\rightarrow \bar{u}'\rightarrow x\rightarrow K_{2}\rightarrow v''\rightarrow v$. Then $|P_{2r+1}|=3+|Q_{1}|+|K_{2}|\leq 3+s+t$. Construct the path $P_{2r+2}$ as $u\rightarrow u''\rightarrow K_{1}\rightarrow \bar{u}''\rightarrow y\rightarrow  Q_{2}\rightarrow v'\rightarrow v$. Then $|P_{2r+2}|=3+|K_{1}|+|Q_{2}|\leq 3+s+t$.

Therefore, we obtain $2r+2$ internally disjoint paths $P_{1},P_{2}, \ldots ,P_{2r+2}$ between
$u$ and $v$ in $E3C(r,s,t)$, with lengths at most $s+t+5$. \qed
\\

\begin{lemma} \label{fenlei4}
		Let $u=ABCd$ and $v=A'B'C'd'$ be two distinct vertices in $E3C(r,s,t)$ such that $B=B'$ and $C=C'$. We discuss three cases.

\item 1. If $d=d'=0$, there exist $2r+2$ internally disjoint paths between $u$ and $v$ with length at most $r+6$.

\item 2. If $d=d'=1$, there exist $2r+2$ internally disjoint paths between $u$ and $v$ with length at most $r+6$.	

\item 3. If $d=d'=2$, there exist $2r+2$ internally disjoint paths between $u$ and $v$ with length at most $r+6$.
	\end{lemma}
\noindent \textbf{Proof.} We consider three cases based on the value of $d$.

Case 1. $d=d'=0$.

\begin{figure}[h]
		\centering
		\includegraphics[width=0.5\textwidth]{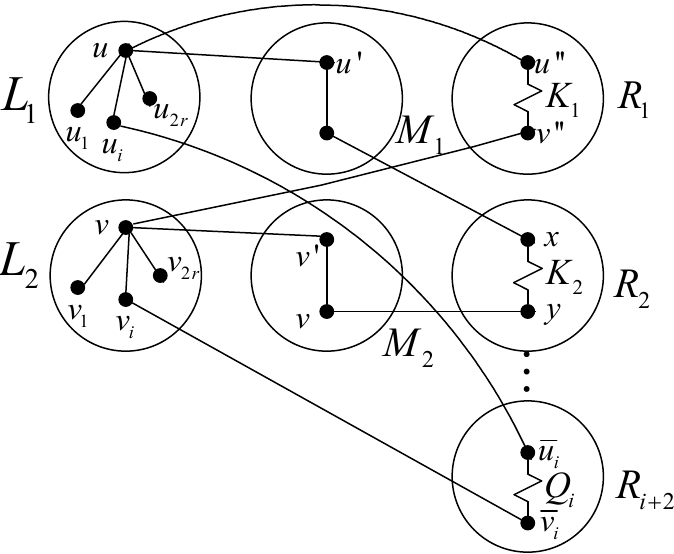}
		\renewcommand{\figurename}{Fig.}
		\caption{The illustration of Case 1 of Lemma \ref{fenlei4}}
		\label{4case1}
	\end{figure}

Let $u=ABC0\in V(L_{1})$ and $v=A'BC0\in V(L_{2})$ (see Fig. \ref{4case1}). Define $u_{i}$ as one of the
$2r$ neighbors of $u$ in $L_{1}$ , where $u_{i}=ABC_{i}0$ for $1\leq i\leq 2r$. Define $v_{i}$ as one of the
$2r$ neighbors of $v$ in $L_{2}$ , where $v_{i}=A'BC_{i}0$ for $1\leq i\leq 2r$. Let $\bar{u}_{i}=ABC_{i}2$ be an outside neighbor of $u_{i}$ in $R_{i+2}$ for $1\leq i\leq 2r$. Since $\bar{v}_{i}=A'BC_{i}2$ is also in $R_{i+2}$, there exists a shortest path $Q_{i}$ between $\bar{u}_{i}$ and $\bar{v}_{i}$ in $R_{i+2}$ for $1\leq i\leq 2r$. Define the path $P_{i}$ as $u\rightarrow u_{i}\rightarrow \bar{u}_{i}\rightarrow Q_{i}\rightarrow \bar{v}_{i}\rightarrow v_{i}\rightarrow v$ for $1\leq i\leq 2r$. Then $|P_{i}|=|Q_{i}|+4\leq D(R_{i+2})+4=r+4$ for $1\leq i\leq 2r$. Let $u''=ABC2$ and $v''=A'BC2$ be in $R_{1}$. Let $K_{1}$ be a shortest path between $u''$ and $v''$ in $R_{1}$. Assume that $u'=ABC1$ and $v'= A'BC1$ are in $M_{1}$ and $M_{2}$, respectively. $u'$ has a neighbor $\bar{u}'=AB_{1}C1\in V(M_{1})$, whose outside neighbor is $x=AB_{1}C2\in V(R_{2})$. $v'$ has a neighbor $\bar{v}'=A'B_{1}C1\in V(M_{2})$,  whose
outside neighbor is $y=A'B_{1}C2\in V(R_{2})$. Let $K_{2}$ be a shortest path between $x$ and $y$ in $R_{2}$. Construct the path $P_{2r+1}$ as $u\rightarrow u''\rightarrow K_{1}\rightarrow v''\rightarrow v$. Then $|P_{2r+1}|=|K_{1}|+2\leq D(R_{1})+2=r+2$.
Define the path $P_{2r+2}$ as $u\rightarrow u'\rightarrow \bar{u}'\rightarrow x\rightarrow K_{2}\rightarrow y\rightarrow \bar{v}'\rightarrow v'\rightarrow v$. $|P_{2r+2}|=|K_{2}|+6\leq r+6$.

Therefore, we obtain $2r+2$ internally disjoint paths $P_{1},P_{2}, \ldots ,P_{2r+2}$ between
$u$ and $v$ in $E3C(r,s,t)$, with lengths at most $r+6$. \\

Case 2. $d=d'=1$.

The proof for this case is analogous to Case 1.  \\

Case 3. $d=d'=2$.

\begin{figure}[h]
		\centering
		\includegraphics[width=0.5\textwidth]{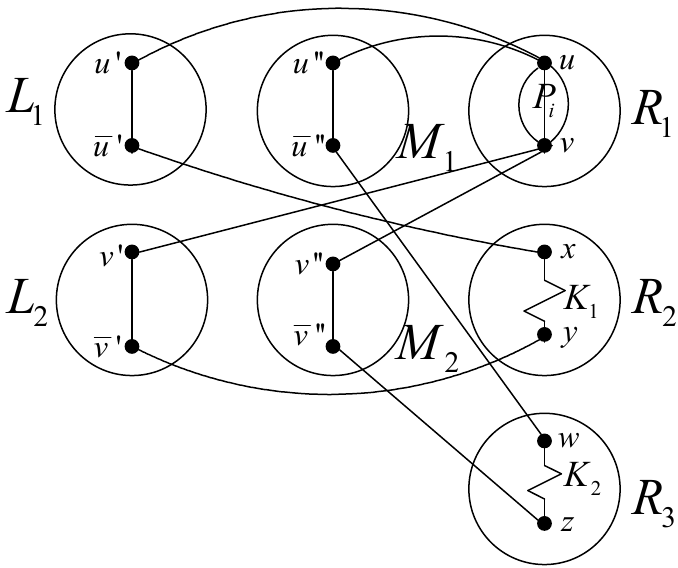}
		\renewcommand{\figurename}{Fig.}
		\caption{The illustration of Case 3 of Lemma \ref{fenlei4}}
		\label{4case3}
	\end{figure}

Let $u=ABC2$ and $v=A'BC2$ be in $R_{1}$ (see Fig. \ref{4case3}). According to Lemma \ref{lemQnklujuli} and its following analysis, there
exist $2r$ internally disjoint paths $P_{1},P_{2}, \ldots ,P_{2r}$ between $u$ and $v$ in $R_{1}$, such that
$|P_{i}| \leq D(R_{1})+2=D(Q_{r}^{3})+2=r+2$ for $1\leq i\leq 2r$. Assume that $u'=ABC0$ and $u''=ABC1$ are in $L_{1}$ and $M_{1}$, respectively. Let $v'=A'BC0$ and $v''=A'BC1$ be in $L_{2}$ and $M_{2}$, respectively.
$u'$ has a neighbor $\bar{u}'=ABC_{1}0$ in $L_{1}$, which has an outside neighbor $x=ABC_{1}2$ in $R_{2}$. $v'$ has a neighbor $\bar{v}'=A'BC_{1}0$ in $L_{2}$, which has an outside neighbor $y=A'BC_{1}2$ also in $R_{2}$. Let $K_{1}$ be a shortest path between $x$ and $y$ in $R_{2}$. $u''$ has a neighbor $\bar{u}''=AB_{1}C1$ in $M_{1}$, which has an outside neighbor $w=AB_{1}C2$ in $R_{3}$. $v''$ has a neighbor $\bar{v}''=A'B_{1}C1$ in $M_{2}$, which has an outside neighbor $z=A'B_{1}C2$ also in $R_{3}$. Let $K_{2}$ be a shortest path between $w$ and $z$ in $R_{3}$. Thus, we construct the path $P_{2r+1}$ as $u\rightarrow u'\rightarrow \bar{u}'\rightarrow x\rightarrow K_{1}\rightarrow y\rightarrow \bar{v}'\rightarrow v'\rightarrow v$. Since $|K_{1}|\leq D(R_{2})= D(Q_{r}^{3})=r$, then $|P_{2r+1}|=|K_{1}|+6\leq r+6$. Define the path $P_{2r+2}$ as $u\rightarrow u''\rightarrow \bar{u}''\rightarrow w\rightarrow K_{2}\rightarrow z\rightarrow \bar{v}''\rightarrow v''\rightarrow v$. Since $|K_{2}|\leq D(R_{3})= D(Q_{r}^{3})=r$, then $|P_{2r+2}|=|K_{2}|+6\leq r+6$.

Therefore, we obtain $2r+2$ internally disjoint paths $P_{1},P_{2}, \ldots ,P_{2r+2}$ between
$u$ and $v$ in $E3C(r,s,t)$, with lengths at most $r+6$.  \qed
\\

\begin{lemma} \label{fenlei5}
		Let $u=ABCd$ and $v=A'B'C'd'$ be two distinct vertices in $E3C(r,s,t)$ such that $A\neq A'$, $B=B'$, $C\neq C'$ and $d=d'$. We discuss three cases.

\item 1. If $d=d'=0$, there exist $2r+2$ internally disjoint paths between $u$ and $v$ with length at most $r+t+7$.

\item 2. If $d=d'=1$, there exist $2r+2$ internally disjoint paths between $u$ and $v$ with length at most $r+t+5$.	

\item 3. If $d=d'=2$, there exist $2r+2$ internally disjoint paths between $u$ and $v$ with length at most $r+t+7$.
	\end{lemma}
\noindent \textbf{Proof.} We consider three cases based on the value of $d$.

Case 1. $d=d'=0$.

\begin{figure}[h]
		\centering
		\includegraphics[width=0.5\textwidth]{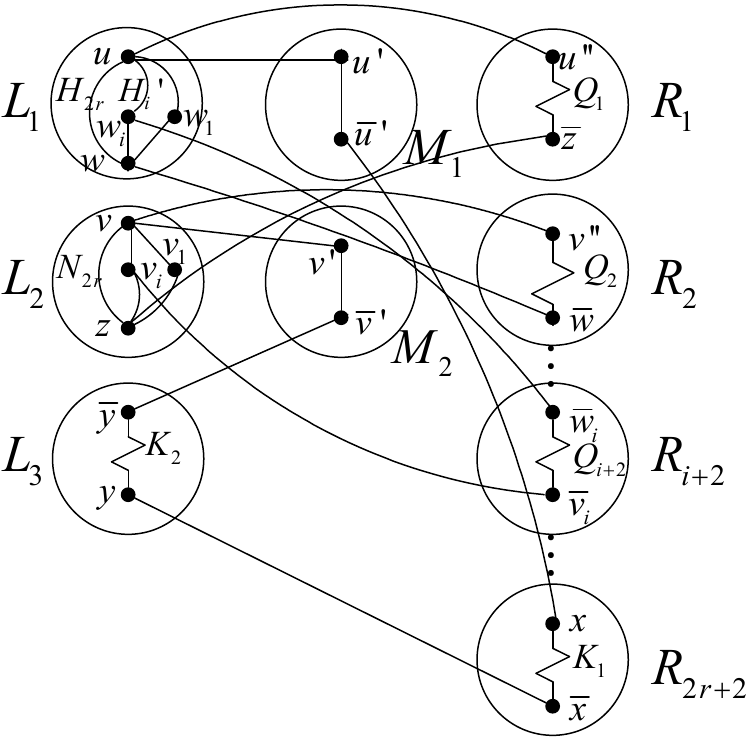}
		\renewcommand{\figurename}{Fig.}
		\caption{The illustration of Case 1 of Lemma \ref{fenlei5}}
		\label{5case1}
	\end{figure}

Let $u=ABC0\in V(L_{1})$ and $v=A'BC'0\in V(L_{2})$ (see Fig. \ref{5case1}). Choose $w=ABC'0$ in $L_{1}$. According to Lemma \ref{lemQnklujuli} and its following analysis, there
exist $2r$ internally disjoint paths $H_{1},H_{2}, \ldots ,H_{2r}$ between $u$ and $w$ in $L_{1}$, such that
$|H_{i}| \leq D(L_{1})+2=D(Q_{t}^{3})+2=t+2$ for $1\leq i\leq 2r$. Without loss of generality, we assume that $|H_{i}|\geq 2$ for $1\leq i\leq 2r-1$. Let $w_{i}=ABC'_{i}0$ be a neighbor of $w$ in $L_{1}$ with $ww_{i}\in E(H_{i})$ for $1\leq i\leq 2r-1$. Define $H'_{i}$ as the subpath from $u$ to $w_{i}$ in $H_{i}$
for $1\leq i\leq 2r-1$. Similarly, choose $z=A'BC0$ in $L_{2}$. There
exist $2r$ internally disjoint paths $N_{1},N_{2}, \ldots ,N_{2r}$ between $v$ and $z$ in $L_{2}$, such that
$|N_{i}| \leq D(L_{2})+2=D(Q_{t}^{3})+2=t+2$ for $1\leq i\leq 2r$. Assume $|N_{i}|\geq 2$ for
$1\leq i\leq 2r-1$. Let $v_{i}=A'BC'_{i}0$ be a neighbor of $v$ in $L_{2}$ with $vv_{i}\in E(N_{i})$ for $1\leq i\leq 2r-1$. Let $\bar{w}_{i}=ABC'_{i}2$, which is an outside neighbor of $w_{i}$ in $R_{i+2}$ for $1\leq i\leq 2r-1$. $v_{i}$ has an outside neighbor $\bar{v}_{i}=A'BC'_{i}2$ also in $R_{i+2}$ for $1\leq i\leq 2r-1$. Let $Q_{i+2}$ be a shortest path between $\bar{w}_{i}$ and $\bar{v}_{i}$ in $R_{i+2}$ for $1\leq i\leq 2r-1$.  Let $\bar{z}=A'BC2$ be an outside neighbor of $z$ in $R_{1}$. Then $u''=ABC2$ is also in $R_{1}$. Let $Q_{1}$ be a shortest path between $\bar{z}$ and $u''$ in $R_{1}$. Let $\bar{w}=ABC'2$ be an outside neighbor of $w$ in $R_{2}$. Then $v''=A'BC'2$ is also in $R_{2}$. Let $Q_{2}$ be a shortest path between $\bar{w}$ and $v''$ in $R_{2}$. Let $u'=ABC1 \in V(M_{1})$ be another outside neighbor of $u$. Then $u'$ has a neighbor $\bar{u}'=AB_{1}C1\in V(M_{1})$, whose outside neighbor is  $x=AB_{1}C2\in V(R_{2r+2})$. Select $\bar{x}=A'B_{1}C2$ in $R_{2r+2}$. Let $K_{1}$ be a shortest path between $x$ and $\bar{x}$ in $R_{2r+2}$. Note that $\bar{x}$ has an outside neighbor $y=A'B_{1}C0$ in $L_{3}$. Let $v'=A'BC'1 \in V(M_{2})$ be another outside neighbor of $v$. $v'$ has a neighbor $\bar{v}'=A'B_{1}C'1$ in $M_{2}$, which has an outside neighbor $\bar{y}=A'B_{1}C'0$ also in $L_{3}$. Let $K_{2}$ be a shortest path between $y$ and $\bar{y}$ in $L_{3}$.
Define the path $P_{i}$ as $u\rightarrow H'_{i}\rightarrow w_{i}\rightarrow \bar{w}_{i}\rightarrow Q_{i+2}\rightarrow \bar{v}_{i}\rightarrow v_{i}\rightarrow v$ for $1\leq i\leq 2r-1$. Since $|H'_{i}|=|H_{i}|-1\leq D(L_{1})-1=t-1$ and $|Q_{i+2}|\leq D(R_{i+2})=D(Q_{r}^{3})=r$, then $|P_{i}|=|H'_{i}|+|Q_{i+2}|+3\leq r+t+2$ for $1\leq i\leq 2r-1$. Construct the path $P_{2r}$ as $u\rightarrow u''\rightarrow Q_{1}\rightarrow \bar{z}\rightarrow z\rightarrow N_{2r}\rightarrow v$. Then $|P_{2r}|=2+|Q_{1}|+|N_{2r}|\leq 2+D(R_{1})+D(L_{2})=r+t+2$. Define the path $P_{2r+1}$ as $u\rightarrow H_{2r}\rightarrow w\rightarrow \bar{w}\rightarrow Q_{2}\rightarrow v''\rightarrow v$. Then $|P_{2r+1}|=|H_{2r}|+|Q_{2}|+2\leq D(L_{1})+D(R_{2})+2=r+t+2$. Construct the path $P_{2r+2}$ as $u\rightarrow u'\rightarrow \bar{u}'\rightarrow x\rightarrow K_{1}\rightarrow \bar{x}\rightarrow y \rightarrow K_{2}\rightarrow \bar{y}\rightarrow \bar{v}'\rightarrow v'\rightarrow v$. Then $|P_{2r+2}|=|K_{1}|+|K_{2}|+7\leq D(R_{2r+2})+D(L_{3})+7=r+t+7$.

Therefore, we obtain $2r+2$ internally disjoint paths $P_{1},P_{2}, \ldots ,P_{2r+2}$ between
$u$ and $v$ in $E3C(r,s,t)$, with lengths at most $r+t+7$. \\

Case 2. $d=d'=1$.

\begin{figure}[h]
		\centering
		\includegraphics[width=0.5\textwidth]{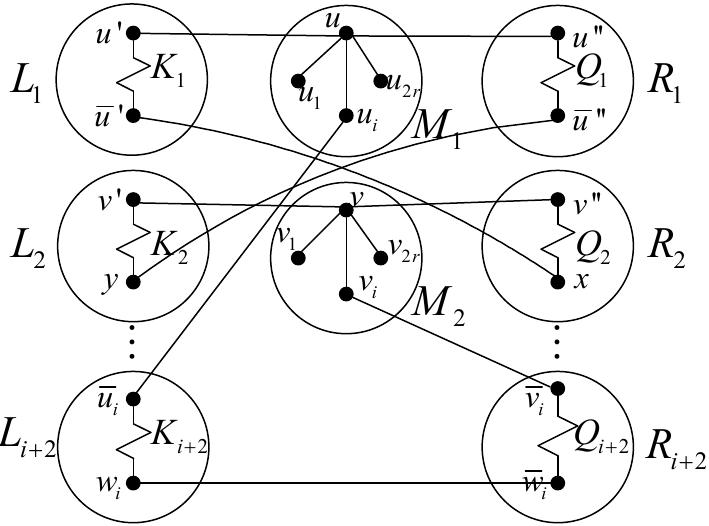}
		\renewcommand{\figurename}{Fig.}
		\caption{The illustration of Case 2 of Lemma \ref{fenlei5}}
		\label{5case2}
	\end{figure}

Let $u=ABC1\in V(M_{1})$ and $v=A'BC'1\in V(M_{2})$ (see Fig. \ref{5case2}). Define $u_{i}$ as one of the
$2r$ neighbors of $u$ in $M_{1}$ , where $u_{i}=AB_{i}C1$ for $1\leq i\leq 2r$. Define $v_{i}$ as one of the
$2r$ neighbors of $v$ in $M_{2}$ , where $v_{i}=A'B_{i}C'1$ for $1\leq i\leq 2r$. Let $\bar{u}_{i}=AB_{i}C0$ be an outside neighbor of $u_{i}$ in $L_{i+2}$ for $1\leq i\leq 2r$. Choose $w_{i}=AB_{i}C'0$ in $L_{i+2}$ for $1\leq i\leq 2r$. Let $K_{i+2}$ be a shortest path between $\bar{u}_{i}$ and $w_{i}$ in $L_{i+2}$ for $1\leq i\leq 2r$. Note that $w_{i}$ has an outside neighbor $\bar{w}_{i}=AB_{i}C'2$ in $R_{i+2}$ for $1\leq i\leq 2r$. Since $\bar{v}_{i}=A'B_{i}C'2$ is also in $R_{i+2}$, there exists a shortest path $Q_{i+2}$ between $\bar{w}_{i}$ and $\bar{v}_{i}$ in $R_{i+2}$ for $1\leq i\leq 2r$. Define the path $P_{i}$ as $u\rightarrow u_{i}\rightarrow \bar{u}_{i}\rightarrow K_{i+2}\rightarrow w_{i}\rightarrow \bar{w}_{i}\rightarrow Q_{i+2}\rightarrow \bar{v}_{i}\rightarrow v_{i}\rightarrow v$ for $1\leq i\leq 2r$. Then $|P_{i}|=5+|K_{i+2}|+|Q_{i+2}|\leq 5+D(L_{i+2})+D(R_{i+2})=r+t+5$ for $1\leq i\leq 2r$.
Assume that $u'= ABC0$ and $v''= A'BC'2$ are in $L_{1}$ and $R_{2}$, respectively. Choose $\bar{u}'=ABC'0$ in $L_{1}$. Let $K_{1}$ be a shortest path between $u'$ and $\bar{u}'$ in $L_{1}$. $\bar{u}'$ has an outside neighbor $x=ABC'2$ also in $R_{2}$. Then there exists a shortest path $Q_{2}$ between $x$ and $v''$ in $R_{2}$. Assume that $u''= ABC2$ and $v'=A'BC'0$ are in $R_{1}$ and $L_{2}$, respectively. Choose $\bar{u}''=A'BC2$ in $R_{1}$. Let $Q_{1}$ be a shortest path between $u''$ and $\bar{u}''$ in $R_{1}$. $\bar{u}''$ has an outside neighbor $y=A'BC0$ also in $L_{2}$. Then there exists a shortest path $K_{2}$ between $y$ and $v'$ in $L_{2}$. Construct the path $P_{2r+1}$ as $u\rightarrow u'\rightarrow K_{1}\rightarrow \bar{u}'\rightarrow x\rightarrow Q_{2}\rightarrow v''\rightarrow v$. Then $|P_{2r+1}|=3+|K_{1}|+|Q_{2}|\leq 3+r+t$. Construct the path $P_{2r+2}$ as $u\rightarrow u''\rightarrow Q_{1}\rightarrow \bar{u}''\rightarrow y\rightarrow  K_{2}\rightarrow v'\rightarrow v$. Then $|P_{2r+2}|=3+|Q_{1}|+|K_{2}|\leq 3+r+t$.

Therefore, we obtain $2r+2$ internally disjoint paths $P_{1},P_{2}, \ldots ,P_{2r+2}$ between
$u$ and $v$ in $E3C(r,s,t)$, with lengths at most $r+t+5$.
\\

Case 3. $d=d'=2$.

The proof for this case is analogous to Case 1. \qed
\\

\begin{lemma} \label{fenlei6}
		Let $u=ABCd$ and $v=A'B'C'd'$ be two distinct vertices in $E3C(r,s,t)$ such that $A\neq A'$, $B\neq B'$, $C=C'$ and $d=d'$. We discuss three cases.

\item 1. If $d=d'=0$, there exist $2r+2$ internally disjoint paths between $u$ and $v$ with length at most $r+s+5$.

\item 2. If $d=d'=1$, there exist $2r+2$ internally disjoint paths between $u$ and $v$ with length at most $r+s+7$.	

\item 3. If $d=d'=2$, there exist $2r+2$ internally disjoint paths between $u$ and $v$ with length at most $r+s+7$.
	\end{lemma}
\noindent \textbf{Proof.} We consider three cases based on the value of $d$.

Case 1. $d=d'=0$.

\begin{figure}[h]
		\centering
		\includegraphics[width=0.5\textwidth]{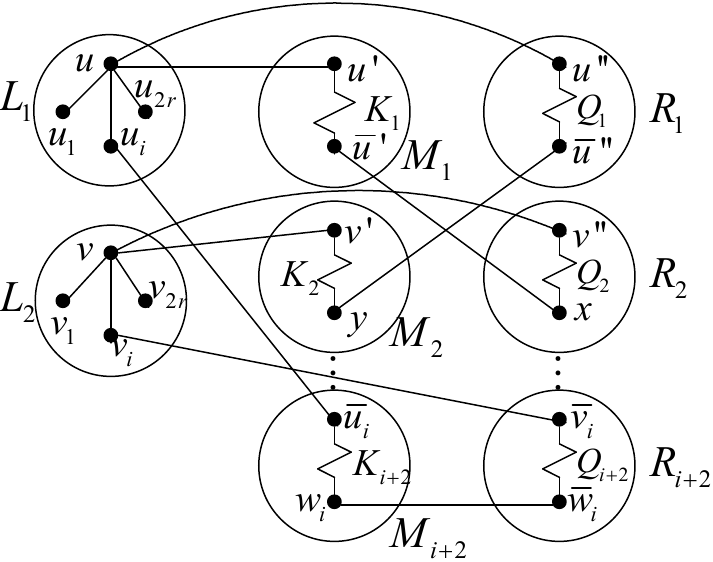}
		\renewcommand{\figurename}{Fig.}
		\caption{The illustration of Case 1 of Lemma \ref{fenlei6}}
		\label{6case1}
	\end{figure}

Let $u=ABC0\in V(L_{1})$ and $v=A'B'C0\in V(L_{2})$ (see Fig. \ref{6case1}). Define $u_{i}$ as one of the
$2r$ neighbors of $u$ in $L_{1}$ , where $u_{i}=ABC_{i}0$ for $1\leq i\leq 2r$. Define $v_{i}$ as one of the
$2r$ neighbors of $v$ in $L_{2}$ , where $v_{i}=A'B'C_{i}0$ for $1\leq i\leq 2r$. Let $\bar{u}_{i}=ABC_{i}1$ be an outside neighbor of $u_{i}$ in $M_{i+2}$ for $1\leq i\leq 2r$. Choose $w_{i}=AB'C_{i}1$ in $M_{i+2}$ for $1\leq i\leq 2r$. Let $K_{i+2}$ be a shortest path between $\bar{u}_{i}$ and $w_{i}$ in $M_{i+2}$ for $1\leq i\leq 2r$. Note that $w_{i}$ has an outside neighbor $\bar{w}_{i}=AB'C_{i}2$ in $R_{i+2}$ for $1\leq i\leq 2r$. Since $\bar{v}_{i}=A'B'C_{i}2$ is also in $R_{i+2}$, there exists a shortest path $Q_{i+2}$ between $\bar{w}_{i}$ and $\bar{v}_{i}$ in $R_{i+2}$ for $1\leq i\leq 2r$. Define the path $P_{i}$ as $u\rightarrow u_{i}\rightarrow \bar{u}_{i}\rightarrow K_{i+2}\rightarrow w_{i}\rightarrow \bar{w}_{i}\rightarrow Q_{i+2}\rightarrow \bar{v}_{i}\rightarrow v_{i}\rightarrow v$ for $1\leq i\leq 2r$. Then $|P_{i}|=5+|K_{i+2}|+|Q_{i+2}|\leq 5+D(M_{i+2})+D(R_{i+2})=r+s+5$ for $1\leq i\leq 2r$.
Assume that $u'= ABC1$ and $v''= A'B'C2$ are in $M_{1}$ and $R_{2}$, respectively. Choose $\bar{u}'=AB'C1$ in $M_{1}$. Let $K_{1}$ be a shortest path between $u'$ and $\bar{u}'$ in $M_{1}$. $\bar{u}'$ has an outside neighbor $x=AB'C2$ also in $R_{2}$. Then there exists a shortest path $Q_{2}$ between $x$ and $v''$ in $R_{2}$. Assume that $u''= ABC2$ and $v'=A'B'C1$ are in $R_{1}$ and $M_{2}$, respectively. Choose $\bar{u}''=A'BC2$ in $R_{1}$. Let $Q_{1}$ be a shortest path between $u''$ and $\bar{u}''$ in $R_{1}$. $\bar{u}''$ has an outside neighbor $y=A'BC1$ also in $M_{2}$. Then there exists a shortest path $K_{2}$ between $y$ and $v'$ in $M_{2}$. Construct the path $P_{2r+1}$ as $u\rightarrow u'\rightarrow K_{1}\rightarrow \bar{u}'\rightarrow x\rightarrow Q_{2}\rightarrow v''\rightarrow v$. Then $|P_{2r+1}|=3+|K_{1}|+|Q_{2}|\leq 3+r+s$. Construct the path $P_{2r+2}$ as $u\rightarrow u''\rightarrow Q_{1}\rightarrow \bar{u}''\rightarrow y\rightarrow  K_{2}\rightarrow v'\rightarrow v$. Then $|P_{2r+2}|=3+|Q_{1}|+|K_{2}|\leq 3+r+s$.

Therefore, we obtain $2r+2$ internally disjoint paths $P_{1},P_{2}, \ldots ,P_{2r+2}$ between
$u$ and $v$ in $E3C(r,s,t)$, with lengths at most $r+s+5$.
\\

Case 2. $d=d'=1$.

\begin{figure}[h]
		\centering
		\includegraphics[width=0.5\textwidth]{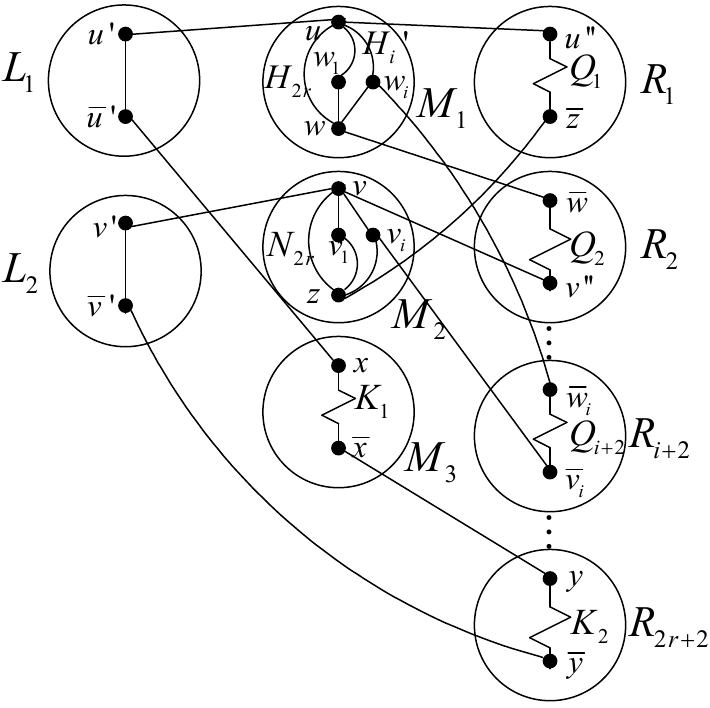}
		\renewcommand{\figurename}{Fig.}
		\caption{The illustration of Case 2 of Lemma \ref{fenlei6}}
		\label{6case2}
	\end{figure}

Let $u=ABC1\in V(M_{1})$ and $v=A'B'C1\in V(M_{2})$ (see Fig. \ref{6case2}). Choose $w=AB'C1$ in $M_{1}$. According to Lemma \ref{lemQnklujuli} and its following analysis, there
exist $2r$ internally disjoint paths $H_{1},H_{2}, \ldots ,H_{2r}$ between $u$ and $w$ in $M_{1}$, such that
$|H_{i}| \leq D(M_{1})+2=D(Q_{s}^{3})+2=s+2$ for $1\leq i\leq 2r$. Without loss of generality, we assume that $|H_{i}|\geq 2$ for $1\leq i\leq 2r-1$. Let $w_{i}=AB_{i}'C1$ be a neighbor of $w$ in $M_{1}$ with $ww_{i}\in E(H_{i})$ for $1\leq i\leq 2r-1$. Define $H'_{i}$ as the subpath from $u$ to $w_{i}$ in $H_{i}$
for $1\leq i\leq 2r-1$. Similarly, choose $z=A'BC1$ in $M_{2}$. There
exist $2r$ internally disjoint paths $N_{1},N_{2}, \ldots ,N_{2r}$ between $v$ and $z$ in $M_{2}$, such that
$|N_{i}| \leq D(M_{2})+2=D(Q_{s}^{3})+2=s+2$ for $1\leq i\leq 2r$. Assume $|N_{i}|\geq 2$ for
$1\leq i\leq 2r-1$. Let $v_{i}=A'B_{i}'C1$ be a neighbor of $v$ in $M_{2}$ with $vv_{i}\in E(N_{i})$ for $1\leq i\leq 2r-1$. Let $\bar{w}_{i}=AB_{i}'C2$, which is an outside neighbor of $w_{i}$ in $R_{i+2}$ for $1\leq i\leq 2r-1$. $v_{i}$ has an outside neighbor $\bar{v}_{i}=A'B_{i}'C2$ also in $R_{i+2}$ for $1\leq i\leq 2r-1$. Let $Q_{i+2}$ be a shortest path between $\bar{w}_{i}$ and $\bar{v}_{i}$ in $R_{i+2}$ for $1\leq i\leq 2r-1$. Let $\bar{z}=A'BC2$ be an outside neighbor of $z$ in $R_{1}$. Then $u''=ABC2$ is also in $R_{1}$. Let $Q_{1}$ be a shortest path between $\bar{z}$ and $u''$ in $R_{1}$. Let $\bar{w}=AB'C2$ be in $R_{2}$. Then $v''=A'B'C2$ is also in $R_{2}$. Let $Q_{2}$ be a shortest path between $\bar{w}$ and $v''$ in $R_{2}$. Let $u'=ABC0 \in V(L_{1})$ be another outside neighbor of $u$. Then $u'$ has a neighbor $\bar{u}'=ABC_{1}0\in V(L_{1})$, whose outside neighbor is  $x=ABC_{1}1\in V(M_{3})$. Select $\bar{x}=AB'C_{1}1$ in $M_{3}$. Let $K_{1}$ be a shortest path between $x$ and $\bar{x}$ in $M_{3}$. Note that $\bar{x}$ has an outside neighbor $y=AB'C_{1}2$ in $R_{2r+2}$. Let $v'=A'B'C0 \in V(L_{2})$ be another outside neighbor of $v$. $v'$ has a neighbor $\bar{v}'=A'B'C_{1}0$ in $L_{2}$. Then $\bar{v}'$ has an outside neighbor $\bar{y}=A'B'C_{1}2$ also in $R_{2r+2}$. Let $K_{2}$ be a shortest path between $y$ and $\bar{y}$ in $R_{2r+2}$.
Define the path $P_{i}$ as $u\rightarrow H'_{i}\rightarrow w_{i}\rightarrow \bar{w}_{i}\rightarrow Q_{i+2}\rightarrow \bar{v}_{i}\rightarrow v_{i}\rightarrow v$ for $1\leq i\leq 2r-1$. Since $|H'_{i}|=|H_{i}|-1\leq D(M_{1})-1=s-1$ and $|Q_{i+2}|\leq D(R_{i+2})=D(Q_{r}^{3})=r$, then $|P_{i}|=|H'_{i}|+|Q_{i+2}|+3=r+s+2$ for $1\leq i\leq 2r-1$. Construct the path $P_{2r}$ as $u\rightarrow u''\rightarrow Q_{1}\rightarrow \bar{z}\rightarrow z\rightarrow N_{2r}\rightarrow v$. Then $|P_{2r}|=2+|Q_{1}|+|N_{2r}|\leq 2+D(R_{1})+D(M_{2})=r+s+2$. Define the path $P_{2r+1}$ as $u\rightarrow H_{2r}\rightarrow w\rightarrow \bar{w}\rightarrow Q_{2}\rightarrow v''\rightarrow v$. Then $|P_{2r+1}|=|H_{2r}|+|Q_{2}|+2\leq D(M_{1})+D(R_{2})+2=r+s+2$. Construct the path $P_{2r+2}$ as $u\rightarrow u'\rightarrow \bar{u}'\rightarrow x\rightarrow K_{1}\rightarrow \bar{x}\rightarrow y \rightarrow K_{2}\rightarrow \bar{y}\rightarrow \bar{v}'\rightarrow v'\rightarrow v$. Then $|P_{2r+2}|=|K_{1}|+|K_{2}|+7\leq D(R_{2r+2})+D(M_{3})+7=r+s+7$.

Therefore, we obtain $2r+2$ internally disjoint paths $P_{1},P_{2}, \ldots ,P_{2r+2}$ between
$u$ and $v$ in $E3C(r,s,t)$, with lengths at most $r+s+7$. \\

Case 3. $d=d'=2$.

The proof for this case is analogous to Case 2. \qed
\\

\begin{lemma} \label{fenlei7}
		Let $u=ABCd$ and $v=A'B'C'd'$ be two distinct vertices in $E3C(r,s,t)$ such that $A\neq A'$, $B\neq B'$, $C\neq C'$ and $d=d'$. We discuss three cases.

\item 1. If $d=d'=0$, there exist $2r+2$ internally disjoint paths between $u$ and $v$ with length at most $r+s+t+4$.

\item 2. If $d=d'=1$, there exist $2r+2$ internally disjoint paths between $u$ and $v$ with length at most $r+s+t+4$.	

\item 3. If $d=d'=2$, there exist $2r+2$ internally disjoint paths between $u$ and $v$ with length at most $r+s+t+4$.
	\end{lemma}
\noindent \textbf{Proof.} We consider three cases based on the value of $d$.

Case 1. $d=d'=0$.

\begin{figure}[h]
		\centering
		\includegraphics[width=0.5\textwidth]{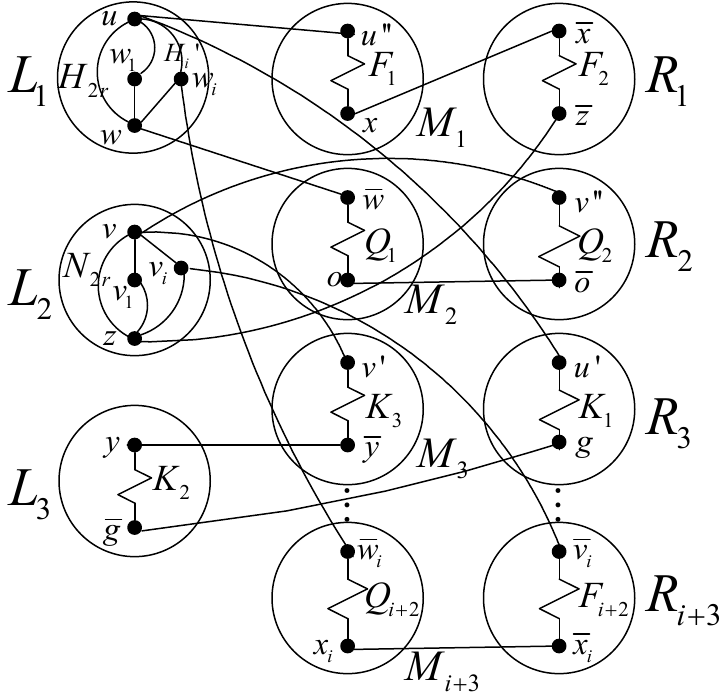}
		\renewcommand{\figurename}{Fig.}
		\caption{The illustration of Case 1 of Lemma \ref{fenlei7}}
		\label{7case1}
	\end{figure}

Let $u=ABC0\in V(L_{1})$ and $v=A'B'C'0\in V(L_{2})$ (see Fig. \ref{7case1}). Choose $w=ABC'0$ in $L_{1}$. According to Lemma \ref{lemQnklujuli} and its following analysis, there
exist $2r$ internally disjoint paths $H_{1},H_{2}, \ldots ,H_{2r}$ between $u$ and $w$ in $L_{1}$, such that
$|H_{i}| \leq D(L_{1})+2=D(Q_{t}^{3})+2=t+2$ for $1\leq i\leq 2r$. Without loss of generality, we assume that $|H_{i}|\geq 2$ for $1\leq i\leq 2r-1$. Let $w_{i}=ABC_{i}'0$ be a neighbor of $w$ in $L_{1}$ with $ww_{i}\in E(H_{i})$ for $1\leq i\leq 2r-1$. Define $H'_{i}$ as the subpath from $u$ to $w_{i}$ in $H_{i}$
for $1\leq i\leq 2r-1$. Similarly, choose $z=A'B'C0$ in $L_{2}$. There
exist $2r$ internally disjoint paths $N_{1},N_{2}, \ldots ,N_{2r}$ between $v$ and $z$ in $L_{2}$, such that
$|N_{i}| \leq D(L_{2})+2=D(Q_{t}^{3})+2=t+2$ for $1\leq i\leq 2r$. Assume $|N_{i}|\geq 2$ for
$1\leq i\leq 2r-1$. Let $v_{i}=A'B'C_{i}'0$ be a neighbor of $v$ in $L_{2}$ with $vv_{i}\in E(N_{i})$ for $1\leq i\leq 2r-1$. Let $\bar{w}_{i}=ABC_{i}'1$, which is an outside neighbor of $w_{i}$ in $M_{i+3}$ for $1\leq i\leq 2r-1$. Select $x_{i}=AB'C_{i}'1$ in $M_{i+3}$ for $1\leq i\leq 2r-1$. Let $Q_{i+2}$ be a shortest path between $\bar{w}_{i}$ and $x_{i}$ in $M_{i+3}$ for $1\leq i\leq 2r-1$. Note that $x_{i}$ has an outside neighbor $\bar{x}_{i}=AB'C_{i}'2$ in $R_{i+3}$ for $1\leq i\leq 2r-1$.  $v_{i}$ has an outside neighbor $\bar{v}_{i}=A'B'C_{i}'2$ also in $R_{i+3}$ for $1\leq i\leq 2r-1$. Let $F_{i+2}$ be a shortest path between $\bar{x}_{i}$ and $\bar{v}_{i}$ in $R_{i+3}$ for $1\leq i\leq 2r-1$. Let $\bar{z}=A'B'C2$ be an outside neighbor of $z$ in $R_{1}$. Select $\bar{x}=AB'C2$ in $R_{1}$. Let $F_{2}$ be a shortest path between $\bar{x}$ and $\bar{z}$ in $R_{1}$. Note that $\bar{x}$ has an outside neighbor $x=AB'C1$ in $M_{1}$.
Then $u''=ABC1$ is also in $M_{1}$. Let $F_{1}$ be a shortest path between $x$ and $u''$ in $M_{1}$.
Let $\bar{w}=ABC'1$ be an outside neighbor of $w$ in $M_{2}$. Select $o=AB'C'1$ in $M_{2}$. Let $Q_{1}$ be a shortest path between $\bar{w}$ and $o$ in $M_{2}$. Note that $o$ has an outside neighbor $\bar{o}=AB'C'2$ in $R_{2}$. Then $v''=A'B'C'2$ is also in $R_{2}$. Let $Q_{2}$ be a shortest path between $\bar{o}$ and $v''$ in $R_{2}$. Let $u'=ABC2 \in V(R_{3})$ be another outside neighbor of $u$. Select $g=A'BC2$ in $R_{3}$. Then there exists a shortest path $K_{1}$ between $u'$ and $g$ in $R_{3}$. Note that $g$ has an outside neighbor $\bar{g}=A'BC0$ in $L_{3}$. Select $y=A'BC'0$ in $L_{3}$. Let $K_{2}$ be a shortest path between $\bar{g}$ and $y$ in $L_{3}$. $y$ has an outside neighbor $\bar{y}=A'BC'1$ in $M_{3}$.
Then $v'=A'B'C'1$ is also in $M_{3}$. Let $K_{3}$ be a shortest path between $\bar{y}$ and $v'$ in $M_{3}$.
Define the path $P_{i}$ as $u\rightarrow H'_{i}\rightarrow w_{i}\rightarrow \bar{w}_{i}\rightarrow Q_{i+2}\rightarrow x_{i}\rightarrow \bar{x}_{i}\rightarrow F_{i+2}\rightarrow \bar{v}_{i}\rightarrow v_{i}\rightarrow v$ for $1\leq i\leq 2r-1$. Since $|H'_{i}|=|H_{i}|-1\leq D(L_{1})-1=t-1$, $|Q_{i+2}|\leq D(M_{i+3})=D(Q_{s}^{3})=s$ and $|F_{i+2}|\leq D(R_{i+3})=D(Q_{r}^{3})=r$, then $|P_{i}|=|H'_{i}|+|Q_{i+2}|+|F_{i+2}|+4\leq r+s+t+3$ for $1\leq i\leq 2r-1$. Construct the path $P_{2r}$ as $u\rightarrow u''\rightarrow F_{1}\rightarrow x\rightarrow \bar{x}\rightarrow F_{2}\rightarrow \bar{z}\rightarrow z\rightarrow N_{2r}\rightarrow v$. Then $|P_{2r}|=3+|F_{1}|+|F_{2}|+|N_{2r}|\leq 3+D(M_{1})+D(R_{1})+D(L_{2})=r+s+t+3$. Define the path $P_{2r+1}$ as $u\rightarrow H_{2r}\rightarrow w\rightarrow \bar{w}\rightarrow Q_{1}\rightarrow o\rightarrow \bar{o}\rightarrow Q_{2}\rightarrow v''\rightarrow v$. Then $|P_{2r+1}|=|H_{2r}|+|Q_{1}|+|Q_{2}|+3\leq D(L_{1})+D(M_{2})+D(R_{2})+3=r+s+t+3$. Construct the path $P_{2r+2}$ as $u\rightarrow u'\rightarrow K_{1}\rightarrow g\rightarrow \bar{g}\rightarrow K_{2}\rightarrow y\rightarrow \bar{y} \rightarrow K_{3}\rightarrow v'\rightarrow v$. Then $|P_{2r+2}|=|K_{1}|+|K_{2}|+|K_{3}|+4\leq D(R_{3})+D(L_{3})+D(M_{3})+4=r+s+t+4$.

Therefore, we obtain $2r+2$ internally disjoint paths $P_{1},P_{2}, \ldots ,P_{2r+2}$ between
$u$ and $v$ in $E3C(r,s,t)$, with lengths at most $r+s+t+4$. \\

Case 2. $d=d'=1$.

The proof for this case is analogous to Case 1.
\\

Case 3. $d=d'=2$.

The proof for this case is analogous to Case 1. \qed
\\

\begin{lemma} \label{fenlei8}
		Let $u=ABCd$ and $v=A'B'C'd'$ be two distinct vertices in $E3C(r,s,t)$ such that $A=A'$, $B=B'$, $C=C'$ and $d\neq d'$. We discuss three cases.

\item 1. If $d=0$ and $d'=1$, there exist $2r+2$ internally disjoint paths between $u$ and $v$ with length at most 7.

\item 2. If $d=0$ and $d'=2$, there exist $2r+2$ internally disjoint paths between $u$ and $v$ with length at most 7.	

\item 3. If $d=1$ and $d'=2$, there exist $2r+2$ internally disjoint paths between $u$ and $v$ with length at most 7.
	\end{lemma}
\noindent \textbf{Proof.} According to the values of $d$ and $d'$, the following three cases are considered.

Case 1. $d=0$ and $d'=1$.

Without loss of generality, let $u=ABC0\in V(L_{1})$ and $v=ABC1\in V(M_{1})$. Define $u_{i}$ as one of the
$2r$ neighbors of $u$ in $L_{1}$ , where $u_{i}=ABC_{i}0$ for $1\leq i\leq 2r$. Define $v_{i}$ as one of the
$2r$ neighbors of $v$ in $M_{1}$ , where $v_{i}=AB_{i}C1$ for $1\leq i\leq 2r$. Assume $u_{i}$ has an outside neighbor $w_{i}=ABC_{i}1$ in $M_{i+1}$ for $1\leq i\leq 2r$. Then $w_{i}$ has a neighbor  $\bar{w}_{i}=AB_{i}C_{i}1\in V(M_{i+1})$, whose outside neighbor is $z_{i}=AB_{i}C_{i}0 \in V(L_{i+1})$ for $1\leq i\leq 2r$. Then $z_{i}$ has a neighbor $\bar{z}_{i}=AB_{i}C0 \in V(L_{i+1})$ for $1\leq i\leq 2r$. $v_{i}$ is adjacent to $\bar{z}_{i}$ by Definition \ref{defE3C}. Construct the path $P_{i}$ as $u\rightarrow u_{i}\rightarrow w_{i}\rightarrow \bar{w}_{i}\rightarrow z_{i}\rightarrow \bar{z}_{i}\rightarrow v_{i}\rightarrow v$ for $1\leq i\leq 2r$. Then $|P_{i}|=7$ for $1\leq i\leq 2r$. Select $u'=ABC2$ in $R_{1}$. Then $u'$ is adjacent to both $u$ and $v$. Define the path $P_{2r+1}$ as $u\rightarrow u'\rightarrow v$. Then $|P_{2r+1}|=2$. $u$ is also adjacent to $v$ by Definition \ref{defE3C}. Define the path $P_{2r+2}$ as $u\rightarrow v$. Then $|P_{2r+2}|=1$.

Therefore, we obtain $2r+2$ internally disjoint paths $P_{1},P_{2}, \ldots ,P_{2r+2}$ between
$u$ and $v$ in $E3C(r,s,t)$, with lengths at most 7. \\

Case 2. $d=0$ and $d'=2$.

The proof for this case is analogous to Case 1.
\\

Case 3. $d=1$ and $d'=2$.

The proof for this case is analogous to Case 1. \qed
\\

\begin{lemma} \label{fenlei9}
		Let $u=ABCd$ and $v=A'B'C'd'$ be two distinct vertices in $E3C(r,s,t)$ such that $A=A'$, $B=B'$, $C\neq C'$ and $d\neq d'$. We discuss three cases.

\item 1. If $d=0$ and $d'=1$, there exist $2r+2$ internally disjoint paths between $u$ and $v$ with length at most $t+6$.

\item 2. If $d=0$ and $d'=2$, there exist $2r+2$ internally disjoint paths between $u$ and $v$ with length at most $t+6$.	

\item 3. If $d=1$ and $d'=2$, there exist $2r+2$ internally disjoint paths between $u$ and $v$ with length at most $t+8$.
	\end{lemma}
\noindent \textbf{Proof.} According to the values of $d$ and $d'$, the following three cases are considered.

Case 1. $d=0$ and $d'=1$.

\begin{figure}[h]
		\centering
		\includegraphics[width=0.5\textwidth]{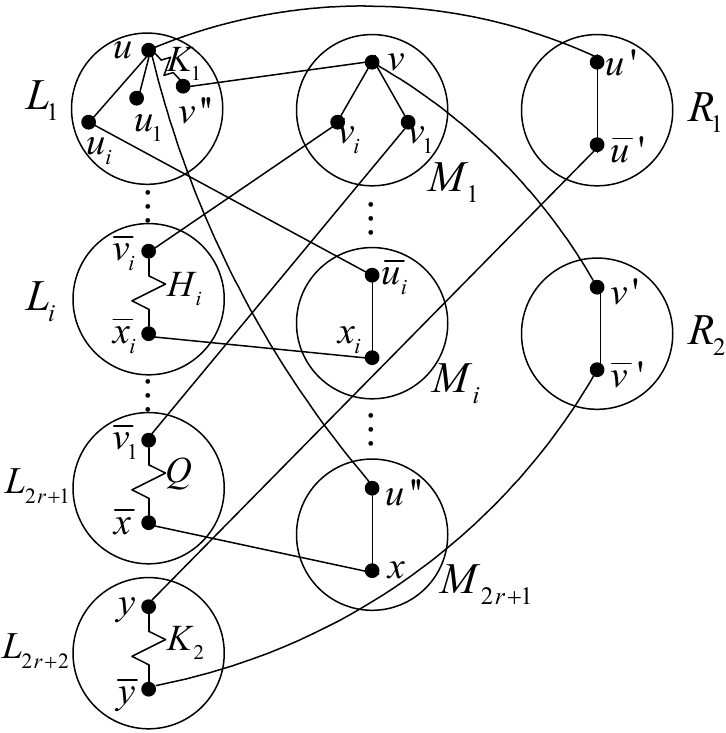}
		\renewcommand{\figurename}{Fig.}
		\caption{The illustration of Case 1 of Lemma \ref{fenlei9}}
		\label{9case1}
	\end{figure}

Without loss of generality, let $u=ABC0\in V(L_{1})$ and $v=ABC'1\in V(M_{1})$ (see Fig. \ref{9case1}). Let $u_{i}=ABC_{i}0$ be the neighbor of $u$ in $L_{1}$ for $1\leq i \leq 2r$, and $v_{i}=AB_{i}C'1$ be the neighbor of $v$ in $M_{1}$ for $1\leq i \leq 2r$. Since $v''=ABC'0$ is also in $L_{1}$, there exists a shortest path $K_{1}$ between $u$ and $v''$ in $L_{1}$. Then at most one $u_{i}\in V(K_{1})$ for $1\leq i \leq 2r$. Without loss of generality, assume $u_{i}\notin V(K_{1})$ for $2\leq i \leq 2r$. Let $\bar{u}_{i}=ABC_{i}1$ be an outside neighbor of $u_{i}$ in $M_{i}$ for $2\leq i\leq 2r$. Then $\bar{u}_{i}$ has a neighbor $x_{i}=AB_{i}C_{i}1 \in V(M_{i})$, whose outside neighbor is $\bar{x}_{i}=AB_{i}C_{i}0\in V(L_{i})$ for $2\leq i \leq 2r$. $v_{i}$ has an outside neighbor $\bar{v}_{i}=AB_{i}C'0$ also in $L_{i}$ for $2\leq i \leq 2r$. Let $H_{i}$ be a shortest path between $\bar{x}_{i}$ and $\bar{v}_{i}$ in $L_{i}$ for $2\leq i \leq 2r$. Define the path $P_{1}$ as $u\rightarrow K_{1}\rightarrow v''\rightarrow v$. Then $|P_{1}|=|K_{1}|+1\leq D(L_{1})+1=D(Q_{t}^{3})+1=t+1$. Construct the path $P_{i}$ as $u\rightarrow u_{i}\rightarrow \bar{u}_{i}\rightarrow x_{i}\rightarrow \bar{x}_{i}\rightarrow H_{i}\rightarrow \bar{v}_{i}\rightarrow v_{i}\rightarrow v$ for $2\leq i \leq 2r$. Then $|P_{i}|=|H_{i}|+6\leq D(L_{i})+6=D(Q_{t}^{3})+6=t+6$ for $2\leq i \leq 2r$. Let $u''=ABC1\in V(M_{2r+1})$. Then $u''$ has a neighbor $x=AB_{1}C1 \in V(M_{2r+1})$, whose outside neighbor is $\bar{x}=AB_{1}C0\in V(L_{2r+1})$. $v_{1}$ has an outside neighbor $\bar{v}_{1}=AB_{1}C'0$ also in $L_{2r+1}$. Let $Q$ be a shortest path between $\bar{x}$ and $\bar{v}_{1}$ in $L_{2r+1}$. Construct the path $P_{2r+1}$ as $u\rightarrow u''\rightarrow x\rightarrow \bar{x}\rightarrow Q\rightarrow \bar{v}_{1}\rightarrow v_{1}\rightarrow v$. Then $|P_{2r+1}|=|Q|+5\leq D(L_{2r+1})+5=t+5$. Let $u'=ABC2\in V (R_{1})$ be another outside neighbor of $u$. Then $u'$ has a neighbor $\bar{u}'=A_{1}BC2\in V(R_{1})$, whose outside neighbor is $y=A_{1}BC0\in V(L_{2r+2})$. Select $\bar{y}=A_{1}BC'0$ in $L_{2r+2}$. Let $K_{2}$ be a shortest path between $y$ and $\bar{y}$ in $L_{2r+2}$. Let $\bar{v}'=A_{1}BC'2$, which is an outside neighbor of $\bar{y}$ in $R_{2}$. $v$ has another outside neighbor $v'=ABC'2$ also in $R_{2}$. Note that $v'$
is adjacent to $\bar{v}'$ in $R_{2}$. Construct the path $P_{2r+2}$ as $u\rightarrow u'\rightarrow \bar{u}'\rightarrow y\rightarrow K_{2}\rightarrow \bar{y}\rightarrow \bar{v}'\rightarrow v'\rightarrow v$. Then $|P_{2r+2}|=|K_{2}|+6\leq D(L_{2r+2})+6=t+6$.

Therefore, we obtain $2r+2$ internally disjoint paths $P_{1},P_{2}, \ldots ,P_{2r+2}$ between
$u$ and $v$ in $E3C(r,s,t)$, with lengths at most $t+6$. \\

Case 2. $d=0$ and $d'=2$.

The proof for this case is analogous to Case 1.
\\

Case 3. $d=1$ and $d'=2$.

\begin{figure}[h]
		\centering
		\includegraphics[width=0.5\textwidth]{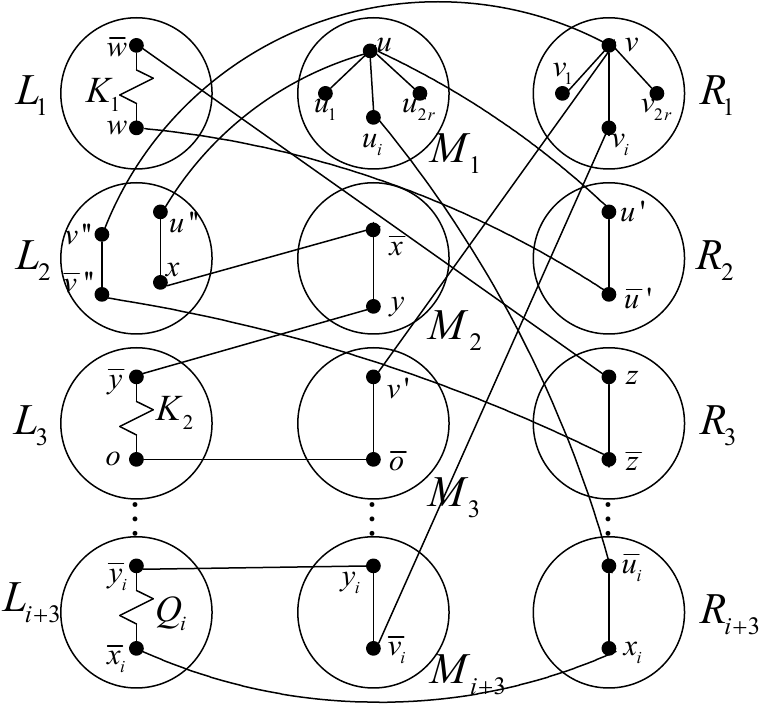}
		\renewcommand{\figurename}{Fig.}
		\caption{The illustration of Case 3 of Lemma \ref{fenlei9}}
		\label{9case3}
	\end{figure}

Without loss of generality, let $u=ABC1\in V(M_{1})$ and $v=ABC'2\in V(R_{1})$ (see Fig. \ref{9case3}). Let $u_{i}=AB_{i}C1$ be the neighbor of $u$ in $M_{1}$ for $1\leq i \leq 2r$, and $v_{i}=A_{i}BC'2$ be the neighbor of $v$ in $R_{1}$ for $1\leq i \leq 2r$. Let $\bar{u}_{i}=AB_{i}C2\in V (R_{i+3})$ be an outside neighbor of $u_{i}$ for $1\leq i \leq 2r$. Then $\bar{u}_{i}$ has a neighbor $x_{i}=A_{i}B_{i}C2\in V(R_{i+3})$, whose outside neighbor is $\bar{x}_{i}=A_{i}B_{i}C0\in V(L_{i+3})$ for $1\leq i \leq 2r$. Select $\bar{y}_{i}=A_{i}B_{i}C'0$ in $L_{i+3}$ for $1\leq i \leq 2r$. Let $Q_{i}$ be a shortest path between $\bar{y}_{i}$ and $\bar{x}_{i}$ in $L_{i+3}$ for $1\leq i \leq 2r$. Let $y_{i}=A_{i}B_{i}C'1$, which is an outside neighbor of $\bar{y}_{i}$ in $M_{i+3}$ for $1\leq i \leq 2r$. $v_{i}$ has an outside neighbor $\bar{v}_{i}=A_{i}BC'1$ also in $M_{i+3}$ for $1\leq i \leq 2r$. Note that $y_{i}$ is adjacent to $\bar{v}_{i}$ in $M_{i+3}$ for $1\leq i \leq 2r$. Construct the path $P_{i}$ as $u\rightarrow u_{i}\rightarrow \bar{u}_{i}\rightarrow x_{i}\rightarrow \bar{x}_{i}\rightarrow Q_{i}\rightarrow \bar{y}_{i}\rightarrow y_{i}\rightarrow \bar{v}_{i}\rightarrow v_{i}\rightarrow v$ for $1\leq i \leq 2r$. Then $|P_{i}|=|Q_{i}|+8\leq D(L_{i+3})+8=t+8$ for $1\leq i \leq 2r$. Let $u'=ABC2\in V (R_{2})$ be an outside neighbor of $u$. Then $u'$ has a neighbor $\bar{u}'=A_{1}BC2\in V(R_{2})$, whose outside neighbor is $w=A_{1}BC0\in V(L_{1})$. Select $\bar{w}=A_{1}BC_{1}'0$ in $L_{1}$. Let $K_{1}$ be a shortest path between $w$ and $\bar{w}$ in $L_{1}$. Let $z=A_{1}BC_{1}'2$, which is an outside neighbor of $\bar{w}$ in $R_{3}$. Then $z$ has a neighbor $\bar{z}=ABC_{1}'2$ in $R_{3}$. $\bar{z}$ has an outside neighbor $\bar{v}''=ABC_{1}'0$ in $L_{2}$. By Lemma \ref{lemE3Clianbian}, assume $v''=ABC'0\in V(L_{2})$. Note that $v''$ is adjacent to $\bar{v}''$ in $L_{2}$. Construct the path $P_{2r+1}$ as $u\rightarrow u'\rightarrow \bar{u}'\rightarrow w\rightarrow K_{1}\rightarrow \bar{w}\rightarrow z\rightarrow \bar{z}\rightarrow \bar{v}''\rightarrow v''\rightarrow v$. Then $|P_{2r+1}|=|K_{1}|+8\leq D(L_{1})+8=t+8$. $u$ has another outside neighbor $u''=ABC0$ also in $L_{2}$. Then $u''$ has a neighbor $x=ABC_{1}0\in V(L_{2})$, whose outside neighbor is $\bar{x}=ABC_{1}1\in V(M_{2})$. $\bar{x}$ has a neighbor $y=AB_{1}C_{1}1\in V(M_{2})$, whose outside neighbor is $\bar{y}=AB_{1}C_{1}0\in V(L_{3})$. Select $o=AB_{1}C'0$ in $L_{3}$. Let $K_{2}$ be a shortest path between $o$ and $\bar{y}$ in $L_{3}$. Let $\bar{o}=AB_{1}C'1$, which is an outside neighbor of $o$ in $M_{3}$. By Lemma \ref{lemE3Clianbian}, assume $v'=ABC'1\in V(M_{3})$. Note that $v'$ is adjacent to $\bar{o}$ in $M_{3}$. Construct the path $P_{2r+2}$ as $u\rightarrow u''\rightarrow x\rightarrow \bar{x}\rightarrow y\rightarrow \bar{y}\rightarrow K_{2}\rightarrow o\rightarrow \bar{o}\rightarrow v'\rightarrow v$. Then $|P_{2r+2}|=|K_{2}|+8\leq D(L_{3})+8=t+8$.

Therefore, we obtain $2r+2$ internally disjoint paths $P_{1},P_{2}, \ldots ,P_{2r+2}$ between
$u$ and $v$ in $E3C(r,s,t)$, with lengths at most $t+8$.  \qed
\\

\begin{lemma} \label{fenlei10}
		Let $u=ABCd$ and $v=A'B'C'd'$ be two distinct vertices in $E3C(r,s,t)$ such that $A=A'$, $B\neq B'$, $C=C'$ and $d\neq d'$. We discuss three cases.

\item 1. If $d=0$ and $d'=1$, there exist $2r+2$ internally disjoint paths between $u$ and $v$ with length at most $s+6$.

\item 2. If $d=0$ and $d'=2$, there exist $2r+2$ internally disjoint paths between $u$ and $v$ with length at most $s+8$.	

\item 3. If $d=1$ and $d'=2$, there exist $2r+2$ internally disjoint paths between $u$ and $v$ with length at most $s+6$.
	\end{lemma}
\noindent \textbf{Proof.} According to the values of $d$ and $d'$, the following three cases are considered.

Case 1. $d=0$ and $d'=1$.

\begin{figure}[h]
		\centering
		\includegraphics[width=0.5\textwidth]{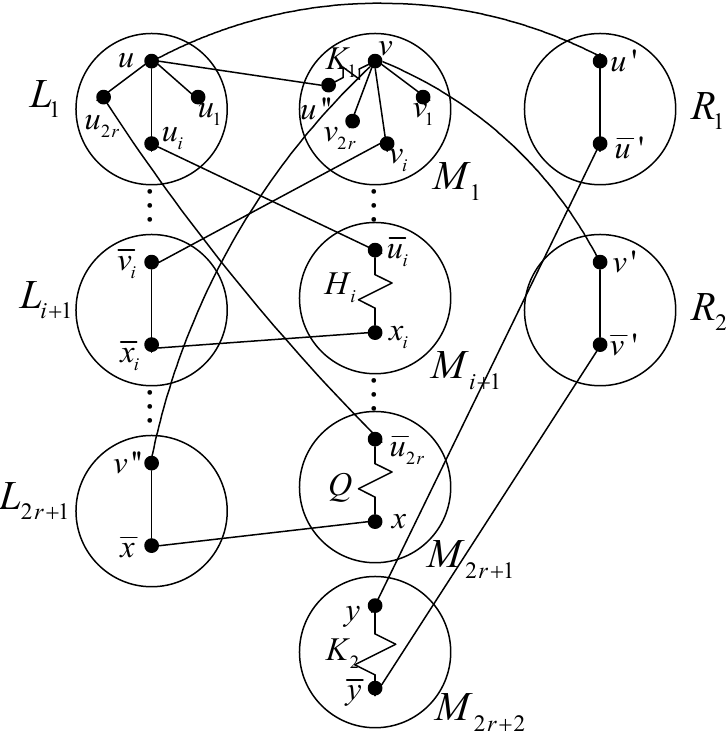}
		\renewcommand{\figurename}{Fig.}
		\caption{The illustration of Case 1 of Lemma \ref{fenlei10}}
		\label{10case1}
	\end{figure}

Without loss of generality, let $u=ABC0\in V(L_{1})$ and $v=AB'C1\in V(M_{1})$ (see Fig. \ref{10case1}). Let $u_{i}=ABC_{i}0$ be the neighbor of $u$ in $L_{1}$ for $1\leq i \leq 2r$, and $v_{i}=AB_{i}'C1$ be the neighbor of $v$ in $M_{1}$ for $1\leq i \leq 2r$. Since $u''=ABC1$ is also in $M_{1}$, there exists a shortest path $K_{1}$ between $v$ and $u''$ in $M_{1}$. Then at most one $v_{i}\in V(K_{1})$ for $1\leq i \leq 2r$. Without loss of generality, assume $v_{i}\notin V(K_{1})$ for $1\leq i \leq 2r-1$. Let $\bar{v}_{i}=AB_{i}'C0$ be an outside neighbor of $v_{i}$ in $L_{i+1}$ for $1\leq i \leq 2r-1$. Then $\bar{v}_{i}$ has a neighbor $\bar{x}_{i}=AB_{i}'C_{i}0 \in V(L_{i+1})$, whose outside neighbor is $x_{i}=AB_{i}'C_{i}1\in V(M_{i+1})$ for $1\leq i \leq 2r-1$. $u_{i}$ has an outside neighbor $\bar{u}_{i}=ABC_{i}1$ also in $M_{i+1}$ for $1\leq i \leq 2r-1$. Let $H_{i}$ be a shortest path between $\bar{u}_{i}$ and $x_{i}$ in $M_{i+1}$ for $1\leq i \leq 2r-1$. Define the path $P_{i}$ as $u\rightarrow u_{i}\rightarrow \bar{u}_{i}\rightarrow H_{i}\rightarrow x_{i}\rightarrow \bar{x}_{i}\rightarrow \bar{v}_{i}\rightarrow v_{i}\rightarrow v$ for $1\leq i \leq 2r-1$. Then $|P_{i}|=|H_{i}|+6\leq D(M_{i+1})+6=D(Q_{s}^{3})+6=s+6$ for $1\leq i \leq 2r-1$. Define the path $P_{2r}$ as $u\rightarrow u''\rightarrow K_{1}\rightarrow v$. Then $|P_{2r}|=|K_{1}|+1\leq D(M_{1})+1=D(Q_{s}^{3})+1=s+1$. Let $u'=ABC2\in V(R_{1})$. Then $u'$ has a neighbor $\bar{u}'=A_{1}BC2 \in V(R_{1})$, whose outside neighbor is $y=A_{1}BC1\in V(M_{2r+2})$. Select $\bar{y}=A_{1}B'C1$ in $M_{2r+2}$. Let $K_{2}$ be a shortest path between $y$ and $\bar{y}$ in $M_{2r+2}$. Let $\bar{v}'=A_{1}B'C2$, which is an outside neighbor of $\bar{y}$ in $R_{2}$. $v$ has an outside neighbor $v'=AB'C2$ also in $R_{2}$. Note that $v'$ is adjacent to $\bar{v}'$ in $R_{2}$. Construct the path $P_{2r+1}$ as $u\rightarrow u'\rightarrow \bar{u}'\rightarrow y\rightarrow K_{2}\rightarrow \bar{y}\rightarrow \bar{v}'\rightarrow v'\rightarrow v$. Then $|P_{2r+1}|=|K_{2}|+6\leq D(M_{2r+2})+6=s+6$. $u_{2r}$ has an outside neighbor $\bar{u}_{2r}=ABC_{2r}1$ in $M_{2r+1}$. Select $x=AB'C_{2r}1$ in $M_{2r+1}$. Let $Q$ be a shortest path between $\bar{u}_{2r}$ and $x$ in $M_{2r+1}$. Let $\bar{x}=AB'C_{2r}0$, which is an outside neighbor of $x$ in $L_{2r+1}$. $v$ has another outside neighbor $v''=AB'C0$ also in $L_{2r+1}$. Note that $v''$ is adjacent to $\bar{x}$ in $L_{2r+1}$. Construct the path $P_{2r+2}$ as $u\rightarrow u_{2r}\rightarrow \bar{u}_{2r}\rightarrow Q\rightarrow x\rightarrow \bar{x}\rightarrow v''\rightarrow v$. Then $|P_{2r+2}|=|Q|+5\leq D(M_{2r+1})+5=s+5$.

Therefore, we obtain $2r+2$ internally disjoint paths $P_{1},P_{2}, \ldots ,P_{2r+2}$ between
$u$ and $v$ in $E3C(r,s,t)$, with lengths at most $s+6$. \\

Case 2. $d=0$ and $d'=2$.

\begin{figure}[h]
		\centering
		\includegraphics[width=0.5\textwidth]{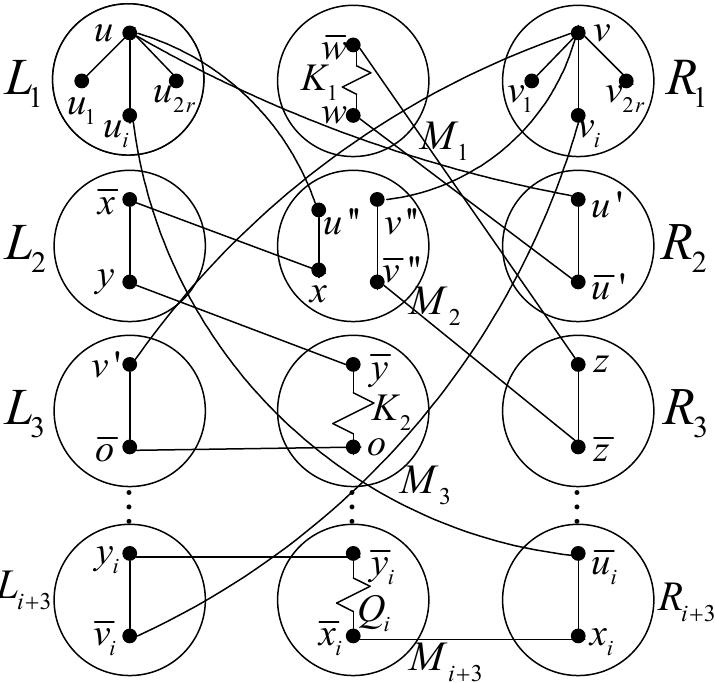}
		\renewcommand{\figurename}{Fig.}
		\caption{The illustration of Case 2 of Lemma \ref{fenlei10}}
		\label{10case2}
	\end{figure}

Without loss of generality, let $u=ABC0\in V(L_{1})$ and $v=AB'C2\in V(R_{1})$ (see Fig. \ref{10case2}). Let $u_{i}=ABC_{i}0$ be the neighbor of $u$ in $L_{1}$ for $1\leq i \leq 2r$, and $v_{i}=A_{i}B'C2$ be the neighbor of $v$ in $R_{1}$ for $1\leq i \leq 2r$. Let $\bar{u}_{i}=ABC_{i}2\in V(R_{i+3})$ be an outside neighbor of $u_{i}$ for $1\leq i \leq 2r$. Then $\bar{u}_{i}$ has a neighbor $x_{i}=A_{i}BC_{i}2\in V(R_{i+3})$, whose outside neighbor is $\bar{x}_{i}=A_{i}BC_{i}1\in V(M_{i+3})$ for $1\leq i \leq 2r$. Select $\bar{y}_{i}=A_{i}B'C_{i}1$ in $M_{i+3}$ for $1\leq i \leq 2r$. Let $Q_{i}$ be a shortest path between $\bar{y}_{i}$ and $\bar{x}_{i}$ in $M_{i+3}$ for $1\leq i \leq 2r$. Let $y_{i}=A_{i}B'C_{i}0$, which is an outside neighbor of $\bar{y}_{i}$ in $L_{i+3}$ for $1\leq i \leq 2r$. $v_{i}$ has an outside neighbor $\bar{v}_{i}=A_{i}B'C0$ also in $L_{i+3}$ for $1\leq i \leq 2r$. Note that $y_{i}$ is adjacent to $\bar{v}_{i}$ in $L_{i+3}$ for $1\leq i \leq 2r$. Construct the path $P_{i}$ as $u\rightarrow u_{i}\rightarrow \bar{u}_{i}\rightarrow x_{i}\rightarrow \bar{x}_{i}\rightarrow Q_{i}\rightarrow \bar{y}_{i}\rightarrow y_{i}\rightarrow \bar{v}_{i}\rightarrow v_{i}\rightarrow v$ for $1\leq i \leq 2r$. Then $|P_{i}|=|Q_{i}|+8\leq D(M_{i+3})+8=s+8$ for $1\leq i \leq 2r$. Let $u'=ABC2\in V (R_{2})$ be an outside neighbor of $u$. Then $u'$ has a neighbor $\bar{u}'=A_{1}BC2\in V(R_{2})$, whose outside neighbor is $w=A_{1}BC1\in V(M_{1})$. Select $\bar{w}=A_{1}B_{1}'C1$ in $M_{1}$. Let $K_{1}$ be a shortest path between $w$ and $\bar{w}$ in $M_{1}$. Let $z=A_{1}B_{1}'C2$, which is an outside neighbor of $\bar{w}$ in $R_{3}$. Then $z$ has a neighbor $\bar{z}=AB_{1}'C2\in V(R_{3})$, whose outside neighbor is $\bar{v}''=AB_{1}'C1\in V(M_{2})$. By Lemma \ref{lemE3Clianbian}, assume $v''=AB'C1\in V(M_{2})$. Note that $v''$ is adjacent to $\bar{v}''$ in $M_{2}$. Construct the path $P_{2r+1}$ as $u\rightarrow u'\rightarrow \bar{u}'\rightarrow w\rightarrow K_{1}\rightarrow \bar{w}\rightarrow z\rightarrow \bar{z}\rightarrow \bar{v}''\rightarrow v''\rightarrow v$. Then $|P_{2r+1}|=|K_{1}|+8\leq D(M_{1})+8=s+8$. $u$ has another outside neighbor $u''=ABC1$ also in $M_{2}$. Then $u''$ has a neighbor $x=AB_{1}C1\in V(M_{2})$, whose outside neighbor is $\bar{x}=AB_{1}C0\in V(L_{2})$. $\bar{x}$ has a neighbor $y=AB_{1}C_{1}0\in V(L_{2})$, whose outside neighbor is $\bar{y}=AB_{1}C_{1}1\in V(M_{3})$. Select $o=AB'C_{1}1$ in $M_{3}$. Let $K_{2}$ be a shortest path between $o$ and $\bar{y}$ in $M_{3}$. Let $\bar{o}=AB'C_{1}0$, which is an outside neighbor of $o$ in $L_{3}$. By Lemma \ref{lemE3Clianbian}, assume $v'=AB'C0\in V(L_{3})$. Note that $v'$ is adjacent to $\bar{o}$ in $L_{3}$. Construct the path $P_{2r+2}$ as $u\rightarrow u''\rightarrow x\rightarrow \bar{x}\rightarrow y\rightarrow \bar{y}\rightarrow K_{2}\rightarrow o\rightarrow \bar{o}\rightarrow v'\rightarrow v$. Then $|P_{2r+2}|=|K_{2}|+8\leq D(M_{3})+8=s+8$.

Therefore, we obtain $2r+2$ internally disjoint paths $P_{1},P_{2}, \ldots ,P_{2r+2}$ between
$u$ and $v$ in $E3C(r,s,t)$, with lengths at most $s+8$.
\\

Case 3. $d=1$ and $d'=2$.

The proof for this case is analogous to Case 1. \qed
\\

\begin{lemma} \label{fenlei11}
		Let $u=ABCd$ and $v=A'B'C'd'$ be two distinct vertices in $E3C(r,s,t)$ such that $A=A'$, $B\neq B'$, $C\neq C'$ and $d\neq d'$. We discuss three cases.

\item 1. If $d=0$ and $d'=1$, there exist $2r+2$ internally disjoint paths between $u$ and $v$ with length at most $s+t+7$.

\item 2. If $d=0$ and $d'=2$, there exist $2r+2$ internally disjoint paths between $u$ and $v$ with length at most $s+t+7$.	

\item 3. If $d=1$ and $d'=2$, there exist $2r+2$ internally disjoint paths between $u$ and $v$ with length at most $s+t+7$.
	\end{lemma}
\noindent \textbf{Proof.} According to the values of $d$ and $d'$, the following three cases are considered.

Case 1. $d=0$ and $d'=1$.

\begin{figure}[h]
		\centering
		\includegraphics[width=0.5\textwidth]{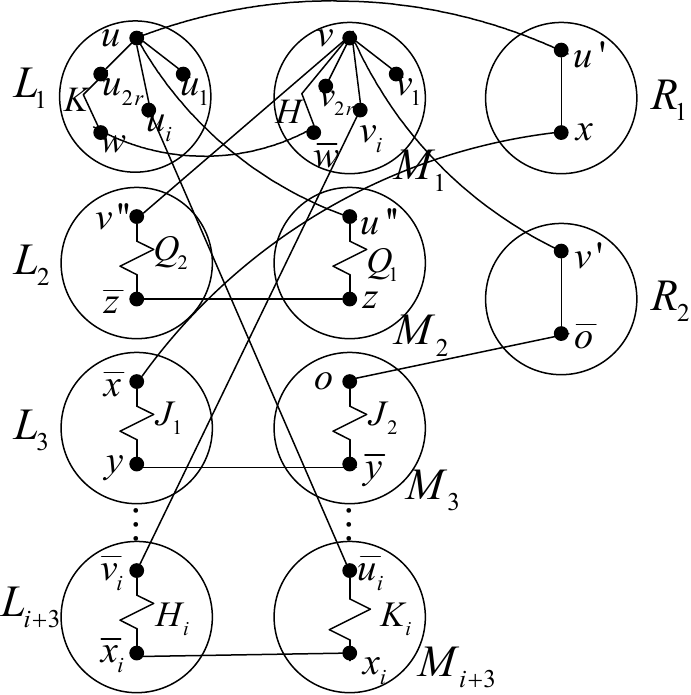}
		\renewcommand{\figurename}{Fig.}
		\caption{The illustration of Case 1 of Lemma \ref{fenlei11}}
		\label{11case1}
	\end{figure}

Without loss of generality, let $u=ABC0\in V(L_{1})$ and $v=AB'C'1\in V(M_{1})$ (see Fig. \ref{11case1}). Let $u_{i}=ABC_{i}0$ be the neighbor of $u$ in $L_{1}$ for $1\leq i \leq 2r$, and $v_{i}=AB_{i}'C'1$ be the neighbor of $v$ in $M_{1}$ for $1\leq i \leq 2r$. Choose $w=ABC'0\in V(L_{1})$. There exists a shortest path $K$ between $u$ and $w$ in $L_{1}$. Then at most one $u_{i}\in V(K)$ for $1\leq i \leq 2r$. Without loss of generality, assume $u_{i}\notin V(K)$ for $1\leq i \leq 2r-1$. Since the outside neighbor $\bar{w}=ABC'1$ of
$w$ also lies in $M_{1}$, there exists a shortest path $H$ between $v$ and $\bar{w}$ in $M_{1}$. Similarly, at most one $v_{i}\in V(H)$ for $1\leq i \leq 2r$. Assume $v_{i}\notin V(H)$ for $1\leq i \leq 2r-1$. Let $\bar{u}_{i}=ABC_{i}1\in V(M_{i+3})$ be an outside neighbor of $u_{i}$ for $1\leq i \leq 2r-1$. Select $x_{i}=AB_{i}'C_{i}1$ in $M_{i+3}$ for $1\leq i \leq 2r-1$. Let $K_{i}$ be a shortest path between $\bar{u}_{i}$ and $x_{i}$ in $M_{i+3}$ for $1\leq i \leq 2r-1$. Note that $x_{i}$ has an outside neighbor $\bar{x}_{i}=AB_{i}'C_{i}0\in V(L_{i+3})$, and $\bar{v}_{i}=AB_{i}'C'0\in V(L_{i+3})$ is the outside neighbor of $v_{i}$ for $1\leq i \leq 2r-1$. Let $H_{i}$ be a shortest path between $\bar{x}_{i}$ and $\bar{v}_{i}$ in $L_{i+3}$ for $1\leq i \leq 2r-1$. Define the path $P_{i}$ as $u\rightarrow u_{i}\rightarrow \bar{u}_{i}\rightarrow K_{i}\rightarrow x_{i}\rightarrow \bar{x}_{i}\rightarrow H_{i}\rightarrow \bar{v}_{i}\rightarrow v_{i}\rightarrow v$ for $1\leq i \leq 2r-1$. Then $|P_{i}|=|H_{i}|+|K_{i}|+5\leq D(M_{i+3})+D(L_{i+3})+5=D(Q_{s}^{3})+D(Q_{t}^{3})+5=s+t+5$ for $1\leq i \leq 2r-1$. Define the path $P_{2r}$ as $u\rightarrow K\rightarrow w\rightarrow \bar{w}\rightarrow H\rightarrow v$. Then $|P_{2r}|=|H|+|K|+1\leq D(M_{1})+D(L_{1})+1=D(Q_{s}^{3})+D(Q_{t}^{3})+1=s+t+1$.
Assume $u''=ABC1\in V(M_{2})$ and $v''=AB'C'0\in V(L_{2})$. Choose $z=AB'C1$ in $M_{2}$. Let $Q_{1}$ be a shortest path between $u''$ and $z$ in $M_{2}$. $z$ has an outside neighbor $\bar{z}=AB'C0$ also in $L_{2}$. Then there exists a shortest path $Q_{2}$ between $v''$ and $\bar{z}$ in $L_{2}$.  Let $u'=ABC2\in V(R_{1})$ with neighbor $x=A_{1}BC2\in V(R_{1})$, whose outside neighbor is $\bar{x}=A_{1}BC0\in V(L_{3})$. Select $y=A_{1}BC'0$ in $L_{3}$. Then there exists a shortest path $J_{1}$ between $\bar{x}$ and $y$ in $L_{3}$. Note that $y$ has an outside neighbor $\bar{y}=A_{1}BC'1$ in $M_{3}$. Select $o=A_{1}B'C'1$ in $M_{3}$. Let $J_{2}$ be a shortest path between $o$ and $\bar{y}$ in $M_{3}$. $o$ has an outside neighbor $\bar{o}=A_{1}B'C'2$ in $R_{2}$. $v$ has another outside neighbor $v'=AB'C'2$ also in $R_{2}$. Note that $v'$ is adjacent to $\bar{o}$ in $R_{2}$. Construct the path $P_{2r+1}$ as $u\rightarrow u''\rightarrow Q_{1}\rightarrow z\rightarrow \bar{z}\rightarrow Q_{2}\rightarrow v''\rightarrow v$. Then $|P_{2r+1}|=|Q_{1}|+|Q_{2}|+3\leq D(M_{2})+D(L_{2})+3=s+t+3$. Construct the path $P_{2r+2}$ as $u\rightarrow u'\rightarrow x\rightarrow  \bar{x}\rightarrow J_{1}\rightarrow y\rightarrow \bar{y}\rightarrow J_{2}\rightarrow o\rightarrow  \bar{o}\rightarrow v'\rightarrow v$. Then $|P_{2r+2}|=|J_{1}|+|J_{2}|+7\leq D(L_{3})+D(M_{3})+7=s+t+7$.

Therefore, we obtain $2r+2$ internally disjoint paths $P_{1},P_{2}, \ldots ,P_{2r+2}$ between
$u$ and $v$ in $E3C(r,s,t)$, with lengths at most $s+t+7$. \\

Case 2. $d=0$ and $d'=2$.

\begin{figure}[h]
		\centering
		\includegraphics[width=0.5\textwidth]{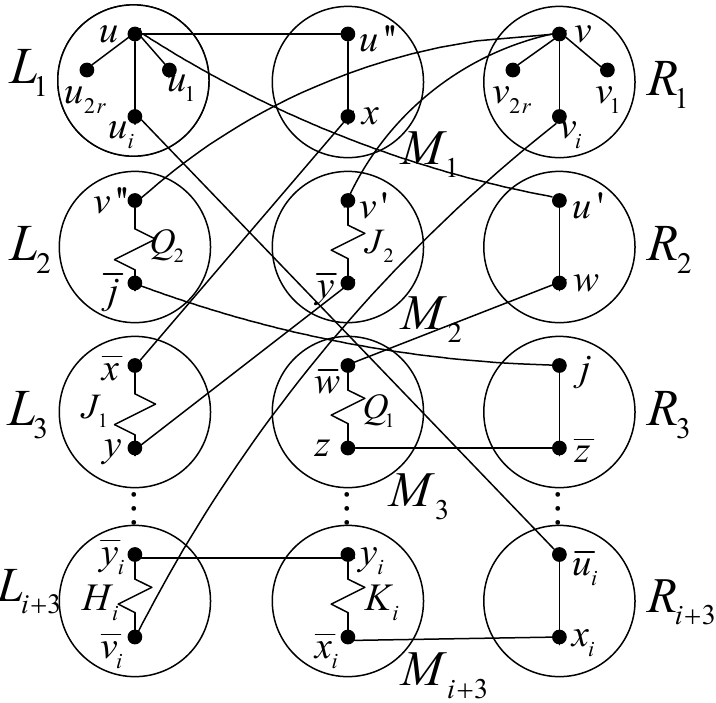}
		\renewcommand{\figurename}{Fig.}
		\caption{The illustration of Case 2 of Lemma \ref{fenlei11}}
		\label{11case2}
	\end{figure}

Without loss of generality, let $u=ABC0\in V(L_{1})$ and $v=AB'C'2\in V(R_{1})$ (see Fig. \ref{11case2}). Define $u_{i}$ as one of the $2r$ neighbors of $u$ in $L_{1}$, where $u_{i}=ABC_{i}0$ for $1\leq i \leq 2r$.
Define $v_{i}$ as one of the $2r$ neighbors of $v$ in $R_{1}$ , where $v_{i}=A_{i}B'C'2$ for $1\leq i \leq 2r$. Let $\bar{u}_{i}=ABC_{i}2\in V(R_{i+3})$ be an outside neighbor of $u_{i}$ for $1\leq i \leq 2r$. Then $\bar{u}_{i}$ has a neighbor  $x_{i}=A_{i}BC_{i}2\in V(R_{i+3})$, whose outside neighbor is $\bar{x}_{i}=A_{i}BC_{i}1\in V(M_{i+3})$ for $1\leq i \leq 2r$. Select $y_{i}=A_{i}B'C_{i}1$ in $M_{i+3}$ for $1\leq i \leq 2r$. Let $K_{i}$ be a shortest path between $\bar{x}_{i}$ and $y_{i}$ in $M_{i+3}$ for $1\leq i \leq 2r$. Let $\bar{y}_{i}=A_{i}B'C_{i}0$, which is an outside neighbor of $y_{i}$ in $L_{i+3}$ for $1\leq i \leq 2r$. $v_{i}$ has an outside neighbor $\bar{v}_{i}=A_{i}B'C'0$ also in $L_{i+3}$ for $1\leq i \leq 2r$. Then there exists a shortest path $H_{i}$ between $\bar{v}_{i}$ and $\bar{y}_{i}$ in $L_{i+3}$ for $1\leq i \leq 2r$. Construct the path $P_{i}$ as $u\rightarrow u_{i}\rightarrow \bar{u}_{i}\rightarrow x_{i}\rightarrow \bar{x}_{i}\rightarrow K_{i}\rightarrow y_{i}\rightarrow \bar{y}_{i}\rightarrow H_{i}\rightarrow \bar{v}_{i}\rightarrow v_{i}\rightarrow v$ for $1\leq i \leq 2r$. Then $|P_{i}|=|K_{i}|+|H_{i}|+7\leq D(M_{i+3})+D(L_{i+3})+7=s+t+7$ for $1\leq i \leq 2r$. Let $u'=ABC2$ be the neighbor of $u$ in $R_{2}$. Then $u'$ has a neighbor  $w=A_{1}BC2\in V(R_{2})$, whose outside neighbor is $\bar{w}=A_{1}BC1\in V(M_{3})$. Select $z=A_{1}B'C1$ in $M_{3}$. Let $Q_{1}$ be a shortest path between $\bar{w}$ and $z$ in $M_{3}$. Let $\bar{z}=A_{1}B'C2$, which is an outside neighbor of $z$ in $R_{3}$. Then $\bar{z}$ has a neighbor  $j=AB'C2\in V(R_{3})$, whose outside neighbor is $\bar{j}=AB'C0\in V(L_{2})$.
 $v$ has an outside neighbor $v''=AB'C'0$ also in $L_{2}$. Let $Q_{2}$ be a shortest path between $v''$ and $\bar{j}$ in $L_{2}$. Define the path $P_{2r+1}$ as $u\rightarrow u'\rightarrow w\rightarrow \bar{w}\rightarrow Q_{1}\rightarrow z\rightarrow \bar{z}\rightarrow j\rightarrow \bar{j}\rightarrow Q_{2}\rightarrow v''\rightarrow v$. This satisfies $|P_{2r+1}|=|Q_{1}|+|Q_{2}|+7\leq D(M_{3})+D(L_{2})+7=s+t+7$. Let $u''=ABC1\in V(M_{1})$ be another outside neighbor of $u$. Then $u''$ has a neighbor  $x=AB_{1}C1\in V(M_{1})$, whose outside neighbor is $\bar{x}=AB_{1}C0\in V(L_{3})$. Select $y=AB_{1}C'0$ in $L_{3}$. Let $J_{1}$ be a shortest path between $\bar{x}$ and $y$ in $L_{3}$. Let $\bar{y}=AB_{1}C'1$, which is an outside neighbor of $y$ in $M_{2}$. $v$ has another outside neighbor
$v'=AB'C'1$ also in $M_{2}$. There exists a shortest path $J_{2}$ between $v'$ and $\bar{y}$ in $M_{2}$.
Construct the path $P_{2r+2}$ as $u\rightarrow u''\rightarrow x\rightarrow \bar{x}\rightarrow J_{1}\rightarrow y\rightarrow \bar{y}\rightarrow J_{2}\rightarrow v'\rightarrow v$. Then $|P_{2r+2}|=|J_{1}|+|J_{2}|+5\leq D(L_{3})+D(M_{2})+5=s+t+5$.

Therefore, we obtain $2r+2$ internally disjoint paths $P_{1},P_{2}, \ldots ,P_{2r+2}$ between
$u$ and $v$ in $E3C(r,s,t)$, with lengths at most $s+t+7$. \\

Case 3. $d=1$ and $d'=2$.

The proof for this case is analogous to Case 2. \qed
\\

\begin{lemma} \label{fenlei12}
		Let $u=ABCd$ and $v=A'B'C'd'$ be two distinct vertices in $E3C(r,s,t)$ such that $A\neq A'$, $B=B'$, $C=C'$ and $d\neq d'$. We discuss three cases.

\item 1. If $d=0$ and $d'=1$, there exist $2r+2$ internally disjoint paths between $u$ and $v$ with length at most $r+8$.

\item 2. If $d=0$ and $d'=2$, there exist $2r+2$ internally disjoint paths between $u$ and $v$ with length at most $r+6$.	

\item 3. If $d=1$ and $d'=2$, there exist $2r+2$ internally disjoint paths between $u$ and $v$ with length at most $r+6$.
	\end{lemma}
\noindent \textbf{Proof.} According to the values of $d$ and $d'$, the following three cases are considered.

Case 1. $d=0$ and $d'=1$.

\begin{figure}[h]
		\centering
		\includegraphics[width=0.5\textwidth]{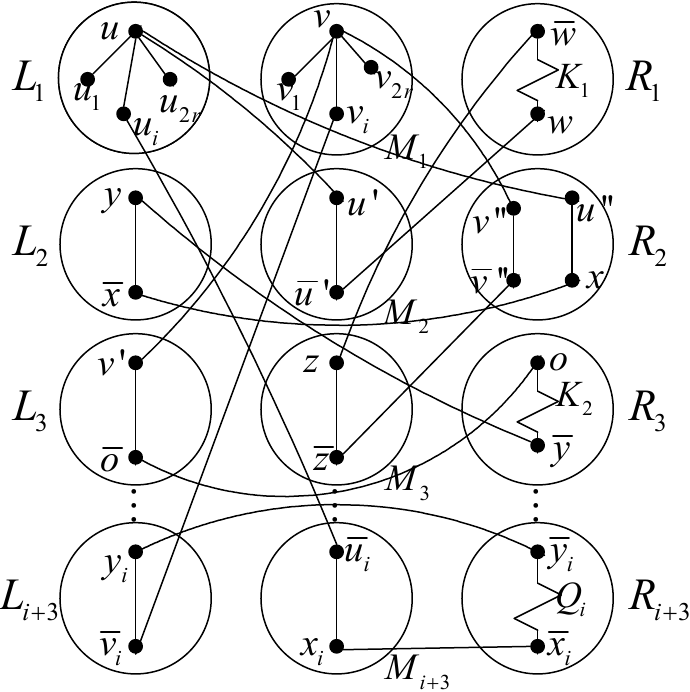}
		\renewcommand{\figurename}{Fig.}
		\caption{The illustration of Case 1 of Lemma \ref{fenlei12}}
		\label{12case1}
	\end{figure}

Without loss of generality, let $u=ABC0\in V(L_{1})$ and $v=A'BC1\in V(M_{1})$ (see Fig. \ref{12case1}). Let $u_{i}=ABC_{i}0$ be the neighbor of $u$ in $L_{1}$ for $1\leq i \leq 2r$, and $v_{i}=A'B_{i}C1$ be the neighbor of $v$ in $M_{1}$ for $1\leq i \leq 2r$. Let $\bar{u}_{i}=ABC_{i}1\in V (M_{i+3})$ be an outside neighbor of $u_{i}$ for $1\leq i \leq 2r$. Then $\bar{u}_{i}$ has a neighbor $x_{i}=AB_{i}C_{i}1\in V(M_{i+3})$, whose outside neighbor is $\bar{x}_{i}=AB_{i}C_{i}2\in V(R_{i+3})$ for $1\leq i \leq 2r$. Select $\bar{y}_{i}=A'B_{i}C_{i}2$ in $R_{i+3}$ for $1\leq i \leq 2r$. Let $Q_{i}$ be a shortest path between $\bar{y}_{i}$ and $\bar{x}_{i}$ in $R_{i+3}$ for $1\leq i \leq 2r$. Let $y_{i}=A'B_{i}C_{i}0$, which is an outside neighbor of $\bar{y}_{i}$ in $L_{i+3}$ for $1\leq i \leq 2r$. $v_{i}$ has an outside neighbor $\bar{v}_{i}=A'B_{i}C0$ also in $L_{i+3}$ for $1\leq i \leq 2r$. Note that $y_{i}$ is adjacent to $\bar{v}_{i}$ in $L_{i+3}$ for $1\leq i \leq 2r$. Construct the path $P_{i}$ as $u\rightarrow u_{i}\rightarrow \bar{u}_{i}\rightarrow x_{i}\rightarrow \bar{x}_{i}\rightarrow Q_{i}\rightarrow \bar{y}_{i}\rightarrow y_{i}\rightarrow \bar{v}_{i}\rightarrow v_{i}\rightarrow v$ for $1\leq i \leq 2r$. Then $|P_{i}|=|Q_{i}|+8\leq D(R_{i+3})+8=r+8$ for $1\leq i \leq 2r$. Let $u'=ABC1\in V(M_{2})$ be an outside neighbor of $u$. Then $u'$ has a neighbor $\bar{u}'=AB_{1}C1\in V(M_{2})$, whose outside neighbor is $w=AB_{1}C2\in V(R_{1})$. Select $\bar{w}=A_{1}'B_{1}C2$ in $R_{1}$. Let $K_{1}$ be a shortest path between $w$ and $\bar{w}$ in $R_{1}$. Let $z=A_{1}'B_{1}C1$, which is an outside neighbor of $\bar{w}$ in $M_{3}$. Then $z$ has a neighbor $\bar{z}=A_{1}'BC1 \in V(M_{3})$, whose outside neighbor is $\bar{v}''=A_{1}'BC2 \in V(R_{2})$. By Lemma \ref{lemE3Clianbian}, assume $v''=A'BC2\in V(R_{2})$. Note that $v''$ is adjacent to $\bar{v}''$ in $R_{2}$. Construct the path $P_{2r+1}$ as $u\rightarrow u'\rightarrow \bar{u}'\rightarrow w\rightarrow K_{1}\rightarrow \bar{w}\rightarrow z\rightarrow \bar{z}\rightarrow \bar{v}''\rightarrow v''\rightarrow v$. Then $|P_{2r+1}|=|K_{1}|+8\leq D(R_{1})+8=r+8$. $u$ has another outside neighbor $u''=ABC2$ also in $R_{2}$. Then $u''$ has a neighbor $x=A_{1}BC2\in V(R_{2})$, whose outside neighbor is $\bar{x}=A_{1}BC0\in V(L_{2})$. $\bar{x}$ has a neighbor $y=A_{1}BC_{1}0\in V(L_{2})$, whose outside neighbor is $\bar{y}=A_{1}BC_{1}2\in V(R_{3})$. Select $o=A'BC_{1}2$ in $R_{3}$. Let $K_{2}$ be a shortest path between $o$ and $\bar{y}$ in $R_{3}$. Let $\bar{o}=A'BC_{1}0$, which is an outside neighbor of $o$ in $L_{3}$. By Lemma \ref{lemE3Clianbian}, assume $v'=A'BC0\in V(L_{3})$. Note that $v'$ is adjacent to $\bar{o}$ in $L_{3}$. Construct the path $P_{2r+2}$ as $u\rightarrow u''\rightarrow x\rightarrow \bar{x}\rightarrow y\rightarrow \bar{y}\rightarrow K_{2}\rightarrow o\rightarrow \bar{o}\rightarrow v'\rightarrow v$. Then $|P_{2r+2}|=|K_{2}|+8\leq D(R_{3})+8=r+8$.

Therefore, we obtain $2r+2$ internally disjoint paths $P_{1},P_{2}, \ldots ,P_{2r+2}$ between
$u$ and $v$ in $E3C(r,s,t)$, with lengths at most $r+8$.
\\

Case 2. $d=0$ and $d'=2$.

\begin{figure}[h]
		\centering
		\includegraphics[width=0.5\textwidth]{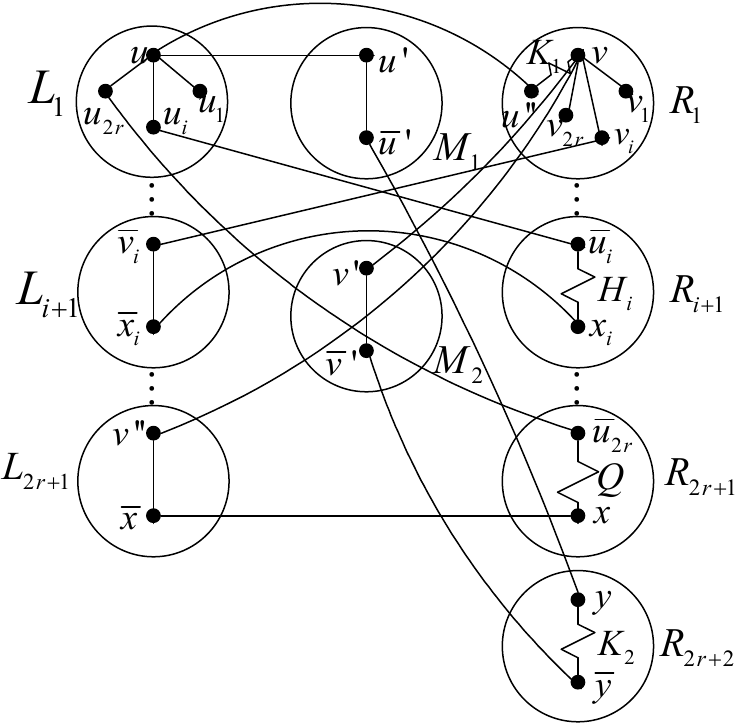}
		\renewcommand{\figurename}{Fig.}
		\caption{The illustration of Case 2 of Lemma \ref{fenlei12}}
		\label{12case2}
	\end{figure}

Without loss of generality, let $u=ABC0\in V(L_{1})$ and $v=A'BC2\in V(R_{1})$ (see Fig. \ref{12case2}). Let $u_{i}=ABC_{i}0$ be the neighbor of $u$ in $L_{1}$ for $1\leq i \leq 2r$, and $v_{i}=A_{i}'BC2$ be the neighbor of $v$ in $R_{1}$ for $1\leq i \leq 2r$. Since $u''=ABC2$ is also in $R_{1}$, there exists a shortest path $K_{1}$ between $v$ and $u''$ in $R_{1}$. Then at most one $v_{i}\in V(K_{1})$ for $1\leq i \leq 2r$. Without loss of generality, assume $v_{i}\notin V(K_{1})$ for $1\leq i \leq 2r-1$. Let  $\bar{u}_{i}=ABC_{i}2\in V(R_{i+1})$ be an outside neighbor of $u_{i}$ for $1\leq i \leq 2r-1$. Select $x_{i}=A_{i}'BC_{i}2$ in $R_{i+1}$ for $1\leq i \leq 2r-1$. Let $H_{i}$ be a shortest path between $\bar{u}_{i}$ and $x_{i}$ in $R_{i+1}$ for $1\leq i \leq 2r-1$. Let $\bar{x}_{i}=A_{i}'BC_{i}0$, which is an outside neighbor of $x_{i}$ in $L_{i+1}$ for $1\leq i \leq 2r-1$. $v_{i}$ has an outside neighbor $\bar{v}_{i}=A_{i}'BC0$ also in $L_{i+1}$ for $1\leq i \leq 2r-1$. Note that $\bar{v}_{i}$ is adjacent to $\bar{x}_{i}$ in $L_{i+1}$ for $1\leq i \leq 2r-1$. Define the path $P_{i}$ as $u\rightarrow u_{i}\rightarrow \bar{u}_{i}\rightarrow H_{i}\rightarrow x_{i}\rightarrow \bar{x}_{i}\rightarrow \bar{v}_{i}\rightarrow v_{i}\rightarrow v$ for $1\leq i \leq 2r-1$. Then $|P_{i}|=|H_{i}|+6\leq D(R_{i+1})+6=D(Q_{r}^{3})+6=r+6$ for $1\leq i \leq 2r-1$. Define the path $P_{2r}$ as $u\rightarrow u''\rightarrow K_{1}\rightarrow v$. Then $|P_{2r}|=|K_{1}|+1\leq D(R_{1})+1=D(Q_{r}^{3})+1=r+1$. Let $u'=ABC1\in V(M_{1})$. Then $u'$ has a neighbor $\bar{u}'=AB_{1}C1 \in V(M_{1})$, whose outside neighbor is $y=AB_{1}C2\in V(R_{2r+2})$. Select $\bar{y}=A'B_{1}C2$ in $R_{2r+2}$. Let $K_{2}$ be a shortest path between $y$ and $\bar{y}$ in $R_{2r+2}$. Let $\bar{v}'=A'B_{1}C1$, which is an outside neighbor of $\bar{y}$ in $M_{2}$. $v$ has an outside neighbor $v'=A'BC1$ also in $M_{2}$. Note that $v'$ is adjacent to $\bar{v}'$ in $M_{2}$. Construct the path $P_{2r+1}$ as $u\rightarrow u'\rightarrow \bar{u}'\rightarrow y\rightarrow K_{2}\rightarrow \bar{y}\rightarrow \bar{v}'\rightarrow v'\rightarrow v$. Then $|P_{2r+1}|=|K_{2}|+6\leq D(R_{2r+2})+6=r+6$. $u_{2r}$ has an outside neighbor $\bar{u}_{2r}=ABC_{2r}2$ in $R_{2r+1}$. Select $x=A'BC_{2r}2$ in $R_{2r+1}$. Let $Q$ be a shortest path between $\bar{u}_{2r}$ and $x$ in $R_{2r+1}$. Let $\bar{x}=A'BC_{2r}0$, which is an outside neighbor of $x$ in $L_{2r+1}$. $v$ has another outside neighbor $v''=A'BC0$ also in $L_{2r+1}$. Note that $v''$ is adjacent to $\bar{x}$ in $L_{2r+1}$. Construct the path $P_{2r+2}$ as $u\rightarrow u_{2r}\rightarrow \bar{u}_{2r}\rightarrow Q\rightarrow x\rightarrow \bar{x}\rightarrow v''\rightarrow v$. Then $|P_{2r+2}|=|Q|+5\leq D(R_{2r+1})+5=r+5$.

Therefore, we obtain $2r+2$ internally disjoint paths $P_{1},P_{2}, \ldots ,P_{2r+2}$ between
$u$ and $v$ in $E3C(r,s,t)$, with lengths at most $r+6$. \\

Case 3. $d=1$ and $d'=2$.

The proof for this case is analogous to Case 2. \qed
\\

\begin{lemma} \label{fenlei13}
		Let $u=ABCd$ and $v=A'B'C'd'$ be two distinct vertices in $E3C(r,s,t)$ such that $A\neq A'$, $B=B'$, $C\neq C'$ and $d\neq d'$. We discuss three cases.

\item 1. If $d=0$ and $d'=1$, there exist $2r+2$ internally disjoint paths between $u$ and $v$ with length at most $r+t+7$.

\item 2. If $d=0$ and $d'=2$, there exist $2r+2$ internally disjoint paths between $u$ and $v$ with length at most $r+t+7$.	

\item 3. If $d=1$ and $d'=2$, there exist $2r+2$ internally disjoint paths between $u$ and $v$ with length at most $r+t+7$.
	\end{lemma}
\noindent \textbf{Proof.} According to the values of $d$ and $d'$, the following three cases are considered.

Case 1. $d=0$ and $d'=1$.

\begin{figure}[h]
		\centering
		\includegraphics[width=0.5\textwidth]{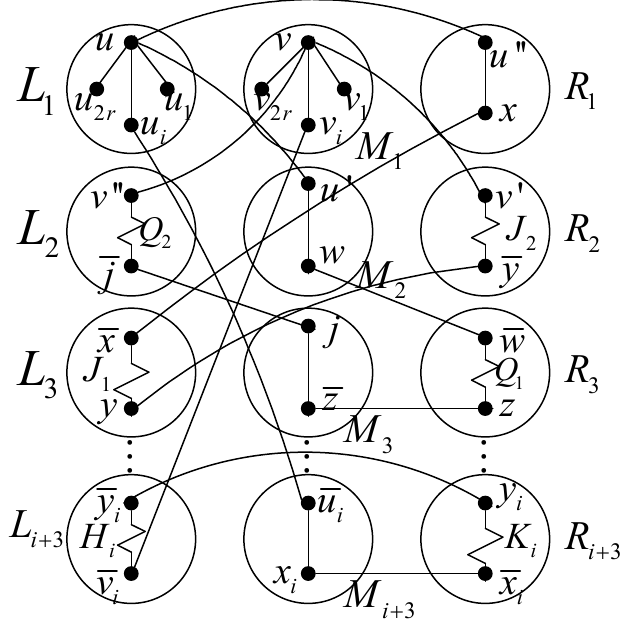}
		\renewcommand{\figurename}{Fig.}
		\caption{The illustration of Case 1 of Lemma \ref{fenlei13}}
		\label{13case1}
	\end{figure}

Without loss of generality, let $u=ABC0\in V(L_{1})$ and $v=A'BC'1\in V(M_{1})$ (see Fig. \ref{13case1}). Define $u_{i}$ as one of the $2r$ neighbors of $u$ in $L_{1}$, where $u_{i}=ABC_{i}0$ for $1\leq i \leq 2r$.
Define $v_{i}$ as one of the $2r$ neighbors of $v$ in $M_{1}$ , where $v_{i}=A'B_{i}C'1$ for $1\leq i \leq 2r$. Let $\bar{u}_{i}=ABC_{i}1\in V(M_{i+3})$ be an outside neighbor of $u_{i}$ for $1\leq i \leq 2r$. Then $\bar{u}_{i}$ has a neighbor  $x_{i}=AB_{i}C_{i}1\in V(M_{i+3})$, whose outside neighbor is $\bar{x}_{i}=AB_{i}C_{i}2\in V(R_{i+3})$ for $1\leq i \leq 2r$. Select $y_{i}=A'B_{i}C_{i}2$ in $R_{i+3}$ for $1\leq i \leq 2r$. Let $K_{i}$ be a shortest path between $\bar{x}_{i}$ and $y_{i}$ in $R_{i+3}$ for $1\leq i \leq 2r$. Let $\bar{y}_{i}=A'B_{i}C_{i}0$, which is an outside neighbor of $y_{i}$ in $L_{i+3}$ for $1\leq i \leq 2r$. $v_{i}$ has an outside neighbor $\bar{v}_{i}=A'B_{i}C'0$ also in $L_{i+3}$ for $1\leq i \leq 2r$. Then there exists a shortest path $H_{i}$ between $\bar{v}_{i}$ and $\bar{y}_{i}$ in $L_{i+3}$ for $1\leq i \leq 2r$. Construct the path $P_{i}$ as $u\rightarrow u_{i}\rightarrow \bar{u}_{i}\rightarrow x_{i}\rightarrow \bar{x}_{i}\rightarrow K_{i}\rightarrow y_{i}\rightarrow \bar{y}_{i}\rightarrow H_{i}\rightarrow \bar{v}_{i}\rightarrow v_{i}\rightarrow v$ for $1\leq i \leq 2r$. Then $|P_{i}|=|K_{i}|+|H_{i}|+7\leq D(R_{i+3})+D(L_{i+3})+7=r+t+7$ for $1\leq i \leq 2r$. Let $u'=ABC1$ be the neighbor of $u$ in $M_{2}$. Then $u'$ has a neighbor  $w=AB_{1}C1\in V(M_{2})$, whose outside neighbor is $\bar{w}=AB_{1}C2\in V(R_{3})$. Select $z=A'B_{1}C2$ in $R_{3}$. Let $Q_{1}$ be a shortest path between $\bar{w}$ and $z$ in $R_{3}$. Let $\bar{z}=A'B_{1}C1$, which is an outside neighbor of $z$ in $M_{3}$. Then $\bar{z}$ has a neighbor  $j=A'BC1\in V(M_{3})$, whose outside neighbor is $\bar{j}=A'BC0\in V(L_{2})$.
 $v$ has an outside neighbor $v''=A'BC'0$ also in $L_{2}$. Let $Q_{2}$ be a shortest path between $v''$ and $\bar{j}$ in $L_{2}$. Define the path $P_{2r+1}$ as $u\rightarrow u'\rightarrow w\rightarrow \bar{w}\rightarrow Q_{1}\rightarrow z\rightarrow \bar{z}\rightarrow j\rightarrow \bar{j}\rightarrow Q_{2}\rightarrow v''\rightarrow v$. This satisfies $|P_{2r+1}|=|Q_{1}|+|Q_{2}|+7\leq D(R_{3})+D(L_{2})+7=r+t+7$. Let $u''=ABC2\in V(R_{1})$ be another outside neighbor of $u$. Then $u''$ has a neighbor  $x=A_{1}BC2\in V(R_{1})$, whose outside neighbor is $\bar{x}=A_{1}BC0\in V(L_{3})$. Select $y=A_{1}BC'0$ in $L_{3}$. Let $J_{1}$ be a shortest path between $\bar{x}$ and $y$ in $L_{3}$. Let $\bar{y}=A_{1}BC'2$, which is an outside neighbor of $y$ in $R_{2}$. $v$ has another outside neighbor
$v'=A'BC'2$ also in $R_{2}$. There exists a shortest path $J_{2}$ between $v'$ and $\bar{y}$ in $R_{2}$.
Construct the path $P_{2r+2}$ as $u\rightarrow u''\rightarrow x\rightarrow \bar{x}\rightarrow J_{1}\rightarrow y\rightarrow \bar{y}\rightarrow J_{2}\rightarrow v'\rightarrow v$. Then $|P_{2r+2}|=|J_{1}|+|J_{2}|+5\leq D(L_{3})+D(R_{2})+5=r+t+5$.

Therefore, we obtain $2r+2$ internally disjoint paths $P_{1},P_{2}, \ldots ,P_{2r+2}$ between
$u$ and $v$ in $E3C(r,s,t)$, with lengths at most $r+t+7$. \\

Case 2. $d=0$ and $d'=2$.

\begin{figure}[h]
		\centering
		\includegraphics[width=0.5\textwidth]{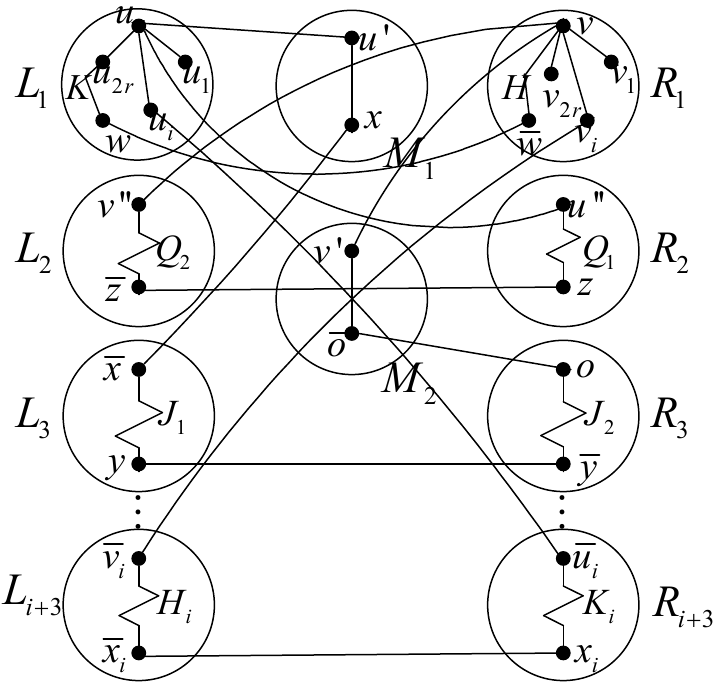}
		\renewcommand{\figurename}{Fig.}
		\caption{The illustration of Case 2 of Lemma \ref{fenlei13}}
		\label{13case2}
	\end{figure}

Without loss of generality, let $u=ABC0\in V(L_{1})$ and $v=A'BC'2\in V(R_{1})$ (see Fig. \ref{13case2}). Let $u_{i}=ABC_{i}0$ be the neighbor of $u$ in $L_{1}$ for $1\leq i \leq 2r$, and $v_{i}=A_{i}'BC'2$ be the neighbor of $v$ in $R_{1}$ for $1\leq i \leq 2r$. Choose $w=ABC'0\in V(L_{1})$. There exists a shortest path $K$ between $u$ and $w$ in $L_{1}$. Then at most one $u_{i}\in V(K)$ for $1\leq i \leq 2r$. Without loss of generality, assume $u_{i}\notin V(K)$ for $1\leq i \leq 2r-1$. Since the outside neighbor $\bar{w}=ABC'2$ of
$w$ also lies in $R_{1}$, there exists a shortest path $H$ between $v$ and $\bar{w}$ in $R_{1}$. Similarly, at most one $v_{i}\in V(H)$ for $1\leq i \leq 2r$. Assume $v_{i}\notin V(H)$ for $1\leq i \leq 2r-1$. Let $\bar{u}_{i}=ABC_{i}2\in V(R_{i+3})$ be an outside neighbor of $u_{i}$ for $1\leq i \leq 2r-1$. Select $x_{i}=A_{i}'BC_{i}2$ in $R_{i+3}$ for $1\leq i \leq 2r-1$. Let $K_{i}$ be a shortest path between $\bar{u}_{i}$ and $x_{i}$ in $R_{i+3}$ for $1\leq i \leq 2r-1$. Note that $x_{i}$ has an outside neighbor $\bar{x}_{i}=A_{i}'BC_{i}0\in V(L_{i+3})$, and $\bar{v}_{i}=A_{i}'BC'0\in V(L_{i+3})$ is the outside neighbor of $v_{i}$ for $1\leq i \leq 2r-1$. Let $H_{i}$ be a shortest path between $\bar{x}_{i}$ and $\bar{v}_{i}$ in $L_{i+3}$ for $1\leq i \leq 2r-1$. Define the path $P_{i}$ as $u\rightarrow u_{i}\rightarrow \bar{u}_{i}\rightarrow K_{i}\rightarrow x_{i}\rightarrow \bar{x}_{i}\rightarrow H_{i}\rightarrow \bar{v}_{i}\rightarrow v_{i}\rightarrow v$ for $1\leq i \leq 2r-1$. Then $|P_{i}|=|H_{i}|+|K_{i}|+5\leq D(R_{i+3})+D(L_{i+3})+5=D(Q_{r}^{3})+D(Q_{t}^{3})+5=r+t+5$ for $1\leq i \leq 2r-1$. Define the path $P_{2r}$ as $u\rightarrow K\rightarrow w\rightarrow \bar{w}\rightarrow H\rightarrow v$. Then $|P_{2r}|=|H|+|K|+1\leq D(R_{1})+D(L_{1})+1=D(Q_{r}^{3})+D(Q_{t}^{3})+1=r+t+1$.
Assume $u''=ABC2\in V(R_{2})$ and $v''=A'BC'0\in V(L_{2})$. Choose $z=A'BC2$ in $R_{2}$. Let $Q_{1}$ be a shortest path between $u''$ and $z$ in $R_{2}$. $z$ has an outside neighbor $\bar{z}=A'BC0$ also in $L_{2}$. Then there exists a shortest path $Q_{2}$ between $v''$ and $\bar{z}$ in $L_{2}$.  Let $u'=ABC1\in V(M_{1})$ with neighbor $x=AB_{1}C1\in V(M_{1})$, whose outside neighbor is $\bar{x}=AB_{1}C0\in V(L_{3})$. Select $y=AB_{1}C'0$ in $L_{3}$. Then there exists a shortest path $J_{1}$ between $\bar{x}$ and $y$ in $L_{3}$. Note that $y$ has an outside neighbor $\bar{y}=AB_{1}C'2$ in $R_{3}$. Select $o=A'B_{1}C'2$ in $R_{3}$. Let $J_{2}$ be a shortest path between $o$ and $\bar{y}$ in $R_{3}$. $o$ has an outside neighbor $\bar{o}=A'B_{1}C'1$ in $M_{2}$. $v$ has another outside neighbor $v'=A'BC'1$ also in $M_{2}$. Note that $v'$ is adjacent to $\bar{o}$ in $M_{2}$. Construct the path $P_{2r+1}$ as $u\rightarrow u''\rightarrow Q_{1}\rightarrow z\rightarrow \bar{z}\rightarrow Q_{2}\rightarrow v''\rightarrow v$. Then $|P_{2r+1}|=|Q_{1}|+|Q_{2}|+3\leq D(R_{2})+D(L_{2})+3=r+t+3$. Construct the path $P_{2r+2}$ as $u\rightarrow u'\rightarrow x\rightarrow  \bar{x}\rightarrow J_{1}\rightarrow y\rightarrow \bar{y}\rightarrow J_{2}\rightarrow o\rightarrow  \bar{o}\rightarrow v'\rightarrow v$. Then $|P_{2r+2}|=|J_{1}|+|J_{2}|+7\leq D(L_{3})+D(R_{3})+7=r+t+7$.

Therefore, we obtain $2r+2$ internally disjoint paths $P_{1},P_{2}, \ldots ,P_{2r+2}$ between
$u$ and $v$ in $E3C(r,s,t)$, with lengths at most $r+t+7$. \\

Case 3. $d=1$ and $d'=2$.

The proof for this case is analogous to Case 1. \qed
\\

\begin{lemma} \label{fenlei14}
		Let $u=ABCd$ and $v=A'B'C'd'$ be two distinct vertices in $E3C(r,s,t)$ such that $A\neq A'$, $B\neq B'$, $C=C'$ and $d\neq d'$. We discuss three cases.

\item 1. If $d=0$ and $d'=1$, there exist $2r+2$ internally disjoint paths between $u$ and $v$ with length at most $r+s+7$.

\item 2. If $d=0$ and $d'=2$, there exist $2r+2$ internally disjoint paths between $u$ and $v$ with length at most $r+s+7$.	

\item 3. If $d=1$ and $d'=2$, there exist $2r+2$ internally disjoint paths between $u$ and $v$ with length at most $r+s+7$.
	\end{lemma}
\noindent \textbf{Proof.} According to the values of $d$ and $d'$, the following three cases are considered.

Case 1. $d=0$ and $d'=1$.

\begin{figure}[h]
		\centering
		\includegraphics[width=0.5\textwidth]{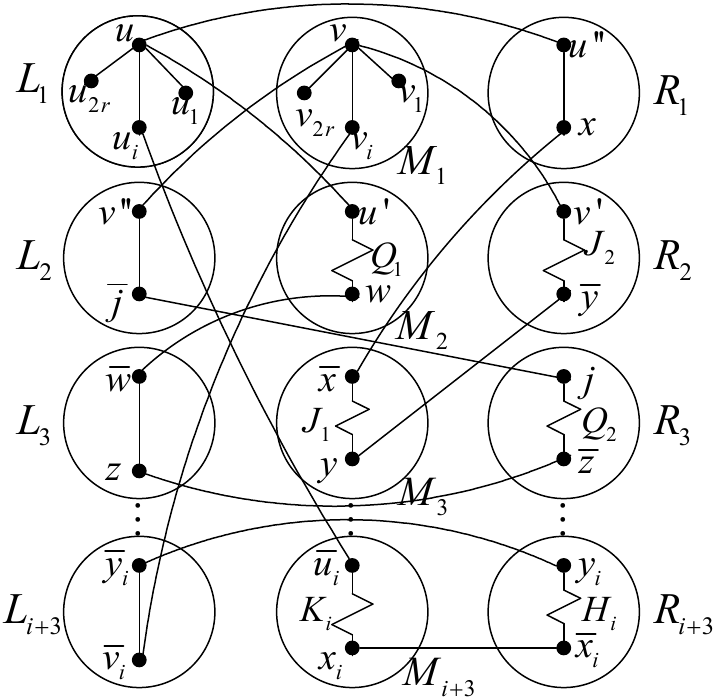}
		\renewcommand{\figurename}{Fig.}
		\caption{The illustration of Case 1 of Lemma \ref{fenlei14}}
		\label{14case1}
	\end{figure}

Without loss of generality, let $u=ABC0\in V(L_{1})$ and $v=A'B'C1\in V(M_{1})$ (see Fig. \ref{14case1}). Define $u_{i}$ as one of the $2r$ neighbors of $u$ in $L_{1}$, where $u_{i}=ABC_{i}0$ for $1\leq i \leq 2r$.
Define $v_{i}$ as one of the $2r$ neighbors of $v$ in $M_{1}$ , where $v_{i}=A'B_{i}'C1$ for $1\leq i \leq 2r$. Let $\bar{u}_{i}=ABC_{i}1\in V(M_{i+3})$ be an outside neighbor of $u_{i}$ for $1\leq i \leq 2r$.  Select $x_{i}=AB_{i}'C_{i}1$ in $M_{i+3}$ for $1\leq i \leq 2r$. Let $K_{i}$ be a shortest path between $\bar{u}_{i}$ and $x_{i}$ in $M_{i+3}$ for $1\leq i \leq 2r$. Let $\bar{x}_{i}=AB_{i}'C_{i}2$, which is an outside neighbor of $x_{i}$ in $R_{i+3}$ for $1\leq i \leq 2r$. Choose $y_{i}=A'B_{i}'C_{i}2$ in $R_{i+3}$ for $1\leq i \leq 2r$. Then there exists a shortest path $H_{i}$ between $\bar{x}_{i}$ and $y_{i}$ in $R_{i+3}$ for $1\leq i \leq 2r$. Let $\bar{y}_{i}=A'B_{i}'C_{i}0$, which is an outside neighbor of
$y_{i}$ in $L_{i+3}$ for $1\leq i \leq 2r$.  $v_{i}$ has an outside neighbor $\bar{v}_{i}=A'B_{i}'C0$ also in $L_{i+3}$ for $1\leq i \leq 2r$. Note that $\bar{v}_{i}$ is adjacent to $\bar{y}_{i}$ in $L_{i+3}$ for $1\leq i \leq 2r$. Construct the path $P_{i}$ as $u\rightarrow u_{i}\rightarrow \bar{u}_{i}\rightarrow K_{i}\rightarrow x_{i}\rightarrow \bar{x}_{i}\rightarrow H_{i}\rightarrow y_{i}\rightarrow \bar{y}_{i}\rightarrow \bar{v}_{i}\rightarrow v_{i}\rightarrow v$ for $1\leq i \leq 2r$. Then $|P_{i}|=|K_{i}|+|H_{i}|+7\leq D(M_{i+3})+D(R_{i+3})+7=r+s+7$ for $1\leq i \leq 2r$. Let $u'=ABC1$ be the neighbor of $u$ in $M_{2}$. Select $w=AB'C1$ in $M_{2}$. Let $Q_{1}$ be a shortest path between $u'$ and $w$ in $M_{2}$. Note that $w$ has an outside neighbor $\bar{w}=AB'C0$ in $L_{3}$. Then $\bar{w}$ has a neighbor  $z=AB'C_{1}0\in V(L_{3})$, whose outside neighbor is $\bar{z}=AB'C_{1}2\in V(R_{3})$. Select $j=A'B'C_{1}2$ in $R_{3}$. Then there exists a shortest path $Q_{2}$ between $\bar{z}$ and $j$ in $R_{3}$. Let $\bar{j}=A'B'C_{1}0$, which is an outside neighbor of $j$ in $L_{2}$. $v$ has an outside neighbor $v''=A'B'C0$ also in $L_{2}$. Note that $v''$ is adjacent to $\bar{j}$ in $L_{2}$. Define the path $P_{2r+1}$ as $u\rightarrow u'\rightarrow Q_{1}\rightarrow w\rightarrow \bar{w}\rightarrow z\rightarrow \bar{z}\rightarrow Q_{2}\rightarrow j\rightarrow \bar{j}\rightarrow v''\rightarrow v$. This satisfies $|P_{2r+1}|=|Q_{1}|+|Q_{2}|+7\leq D(M_{2})+D(R_{3})+7=r+s+7$. Let $u''=ABC2\in V(R_{1})$ be another outside neighbor of $u$. Then $u''$ has a neighbor  $x=A_{1}BC2\in V(R_{1})$, whose outside neighbor is $\bar{x}=A_{1}BC1\in V(M_{3})$. Select $y=A_{1}B'C1$ in $M_{3}$. Let $J_{1}$ be a shortest path between $\bar{x}$ and $y$ in $M_{3}$. Let $\bar{y}=A_{1}B'C2$, which is an outside neighbor of $y$ in $R_{2}$. $v$ has another outside neighbor $v'=A'B'C2$ also in $R_{2}$. There exists a shortest path $J_{2}$ between $v'$ and $\bar{y}$ in $R_{2}$. Construct the path $P_{2r+2}$ as $u\rightarrow u''\rightarrow x\rightarrow \bar{x}\rightarrow J_{1}\rightarrow y\rightarrow \bar{y}\rightarrow J_{2}\rightarrow v'\rightarrow v$. Then $|P_{2r+2}|=|J_{1}|+|J_{2}|+5\leq D(R_{2})+D(M_{3})+5=r+s+5$.

Therefore, we obtain $2r+2$ internally disjoint paths $P_{1},P_{2}, \ldots ,P_{2r+2}$ between
$u$ and $v$ in $E3C(r,s,t)$, with lengths at most $r+s+7$. \\

Case 2. $d=0$ and $d'=2$.

The proof for this case is analogous to Case 1. \\

Case 3. $d=1$ and $d'=2$.

\begin{figure}[h]
		\centering
		\includegraphics[width=0.5\textwidth]{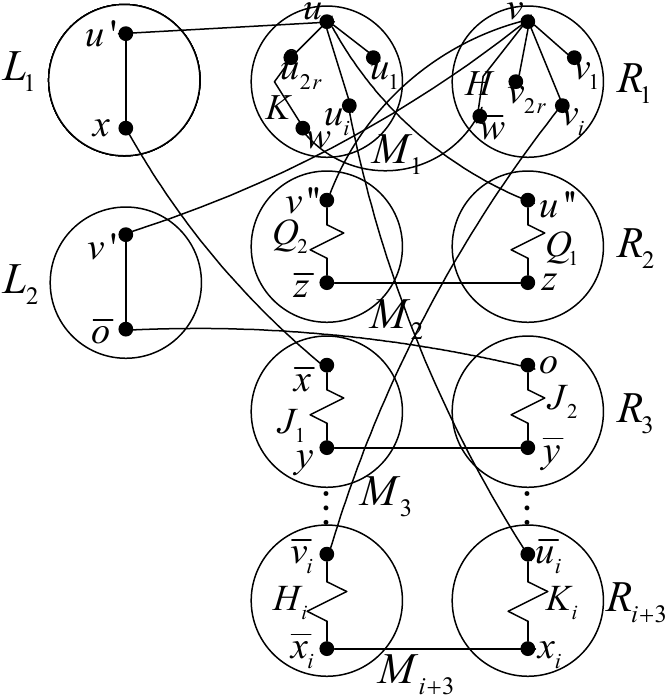}
		\renewcommand{\figurename}{Fig.}
		\caption{The illustration of Case 3 of Lemma \ref{fenlei14}}
		\label{14case3}
	\end{figure}

Without loss of generality, let $u=ABC1\in V(M_{1})$ and $v=A'B'C2\in V(R_{1})$ (see Fig. \ref{14case3}). Let $u_{i}=AB_{i}C1$ be the neighbor of $u$ in $M_{1}$ for $1\leq i \leq 2r$, and $v_{i}=A_{i}'B'C2$ be the neighbor of $v$ in $R_{1}$ for $1\leq i \leq 2r$. Choose $w=AB'C1\in V(M_{1})$. There exists a shortest path $K$ between $u$ and $w$ in $M_{1}$. Then at most one $u_{i}\in V(K)$ for $1\leq i \leq 2r$. Without loss of generality, assume $u_{i}\notin V(K)$ for $1\leq i \leq 2r-1$. Since the outside neighbor $\bar{w}=AB'C2$ of
$w$ also lies in $R_{1}$, there exists a shortest path $H$ between $v$ and $\bar{w}$ in $R_{1}$. Similarly, at most one $v_{i}\in V(H)$ for $1\leq i \leq 2r$. Assume $v_{i}\notin V(H)$ for $1\leq i \leq 2r-1$. Let $\bar{u}_{i}=AB_{i}C2\in V(R_{i+3})$ be an outside neighbor of $u_{i}$ for $1\leq i \leq 2r-1$. Select $x_{i}=A_{i}'B_{i}C2$ in $R_{i+3}$ for $1\leq i \leq 2r-1$. Let $K_{i}$ be a shortest path between $\bar{u}_{i}$ and $x_{i}$ in $R_{i+3}$ for $1\leq i \leq 2r-1$. Note that $x_{i}$ has an outside neighbor $\bar{x}_{i}=A_{i}'B_{i}C1\in V(M_{i+3})$, and $\bar{v}_{i}=A_{i}'B'C1\in V(M_{i+3})$ is the outside neighbor of $v_{i}$ for $1\leq i \leq 2r-1$. Let $H_{i}$ be a shortest path between $\bar{x}_{i}$ and $\bar{v}_{i}$ in $M_{i+3}$ for $1\leq i \leq 2r-1$. Define the path $P_{i}$ as $u\rightarrow u_{i}\rightarrow \bar{u}_{i}\rightarrow K_{i}\rightarrow x_{i}\rightarrow \bar{x}_{i}\rightarrow H_{i}\rightarrow \bar{v}_{i}\rightarrow v_{i}\rightarrow v$ for $1\leq i \leq 2r-1$. Then $|P_{i}|=|H_{i}|+|K_{i}|+5\leq D(M_{i+3})+D(R_{i+3})+5=D(Q_{s}^{3})+D(Q_{r}^{3})+5=r+s+5$ for $1\leq i \leq 2r-1$. Define the path $P_{2r}$ as $u\rightarrow K\rightarrow w\rightarrow \bar{w}\rightarrow H\rightarrow v$. Then $|P_{2r}|=|H|+|K|+1\leq D(M_{1})+D(R_{1})+1=D(Q_{s}^{3})+D(Q_{r}^{3})+1=r+s+1$.
Assume $u''=ABC2\in V(R_{2})$ and $v''=A'B'C1\in V(M_{2})$. Choose $z=A'BC2$ in $R_{2}$. Let $Q_{1}$ be a shortest path between $u''$ and $z$ in $R_{2}$. $z$ has an outside neighbor $\bar{z}=A'BC1$ also in $M_{2}$. Then there exists a shortest path $Q_{2}$ between $v''$ and $\bar{z}$ in $M_{2}$.  Let $u'=ABC0\in V(L_{1})$ with neighbor $x=ABC_{1}0\in V(L_{1})$, whose outside neighbor is $\bar{x}=ABC_{1}1\in V(M_{3})$. Select $y=AB'C_{1}1$ in $M_{3}$. Then there exists a shortest path $J_{1}$ between $\bar{x}$ and $y$ in $M_{3}$. Note that $y$ has an outside neighbor $\bar{y}=AB'C_{1}2$ in $R_{3}$. Select $o=A'B'C_{1}2$ in $R_{3}$. Let $J_{2}$ be a shortest path between $o$ and $\bar{y}$ in $R_{3}$. $o$ has an outside neighbor $\bar{o}=A'B'C_{1}0$ in $L_{2}$. $v$ has another outside neighbor $v'=A'B'C0$ also in $L_{2}$. Note that $v'$ is adjacent to $\bar{o}$ in $L_{2}$. Construct the path $P_{2r+1}$ as $u\rightarrow u''\rightarrow Q_{1}\rightarrow z\rightarrow \bar{z}\rightarrow Q_{2}\rightarrow v''\rightarrow v$. Then $|P_{2r+1}|=|Q_{1}|+|Q_{2}|+3\leq D(R_{2})+D(M_{2})+3=r+s+3$. Construct the path $P_{2r+2}$ as $u\rightarrow u'\rightarrow x\rightarrow  \bar{x}\rightarrow J_{1}\rightarrow y\rightarrow \bar{y}\rightarrow J_{2}\rightarrow o\rightarrow  \bar{o}\rightarrow v'\rightarrow v$. Then $|P_{2r+2}|=|J_{1}|+|J_{2}|+7\leq D(R_{3})+D(M_{3})+7=r+s+7$.

Therefore, we obtain $2r+2$ internally disjoint paths $P_{1},P_{2}, \ldots ,P_{2r+2}$ between
$u$ and $v$ in $E3C(r,s,t)$, with lengths at most $r+s+7$. \qed
\\

\begin{lemma} \label{fenlei15}
		Let $u=ABCd$ and $v=A'B'C'd'$ be two distinct vertices in $E3C(r,s,t)$ such that $A\neq A'$, $B\neq B'$, $C\neq C'$ and $d\neq d'$. We discuss three cases.

\item 1. If $d=0$ and $d'=1$, there exist $2r+2$ internally disjoint paths between $u$ and $v$ with length at most $r+s+t+6$.

\item 2. If $d=0$ and $d'=2$, there exist $2r+2$ internally disjoint paths between $u$ and $v$ with length at most $r+s+t+6$.	

\item 3. If $d=1$ and $d'=2$, there exist $2r+2$ internally disjoint paths between $u$ and $v$ with length at most $r+s+t+6$.
	\end{lemma}
\noindent \textbf{Proof.} According to the values of $d$ and $d'$, the following three cases are considered.

Case 1. $d=0$ and $d'=1$.

\begin{figure}[h]
		\centering
		\includegraphics[width=0.5\textwidth]{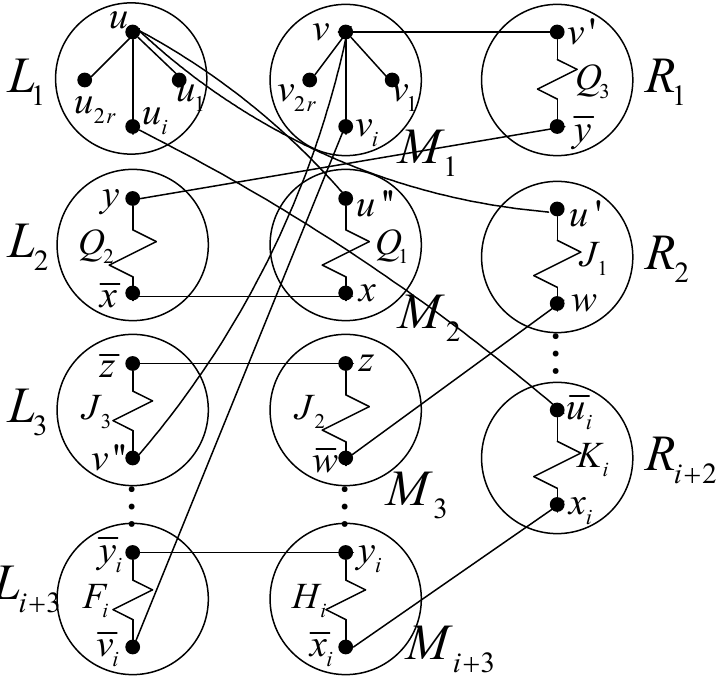}
		\renewcommand{\figurename}{Fig.}
		\caption{The illustration of Case 1 of Lemma \ref{fenlei15}}
		\label{15case1}
	\end{figure}

Without loss of generality, let $u=ABC0\in V(L_{1})$ and $v=A'B'C'1\in V(M_{1})$ (see Fig. \ref{15case1}). Let $u_{i}=ABC_{i}0$ be the neighbor of $u$ in $L_{1}$ for $1\leq i \leq 2r$, and $v_{i}=A'B_{i}'C'1$ be the neighbor of $v$ in $M_{1}$ for $1\leq i \leq 2r$. Let $\bar{u}_{i}=ABC_{i}2\in V(R_{i+2})$ be an outside neighbor of $u_{i}$ for $1\leq i \leq 2r$. Select $x_{i}=A'BC_{i}2$ in $R_{i+2}$ for $1\leq i \leq 2r$. Let $K_{i}$ be a shortest path between $\bar{u}_{i}$ and $x_{i}$ in $R_{i+2}$ for $1\leq i \leq 2r$. Note that $x_{i}$ has an outside neighbor $\bar{x}_{i}=A'BC_{i}1\in V(M_{i+3})$  for $1\leq i \leq 2r$. Choose $y_{i}=A'B_{i}'C_{i}1$ in $M_{i+3}$ for $1\leq i \leq 2r$. Let $H_{i}$ be a shortest path between $\bar{x}_{i}$ and $y_{i}$ in $M_{i+3}$ for $1\leq i \leq 2r$.  Let $\bar{y}_{i}=A'B_{i}'C_{i}0$, which is an outside neighbor of $y_{i}$ in $L_{i+3}$ for $1\leq i \leq 2r$. $v_{i}$ has an outside neighbor $\bar{v}_{i}= A'B_{i}'C'0$ also in $L_{i+3}$ for $1\leq i \leq 2r$. There exists a shortest path $F_{i}$ between $\bar{y}_{i}$ and $\bar{v}_{i}$ in $L_{i+3}$ for $1\leq i \leq 2r$. Define the path $P_{i}$ as $u\rightarrow u_{i}\rightarrow \bar{u}_{i}\rightarrow K_{i}\rightarrow x_{i}\rightarrow \bar{x}_{i}\rightarrow H_{i}\rightarrow y_{i}\rightarrow \bar{y}_{i}\rightarrow F_{i}\rightarrow \bar{v}_{i}\rightarrow v_{i}\rightarrow v$ for $1\leq i \leq 2r$. Then $|P_{i}|=|K_{i}|+|H_{i}|+|F_{i}|+6\leq D(R_{i+2})+D(M_{i+3})+D(L_{i+3})+6=D(Q_{r}^{3})+D(Q_{s}^{3})+D(Q_{t}^{3})+6=r+s+t+6$ for $1\leq i \leq 2r$. Let $u'=ABC2$ be the neighbor of $u$ in $R_{2}$. Select $w=A'BC2$ in $R_{2}$. Let $J_{1}$ be a shortest path between $u'$ and $w$ in $R_{2}$. Note that $w$ has an outside neighbor $\bar{w}=A'BC1$ in $M_{3}$. Select $z=A'B'C1$ in $M_{3}$. Let $J_{2}$ be a shortest path between $\bar{w}$ and $z$ in $M_{3}$.  Let $\bar{z}= A'B'C0$, which is an outside neighbor of $z$ in $L_{3}$. By Lemma \ref{lemE3Clianbian}, $v$ has an outside neighbor $v''=A'B'C'0$ also in $L_{3}$. There exists a shortest path $J_{3}$ between $\bar{z}$ and $v''$ in $L_{3}$.  Construct the path $P_{2r+1}$ as $u\rightarrow u'\rightarrow J_{1}\rightarrow w\rightarrow \bar{w}\rightarrow J_{2}\rightarrow z\rightarrow \bar{z}\rightarrow J_{3}\rightarrow v''\rightarrow v$. Then $|P_{2r+1}|=|J_{1}|+|J_{2}|+|J_{3}|+4\leq D(R_{2})+D(M_{3})+D(L_{3})+4=r+s+t+4$.  Let $u''=ABC1\in V(M_{2})$ be another outside neighbor of $u$. Select $x=AB'C1$ in $M_{2}$. Let $Q_{1}$ be a shortest path between $u''$ and $x$ in $M_{2}$. Note that $x$ has an outside neighbor $\bar{x}=AB'C0$ in $L_{2}$. Select $y=AB'C'0$ in $L_{2}$. Let $Q_{2}$ be a shortest path between $\bar{x}$ and $y$ in $L_{2}$.  Let $\bar{y}=AB'C'2$, which is an outside neighbor of $y$ in $R_{1}$. By Lemma \ref{lemE3Clianbian}, suppose another outside neighbor
$v'=A'B'C'2$ of $v$ is also in $R_{1}$.  Then there exists a shortest path $Q_{3}$ between $v'$ and $\bar{y}$ in $R_{1}$. Construct the path $P_{2r+2}$ as $u\rightarrow u''\rightarrow Q_{1}\rightarrow x\rightarrow \bar{x}\rightarrow Q_{2}\rightarrow y\rightarrow \bar{y}\rightarrow Q_{3}\rightarrow v'\rightarrow v$. Then $|P_{2r+2}|=|Q_{1}|+|Q_{2}|+|Q_{3}|+4\leq D(M_{2})+D(L_{2})+D(R_{1})+4=r+s+t+4$.

Therefore, we obtain $2r+2$ internally disjoint paths $P_{1},P_{2}, \ldots ,P_{2r+2}$ between
$u$ and $v$ in $E3C(r,s,t)$, with lengths at most $r+s+t+6$.
\\

Case 2. $d=0$ and $d'=2$.

The proof for this case is analogous to Case 1. \\

Case 3. $d=1$ and $d'=2$.

The proof for this case is analogous to Case 1. \qed
\\

From Lemma \ref{fenlei1} to Lemma \ref{fenlei15}, we obtain $2r+2$ internally disjoint paths
between any two distinct vertices in $E3C(r,s,t)$, all of which have lengths at most $r+s+t+6$ for $1\leq r\leq s\leq t$. Therefore, we obtain the following theorem.

\begin{theorem}\label{thwidediameter}
		
$D_{2r+2}(E3C(r,s,t))\leq r+s+t+6=n+5$ for $1\leq r\leq s\leq t$.
	\end{theorem}	
	
Combining Theorem \ref{th1} and Theorem \ref{thwidediameter}, we obtain the following main result.

	\begin{theorem}\label{thzongjie}
		 $n+3=r+s+t+4\leq D_{2r+1}^{f}(E3C(r,s,t))\leq D_{2r+2}(E3C(r,s,t))\leq r+s+t+6=n+5$ for $1\leq r\leq s\leq t$.

	\end{theorem}

\section{Conclusion}

Fault diameter and wide diameter are two critical parameters for evaluating communication performance in interconnection networks. They measure the fault tolerance and transmission efficiency of such networks. In this paper, we discuss the fault diameter and wide diameter of the exchanged 3-ary $n$-cube $E3C(r,s,t)$ by constructing $2r+2$ internally disjoint paths between any two distinct vertices. We obtain that the ($2r+1$)-fault diameter and ($2r+2$)-wide diameter of $E3C(r,s,t)$ are bounded between $n+3$ and $n+5$ for $1\leq r\leq s\leq t$. These properties demonstrate that $E3C(r,s,t)$ has high fault tolerance and excellent parallel transmission efficiency.


\begin{thebibliography}{99}
		
		\bibitem{Bose+Broeg+Kwon+Ashir} B. Bose, B. Broeg, Y. Kwon, Y. Ashir, Lee Distance and Topological Properties of $k$-ary $n$-cubes, IEEE Trans. Comput. 44 (1995) 1021-1030.
		

        \bibitem{Bondy+Murty} J.A. Bondy, U.S.R. Murty, Graph Theory, Springer, Berlin, 2008.
		
		\bibitem{Chang+Sung+Hsu} C.P. Chang, T.Y. Sung, L.H. Hsu, Edge congestion and topological
properties of crossed cubes, IEEE Trans. Parallel Distrib. Syst. 11 (2000) 64-80.

       \bibitem{Day+Al-Ayyoub} K. Day, A.E. Al-Ayyoub, Fault diameter of $k$-ary $n$-cube networks, IEEE Trans. Parallel Distrib. Syst. 8 (1997) 903-907.


\bibitem{Efe} K. Efe, The crossed cube architecture for parallel computation, IEEE Trans. Parallel Distrib. Syst. 3 (5) (1992) 513-524.
		
        \bibitem{Fu+Chen+Duh}  J.S. Fu, G.H. Chen, D.R. Duh, Node-disjoint paths and related problems on hierarchical cubic networks, Networks 40 (2002) 142-154.

        \bibitem{Flandrin+Li}  E. Flandrin, H. Li, Mengerian properties, hamiltonicity, and clawfree graphs, Networks 24 (1994) 177-183.

         \bibitem{Faudree+Ordman+Shelp+Jacobson}  R.J. Faudree, E.T. Ordman, R.H. Shelp, M.S. Jacobson, Z. Tuza, Menger's theorem and short paths, J. Combin. Math. Combin. Comput 2 (1987) 235-253.

         \bibitem{Hsu}  D.F. Hsu, On container width and length in graphs, groups and networks, IEICE Transactions on Fundamentals of Electronics Communications and Computer Sciences 4 (1994) 668-680.

          \bibitem{Hsu+Lyuu}   D.F. Hsu, Y.D. Lyuu, A graph-theoretical study of transmission delay and fault tolerance, Proceedings of the 4th ISMM International Conference on Parallel and Distributed Computing and Systems (1991) 20-24.

		\bibitem{Hibers+Koopman+Snepscheut} P.A.J. Hibers, M.R.J. Koopman, J.V.D. Snepscheut, The twisted cube, Proc. Conf. Parallel Archit. Lang. Eur. 258 (1) (1987) 152-159.
		
		\bibitem{Han+You+Lin+Fan} Y.J. Han, L.T. You, C.K. Lin, J.X. Fan, Communication performance evaluation of the locally twisted cube, Int. J. Found. Comput. S. 31 (2) (2020) 233-252.
		
		
        \bibitem{Krishnamoorthy+Krishnamurthy}  M.S. Krishnamoorthy, B. Krishnamurthy, Fault diameter of interconnection networks, Comput. Math. Appl. 13 (1987) 577-582.

		 \bibitem{Latifi}  S. Latifi, Combinatorial analysis of fault-diameter of the $n$-cube, IEEE Trans. Comput. 42 (1993) 27-33.

         \bibitem{Lou+Hsu+Pan} P.K.K. Lou, W.J. Hsu, Y. Pan, The exchanged hypercube, IEEE Trans. Parallel Distrib. Syst. 16 (9) (2005) 866-874.
		
		\bibitem{Lv+Lin+Wang} Y.L. Lv, C.K. Lin, G.J. Wang, An exchanged 3-ary $n$-cube interconnection network for parallel computation, Int. J. Found. Comput. Sci. 32 (3) (2021) 235-252.


\bibitem{Li+Mu+Li+Min} K.Q. Li, Y.P. Mu, K.Q. Li, G.Y. Min, Exchanged crossed cube: a novel interconnection network for parallel computation, IEEE Trans. Parallel Distrib. Syst. 24 (11) (2013) 2211-2219.
		
		\bibitem{Ning+Guo}W.T. Ning, L.T. Guo, Connectivity and super connectivity of the exchanged 3-ary $n$-cube, Theor. Comput. Sci.  923 (2022) 160-166.

\bibitem{Niu+Zhou+Tian+Zhang}  B.H. Niu, S.M. Zhou, T. Tian, Q.F. Zhang, The wide diameter and
fault diameter of exchanged crossed cube, Int. J. Found. Comput. Sci. (2023) 1-17.
	
		
	    \bibitem{Qi+Zhu}  H. Qi, X.D. Zhu, The fault-diameter and wide-diameter of twisted
hypercubes, Discrete Appl. Math. 235 (2018) 154-160.
		
		\bibitem{Rouskov+Srimani}  Y. Rouskov, P.K. Srimani, Fault diameter of star graphs, Inf. Process. Lett. 48 (1993) 243-251.

        \bibitem{Tsai+Chen+Tan}  T.H. Tsai, Y.C. Chen, J.J.M. Tan, Topological properties on the
wide and fault diameters of exchanged hypercubes, IEEE Trans. Parallel Distrib. Syst. 25 (2014) 3317-3327.

        \bibitem{Wu+Chen+Kuo+Chang}  R.Y. Wu, G.H. Chen, Y.L. Kuo, G.J. Chang, Node-disjoint paths in
hierarchical hypercube networks, Inf. Sci. 177 (2007) 4200-4207.


        \bibitem{Yin+Li+Chen+Zhong}   J.H. Yin, J.S. Li, G.L. Chen, C. Zhong, On the fault-tolerant diameter and wide diameter of $\omega$-connected graph, Networks 45 (2005) 88-94.




	\end{thebibliography}
\end{document}